\documentclass[11pt]{article}
\usepackage{graphicx,latexsym}
\usepackage{amssymb}
\usepackage{cite}
\newcommand{\Treal}{\ensuremath{T_\mathrm{real}}}
\newcommand{\Tadj}{\ensuremath{T_\mathrm{adj}}}
\newcommand{\NAS}{\ensuremath{N_\mathrm{EAS}}}
\newcommand{\Ncl}{\ensuremath{N_\mathrm{cl}}}
\newcommand{\Padj}{\ensuremath{\mathcal{P}_\mathrm{adj}}}
\newcommand{\Nexp}{\ensuremath{N_\mathrm{exp}}}

\begin{document}
\title{%
     \begin{flushright}
          \small
          SINP MSU 2002-9/693 \\
          \texttt{astro-ph/0203478; v.3}
	\end{flushright}
	\bigskip\bigskip
	\textbf{
     Clusters of EAS with electron number $\mathbf{\gtrsim10^4}$}
}

\author{%
     Yu.~A. Fomin, G. V. Kulikov, M. Yu.~Zotov%
     \footnote{\texttt{fomin}, \texttt{kulikov},
     \texttt{zotov@eas.sinp.msu.ru}}\\[3mm]
     \slshape
     Ultrahigh Energy Particles Laboratory \\
     \slshape
     D. V. Skobeltsyn Institute of Nuclear Physics \\
     \slshape
     M. V. Lomonosov Moscow State University \\
     \slshape
     Moscow 119992, Russia
}
\date{v.1: March 27, 2002; v.2: April 9, 2002; v.3: March 4, 2003}
\maketitle
\begin{abstract}
\sloppy
     We perform cluster analysis of arrival times of extensive air
     showers (EAS) registered with the \hbox{EAS--1000} Prototype array
     in the period from August, 1997 till February, 1999.
     We present twenty cluster events each consisting of one or
     several EAS clusters, study the dynamics of their
     development, and analyze the angular distribution of EAS in
     clusters.
     We find that there may be certain correlation between EAS
     clusters with mean electron number~$\sim10^5$ on the one
     hand and gamma-ray bursts and ultrahigh energy cosmic rays
     on the other.
\end{abstract}


\section{Introduction}

     There are a considerable number of articles devoted
     to the analysis of arrival times of extensive air showers (EAS)
     (see, e.g.,~\cite{Jap91,Jap93,Jap95,Jap01,Ochi:NC,Dubna,Hamburg,Taro}
     and references therein).
     In this paper, we address this problem using the experimental
     data obtained with the EAS--1000 Prototype array.
     This array operates at the Skobeltsyn Institute of Nuclear
     Physics of Moscow State University.
     It consists of eight detectors situated in the central
     part of the EAS MSU array along longer sides of the
     $64 \mbox{ m}\times22$~m rectangle.
     The array registers EAS generated by cosmic rays with
     the energy of a primary particle more than
     $10^{14}$~eV~\cite{Durban,SLC}.

     For the purposes of the present investigation, we selected
     203 days of regular operation of the array in
     the period from August~30, 1997 till February~1, 1999.
     The total EAS number in this data set equals 1~668~489.
     The mean interval between consecutive EAS arrival times is
     equal to 10.5~sec.
     The discreteness in the moments of EAS registration is
     approximately 0.055~sec (one tic of the PC clock).
     Parameters of EAS were found using a program written by
     V.~P.~Sulakov~\cite{preprint}.
     The mean electron number in the EAS under consideration
     is of the order of $1.2\times10^5$ particles.

     As we have mentioned above, this paper is mainly devoted
     to the analysis of EAS arrival times.
     Certain results of our investigation have already been
     presented briefly in~\cite{Dubna, Hamburg,IzvRAN01}.
     Here we shall go into some details concentrating on EAS clusters.

     One may study arrival times either as a sequence of consecutive
     moments of time $t_1$, $t_2$,\dots,~$t_n$, where~$t_i$ is the
     moment of registration of the $i$th EAS, or as a sequence
     $x_1$, $x_2$,\dots,~$x_n$ of time intervals (``delays")
     between consecutive EAS:  $x_i = t_i - t_{i-1}$, $i=1,\dots,n$,
     where $t_0=0$ corresponds to midnight.
     Since we are interested in EAS clusters, let us discuss their
     possible formal definitions.

\newpage
     One can define an EAS cluster at least in two different ways:
\begin{enumerate}
\item
     a cluster is a group of consecutive EAS in which all~$x_i$ are
     less than some predefined value~$\xi$ which in turn is much less
     than the mean delay $\bar X = \frac1n \sum_1^n x_i$;

\item
     a cluster is a group of $k$ consecutive EAS such that the full
     ``length" $L_{j,k} = x_j + x_{j+1} + \cdots + x_{j+k-1}$
     of this group is much less than the expected length
     that is equal to~$k \bar X$.
\end{enumerate}

     When one uses any of these definitions, a natural question
     appears: How one should choose~$\xi$ or~$k$?
     In general, as soon as the mean delay~$\bar X$ is known,
     one has considerable freedom in the choice of~$\xi$ or~$k$.
     Actually, if one uses the second definition, then it is
     possible to select clusters as groups such that
     $L_{j,k}\leqslant k \bar X/a$, $a=2$,~$3$,~$4$,\dots{}
     But this way makes it difficult to estimate the probability of
     appearance of the corresponding cluster since the statistical
     distribution of~$x_i$ is not known a~priori.
     Thus it seems more reasonable to study the distribution of~$x_i$
     first.

     As is well known, the vast majority of statistical methods aimed for
     time series analysis are developed for stationary time series.
     Recall that the most common definition of a stationary process found
     in textbooks (often called strong stationarity) is that all
     conditional probabilities are constant in time (see,
     e.g.,~\cite{Anderson,BP}).
     Our previous investigation of the distribution of EAS arrival times
     has revealed that in general case time series obtained during a
     sufficiently long period of time are not
     stationary~\cite{Dubna,Hamburg}.
     We have also found that the main reason that makes
     our time series non-stationary is the barometric effect.
	Namely, it was found that the mean number of EAS registered in a
     time unit is inversely proportional to the atmospheric pressure.
     This dependence can be approximately expressed by a simple formula:
\[
        \ln N = -\beta P + \mathrm{const},
\]
	where $N$ is the number of EAS registered in a time unit,
     $P$ is the atmospheric pressure, mm~Hg, and~$\beta$ is the
	barometric coefficient~\cite{SLC,preprint}.
     For the EAS--1000 Prototype array, the value of~$\beta$ varies in
     the interval (1.08--1.24)$\times10^{-2}$ depending on the concrete
     data sample and chosen time bin.
     For our data set, the statistical error in the value of~$\beta$ is
     of the order of~$10^{-4}$.
     It is easy to see that $\beta=1.08\times10^{-2}$ is the most
     ``pessimistic" for a search of clusters since it gives the
	highest intensity of EAS registration for a fixed atmospheric
	pressure.

     As we have already shown, almost all samples of sufficient
     length in our data set are stationary and satisfy the
     homogeneity hypothesis if we take into account the barometric
     effect, though certain deviations have been observed for
     some samples~\cite{Dubna,Hamburg}.
     If one considers homogeneous samples, then it is possible to point
     out an interval of time delays for which one may accept
     a hypothesis that time delays have exponential distribution.
     In other words, for such a set of time delays the number of EAS
     registered in a unit of time (``the count rate")
     obeys the Poisson distribution.
	This hypothesis has been verified on the basis of
	$\chi^2$--criterion.

	On the other hand, it is known that different statistical criteria
     can lead to different conclusions (see, e.g.,~\cite{BP}).
     This made us to apply one more criterion to the same data set
     to verify the above hypothesis.
	Namely, we have used a criterion that is based on the variable
\[
     A = \sqrt{\frac{n}{2}}\,
         \left|\frac{S^2/\bar X - 1}{\bar X} \right|,
\]
     which has standard Gaussian distribution $\mathcal{N}(0,1)$
     for~$n$ large enough~\cite{TM}; here
     $S^2 = \frac{1}{n} \sum_1^n (x_i - \bar X)^2$.
	We have calculated~$A$ for different values of~$\beta$ and found
	out that the ``best fit" is achieved for $\beta=1.16\times10^{-2}$.
     For this~$\beta$, $A=0.068$ and one may accept the hypothesis that
     the intensity of EAS registration for data set as a whole obeys the
     Poisson distribution (the probability
     $\mathcal{P}(A<0.08)\approx0.532$, see~\cite{BS}).
	At the same time, the ``pessimistic" value of the barometric
     coefficient $\beta\approx1.08\times10^{-2}$ also allows one to
     accept the hypothesis, since $A=0.070$ for this~$\beta$.
	By this reason, in what follows we shall present the results that
	were obtained with the ``pessimistic" barometric coefficient.


\section{The Main Results}

     To guarantee that the time series under consideration is
     stationary, we have adjusted experimental arrival times to the
     atmospheric pressure $P^* = 742$~mm~Hg which is close to the
     average pressure for the whole analyzed data set.
     This adjustment was made by the following formula:%
\footnote{In the first version of the paper, there was
a typo in this formula.}
\[
     x_i = x_i^0 \exp [\, \beta(P^* - P_i)],
\]
     where $x_i^0$ is the experimental time delay, and
     $P_i$ is the atmospheric pressure at the $i$th delay,
     $i=1, 2,\dots,n$.
	The mean delay for~$P^*$ is equal to 10.4~sec.

     Since we have accepted the hypothesis that the distribution of
     the number of EAS registered in a time unit
     obeys the Poisson distribution with the intensity
     $\lambda\approx5.77$~min$^{-1}$, the following algorithm
     of cluster selection was accepted:
\begin{enumerate}
\item
     Fix some sufficiently small number $\epsilon>0$ and some
     duration~$T$ of the time interval.

\item
     Basing on the known intensity~$\lambda$ of the Poisson process,
     find~$K$ such that the probability to register more than~$K$
     showers in time~$T$ is of the order of~$\epsilon$.
     Recall that this probability is given by the formula
\[
     \mathcal{P}(N_T>K) = 1 - \sum_{k=0}^K \frac{(\lambda T)^k}{k!},
\]
     where $N_T$ is the number of EAS registered in time~$T$.

\item
     A sequence of EAS that is registered in a period of time less or
     equal than~$T$ and contains more than~$K$ showers is selected as
     a cluster.

\end{enumerate}
     We emphasize that the above hypothesis is not important for our
     results, but only allows us to estimate the probability of
     appearance of a cluster.

     For the given criterion of cluster selection, we performed
     a search for EAS clusters using GNU Octave~\cite{Octave}
     running in Mandrake Linux.
     The search was arranged in the following way.
     Fix the initial moment of time $t_0 = 0$, add the chosen time
     interval~$T$ to it and count the number~$N_T$ of EAS registered
     in time~$T$.
     If $N_T > K$, then take the set of EAS that belongs to~$T$
     as a cluster; otherwise skip it.
     After this, move the left end of the time interval~$T$
     to~$t_1$ (the arrival time of the first shower in the data
     set) and repeat the analysis.
     The procedure goes on until the end of the data set.
     Notice that this procedure does not exclude the appearance
     of intersecting clusters.
     This is possible when one cluster belongs to the time
     interval $[t_i, t_i+T]$, another one belongs to $[t_j, t_j+T]$,
     and $t_j<t_i+T$.

     One may certainly choose another strategy of cluster
     search.
     For example, one can move along the data set by a sequence of
     steps with a fixed length.
     But in this case there is a chance to loose a sufficiently dense
     cluster that is divided between two intervals.
     One may also spread time intervals~$T$ over the time series in a
     random way.
     This procedure seems to be accurate, but it needs a considerable
     amount of computer time to be performed.

     Parameters~$T$ and~$K$ that were used for the selection of EAS
     clusters are given in Table~1.
     These values of parameters has lead to the discovery of eighteen
     cluster events---sequences of EAS that may be identified as
     clusters.
     Below we shall also discuss two events that were chosen for~$K$
     which was less by a unit than the values given in Table~1.
     We have added them to the list because they demonstrate
     certain interesting features.
     Table~1 also contains an expected number of observed clusters
\[
     \Nexp = \frac{1}{T} \times
             (\mbox{duration of the observation period})
             \times \mathcal{P}(N_T>K).
\]

\newpage
{\narrower
\noindent
Table 1: Parameters of selection of EAS clusters and the expected
number of corresponding clusters in a 203-day period~($\Nexp$).

}

\begin{center}
\begin{tabular}{|c|c|c|c||c|c|c|c|}
\hline
     $T$, sec  &    $K$  &    $\mathcal{P}(N_T>K)$     & $\Nexp$ &
     $T$, min  &    $K$  &    $\mathcal{P}(N_T>K)$     & $\Nexp$ \\
\hline
5   &  7   & $4.6\times10^{-8}$  &0.16&  2   &  30   & $1.5\times10^{-6}$&0.22\\
10  &  9   & $7.8\times10^{-8}$  &0.14&  3   &  39   & $2.1\times10^{-6}$&0.20\\
15  &  11  & $4.5\times10^{-8}$  &0.05&  4   &  48   & $1.8\times10^{-6}$&0.13\\
20  &  12  & $1.3\times10^{-7}$  &0.11&  5   &  57   & $1.2\times10^{-6}$&0.07\\
25  &  13  & $2.6\times10^{-7}$  &0.18&  6   &  65   & $1.3\times10^{-6}$&0.06\\
30  &  15  & $7.4\times10^{-8}$  &0.04&  7   &  73   & $1.3\times10^{-6}$&0.05\\
35  &  16  & $1.1\times10^{-7}$  &0.06&  8   &  81   & $1.2\times10^{-6}$&0.04\\
40  &  17  & $1.4\times10^{-7}$  &0.06&  9   &  89   & $1.0\times10^{-6}$&0.03\\
45  &  18  & $1.7\times10^{-7}$  &0.07&  10  &  96   & $1.5\times10^{-6}$&0.04\\
50  &  19  & $1.9\times10^{-7}$  &0.07&  11  &  104  & $1.1\times10^{-6}$&0.03\\
55  &  20  & $2.0\times10^{-7}$  &0.06&  12  &  111  & $1.4\times10^{-6}$&0.03\\
60  &  21  & $2.0\times10^{-7}$  &0.06&  13  &  119  & $1.0\times10^{-6}$&0.02\\
65  &  22  & $2.0\times10^{-7}$  &0.05&  14  &  126  & $1.2\times10^{-6}$&0.03\\
70  &  23  & $2.0\times10^{-7}$  &0.05&  15  &  133  & $1.4\times10^{-6}$&0.03\\
75  &  24  & $1.8\times10^{-7}$  &0.04&  16  &  141  & $9.6\times10^{-7}$&0.02\\
80  &  25  & $1.7\times10^{-7}$  &0.04&  17  &  148  & $1.0\times10^{-6}$&0.02\\
85  &  26  & $1.6\times10^{-7}$  &0.03&  18  &  155  & $1.1\times10^{-6}$&0.02\\
90  &  27  & $1.4\times10^{-7}$  &0.03&  19  &  162  & $1.1\times10^{-6}$&0.02\\
    &      &                     &    &  20  &  169  & $1.2\times10^{-6}$&0.02\\
\hline
\end{tabular}
\end{center}
\bigskip

     As one can see, in all cases $\Nexp \ll 1$, i.e., if we have
     strictly Poisson distribution of the count rate, we should
     observe no clusters at all.
     But this is not the case, since we have found nearly twenty events
     that contain one or several (partially) intersecting clusters.
     It is possible to find a number of other EAS clusters if less strong
     conditions of their selection are used.

     For the purposes of this discussion, we have conditionally divided
     all events into two groups defined as follows:
\begin{itemize}
\item[a.]
     Events that consist of a single cluster (see Table 2a).

\item[b.]
     Events that contain several clusters (Tables 2b and 2c).

\end{itemize}
     Let us discuss these two types of events consecutively.

\subsection{Single Clusters}

     As one can see from Table~2a, the first group consists of
     eight events.
     Clusters in this group have different length: from approximately
	5~seconds to more than 4~minutes.
	Let us briefly discuss these events beginning with the shortest one.

\begin{figure}
\centerline{\textbf{Figures}}
\medskip
\centerline{
\includegraphics[width=8.4cm]{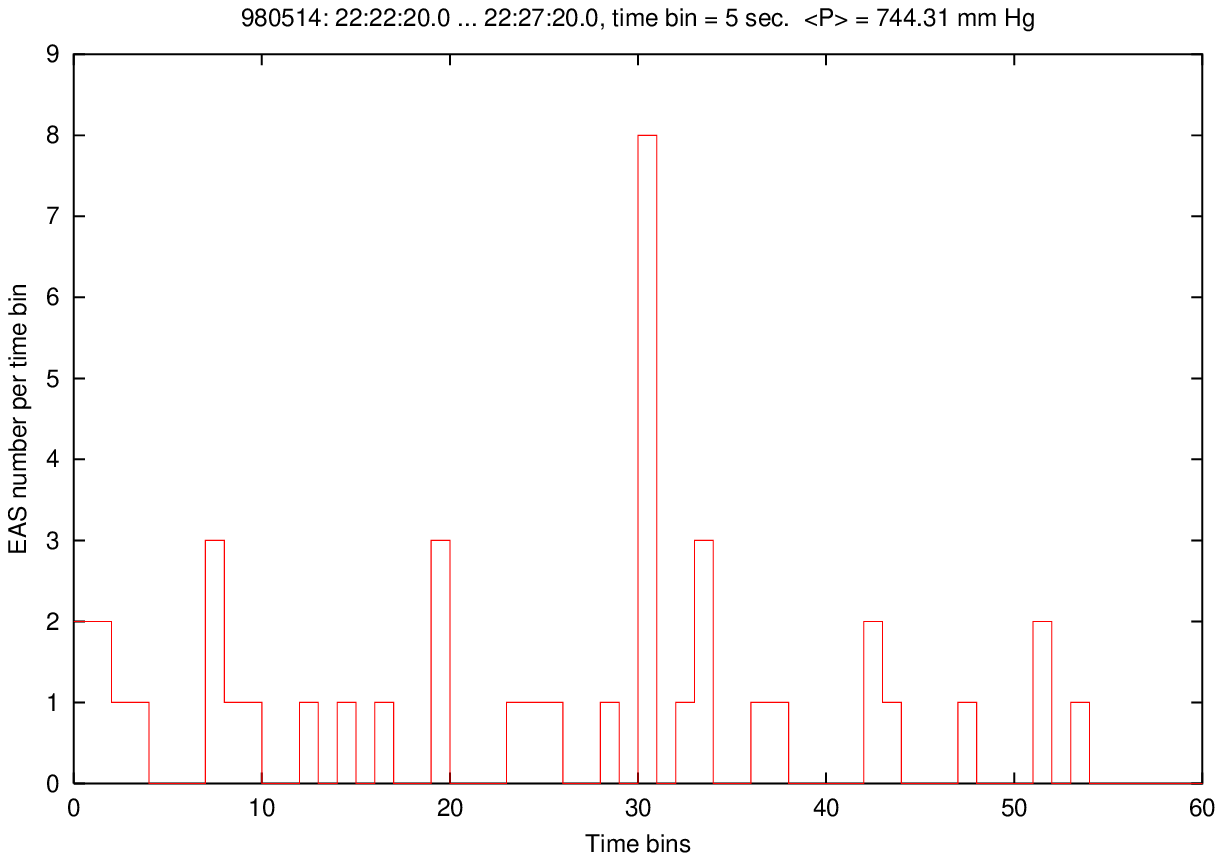}
\includegraphics[width=8.4cm]{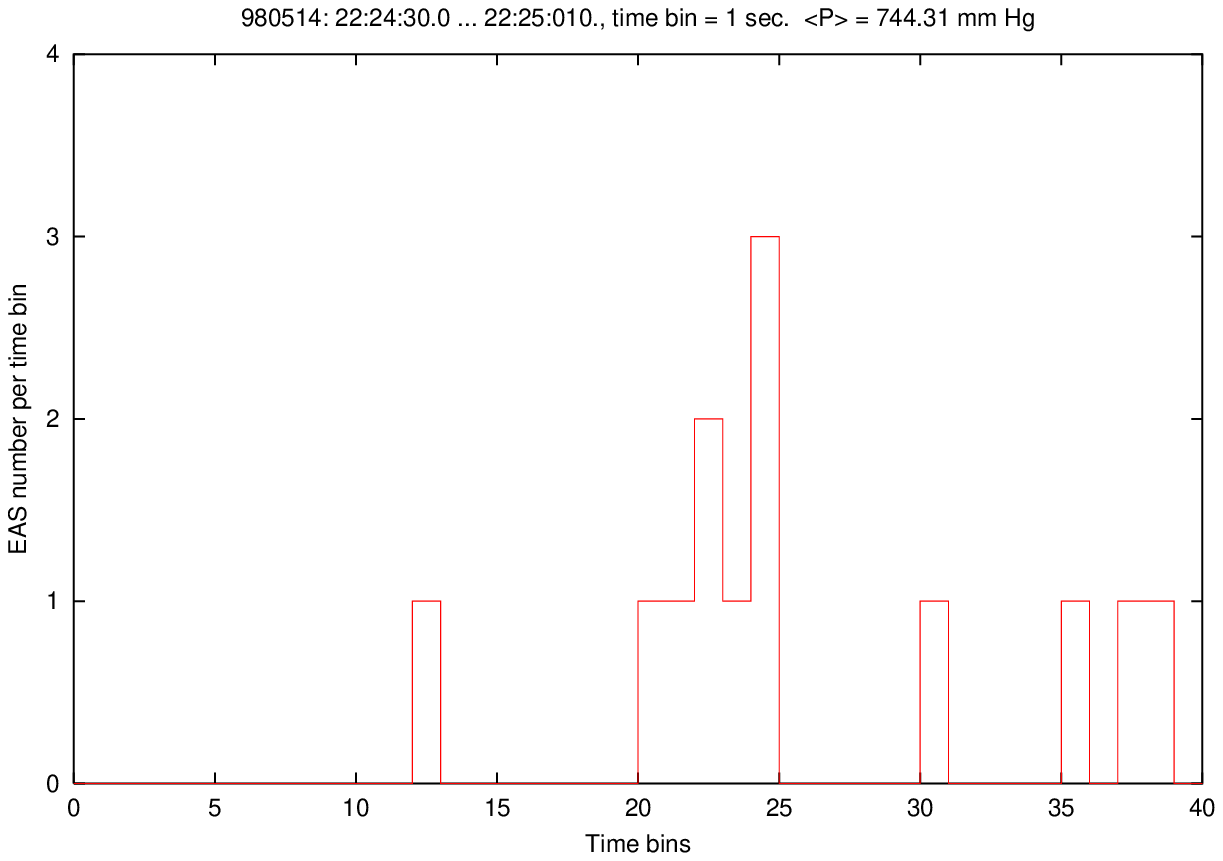}
}
\caption{%
The count rate during an interval that contains a cluster registered
on May~14, 1998 (left), see Table~2a, and the ``interior structure"
of this event (right).
In both cases, the cluster begins in the center of the plot.
Titles of these and the following figures with the count rate
contain the date of the event (yymmdd), the time interval,
the time bin, and the average pressure~$\langle P \rangle$.
}
\label{Fig:980514}
\end{figure}

	Let us begin with a cluster registered on May~14, 1998 (it
     was selected for $T=5$~sec).
	The count rate at an interval that contains this cluster is shown
	in Fig.~\ref{Fig:980514}.
	Notice that the amplitude of the peak that presents the cluster
     is nearly~19 times higher than the average value.
	Remark that though the cluster is very short,
	the distribution of arrival times of EAS that form this cluster
	is not uniform.
	The right plot in Fig.~\ref{Fig:980514} demonstrates that the
	highest count rate is at the end of this event.

\newpage
{\narrower
\noindent
     Table 2a: Single clusters.
     See the parameters of selection in Table~1.
     Notation:
     $P$---the atmospheric pressure;
     \Treal---the real duration of a cluster (for single clusters),~sec;
     \NAS---the number of EAS in a cluster (group of clusters);
     \Ncl---the number of clusters in a group;
     \Tadj---the duration of a cluster adjusted to $P^* = 742$~mm~Hg;
     \Padj---the probability that~\NAS\ are registered in time~\Tadj.
     For events marked by~$*$ in the $T$-column,
     $K$ is less by a unit than the value given in Table~1.
     Moscow local time~$T_M$ is used: $T_M = T_\mathrm{UT} + 3 + \Delta$
     (hours); $T_\mathrm{UT}$ is the UT time, and $\Delta=1$~hour
     during Daylight Saving Time, 0~otherwise.

}
\begin{center}
\begin{tabular}{|c|c|c|c|c|c|c|c|c|c|}
\hline
Date    & $P$,& $T$, &Beginning  &End        &\Treal,&\NAS&\Ncl&\Tadj,&\Padj           \\
dd.mm.yy&mm Hg&  sec &hh:mm:ss   &hh:mm:ss   &sec    &    &    &sec   &                \\
\hline
14.05.98&744.3&  5   &22:24:50.09&22:24:54.70& 4.61 &  8 & 1  &  4.49&$1.9\times10^{-8}$\\
\hline
24.08.98&735.4&  10  &07:09:23.89&07:09:33.07&  9.17& 10 & 1  &  9.86&$6.2\times10^{-8}$\\
\hline
24.12.98&750.5&  10  &04:29:21.91&04:29:30.64&  8.73& 10 & 1  &  7.97&$8.9\times10^{-9}$\\
\hline
06.01.99(a)&735.1& 25  &13:51:57.23&13:52:19.64&22.41 & 14 & 1  & 24.16&$1.5\times10^{-7}$\\
\hline
28.12.98&740.8&45$^*$&15:22:33.18&15:23:17.34& 44.16& 18 & 1  & 44.74&$5.4\times10^{-7}$\\
\hline
05.10.97&717.0& 180  &20:38:00.60&20:40:17.75&137.15& 40 & 1  &179.90&$1.2\times10^{-6}$\\
\hline
03.01.99&748.6& 180  &15:36:06.89&15:39:19.73&192.84& 40 & 1  &179.49&$1.2\times10^{-6}$\\
\hline
16.08.98&744.0& 240  &02:09:42.38&02:13:46.58&244.19& 49 & 1  &239.07&$8.8\times10^{-7}$\\
\hline
\end{tabular}
\end{center}
\bigskip

	Now let us consider longer clusters.

\begin{figure}[!t]
\centerline{
\includegraphics[width=8.4cm]{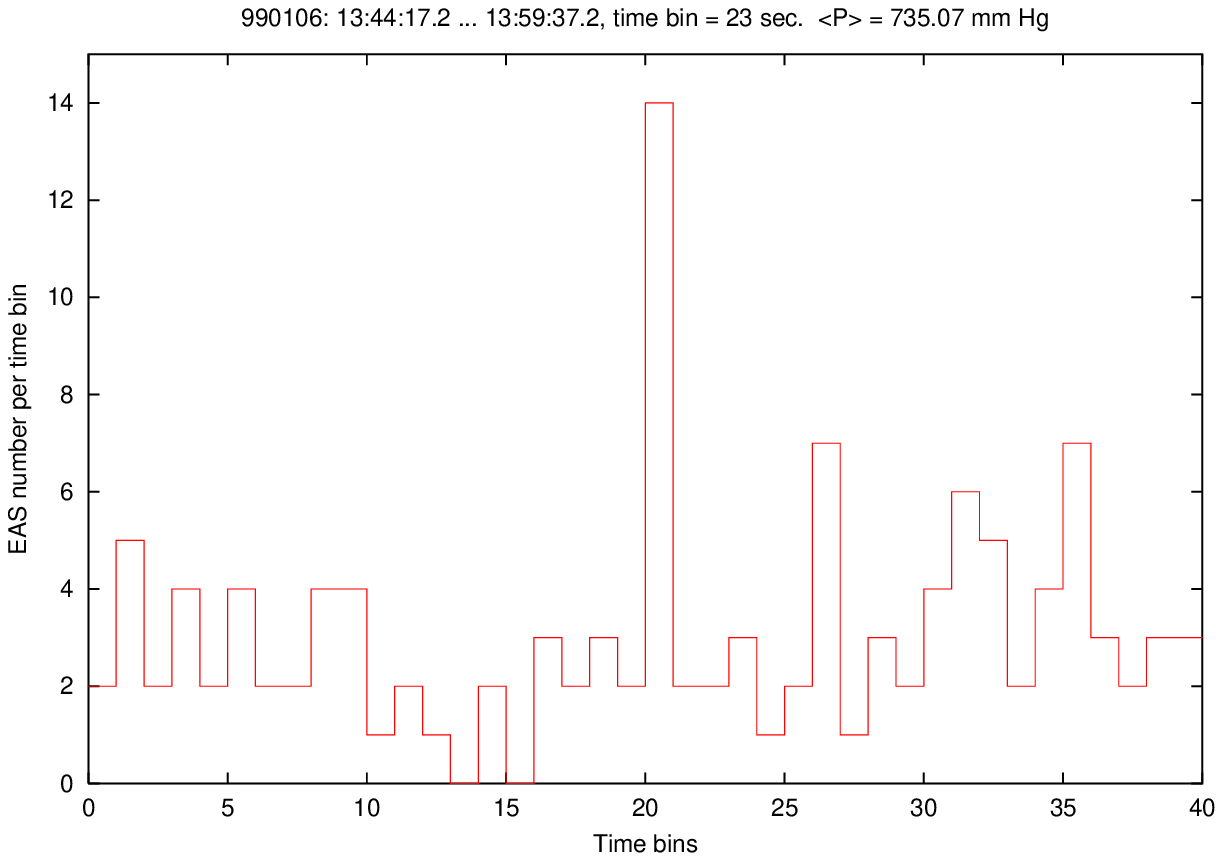}
\includegraphics[width=8.4cm]{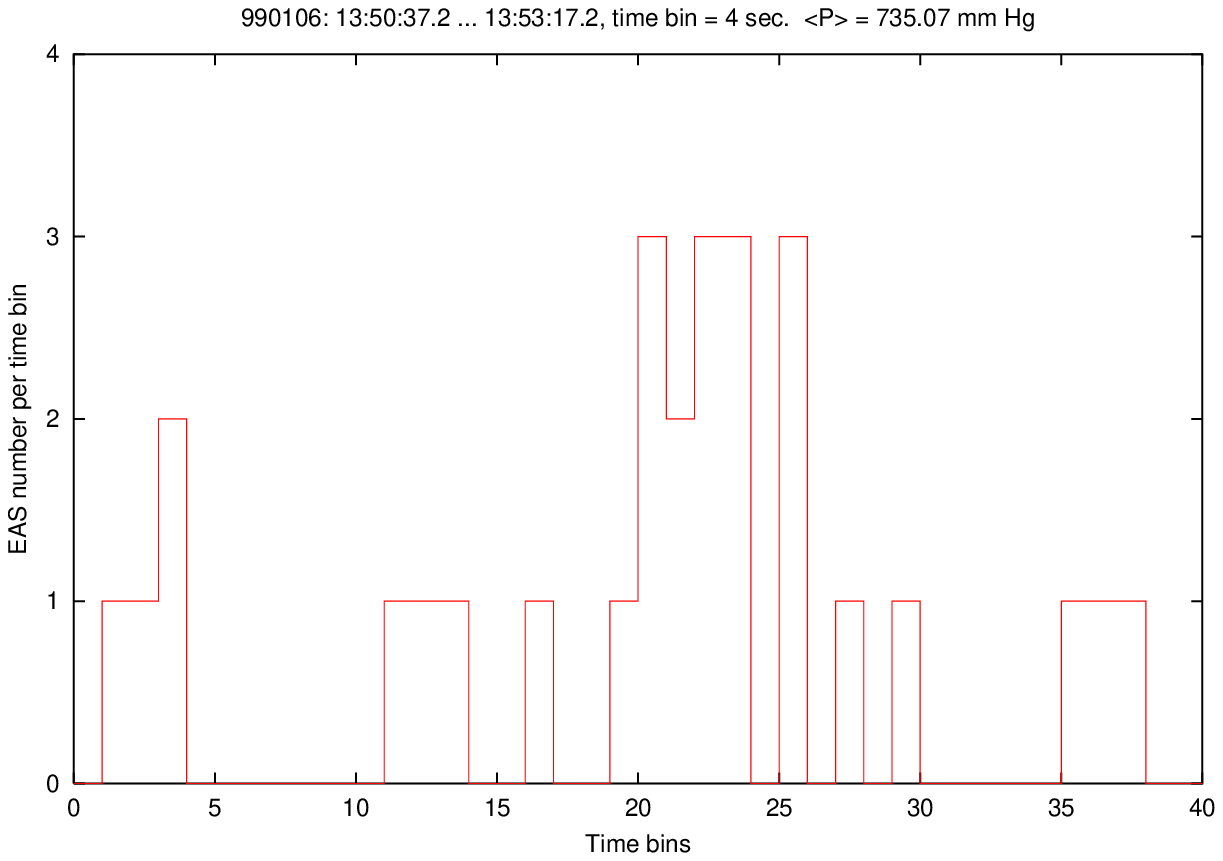}
}
\caption{%
The count rate during time intervals that contain a cluster registered
on January~6, 1999(a), see Table~2a.
The right plot shows the 4-second structure of the event (bins No.~21--26).
}
\label{Fig:990106a}
\end{figure}

\begin{figure}[!h]
\centerline{
\includegraphics[width=8.4cm]{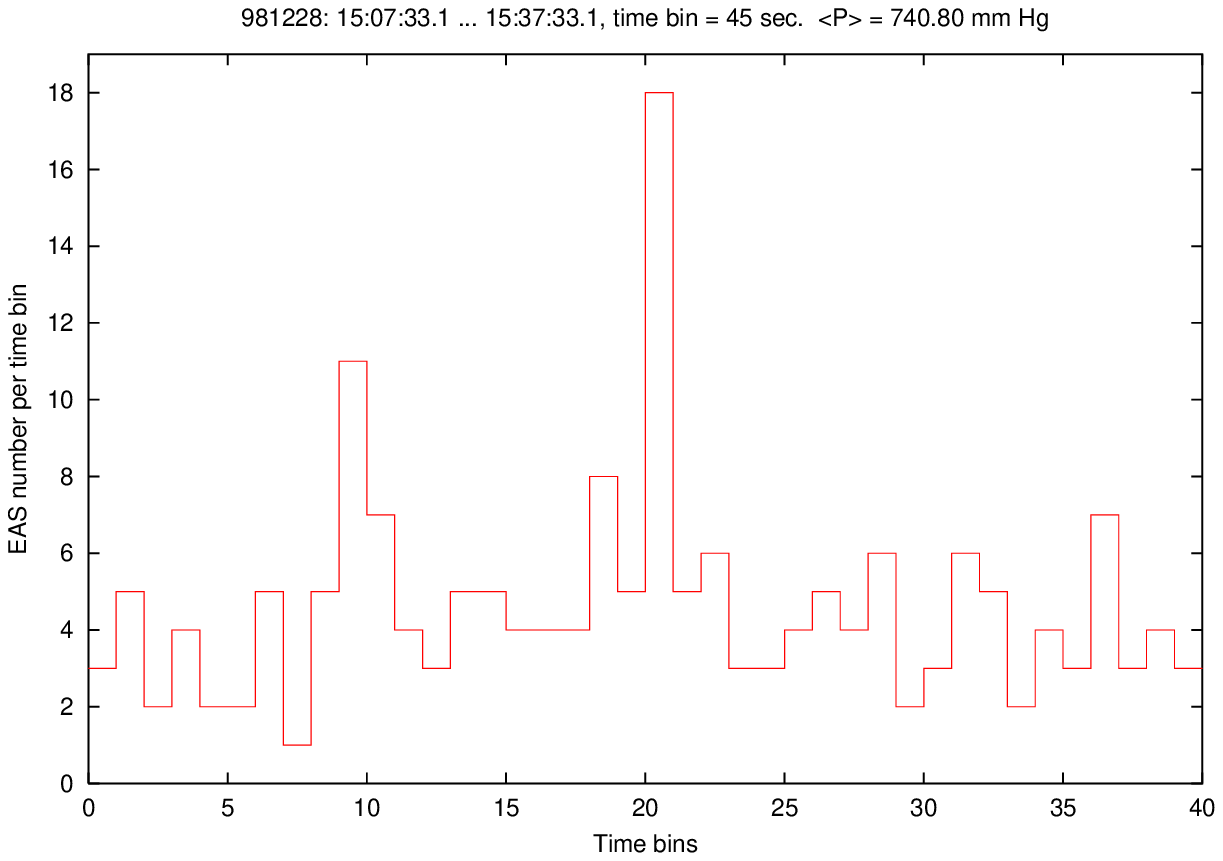}
\includegraphics[width=8.4cm]{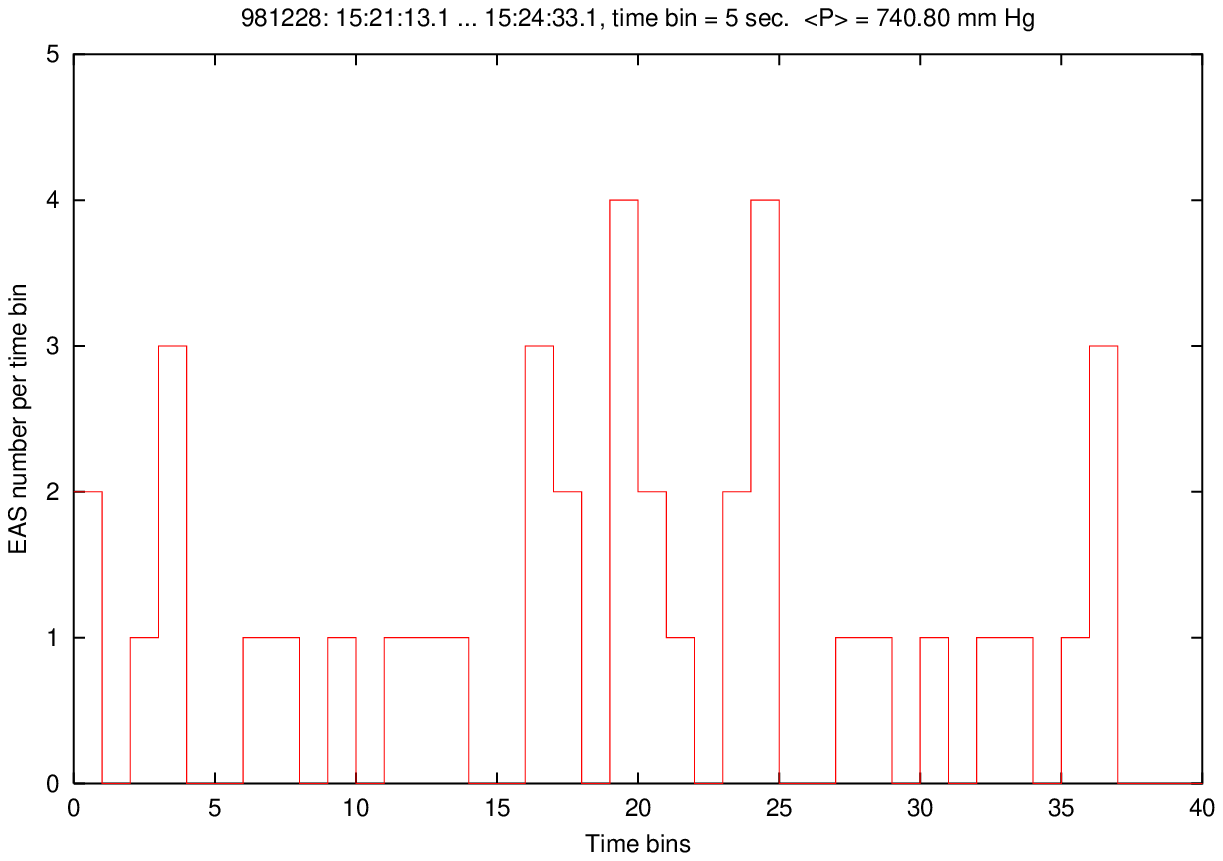}
}
\caption{The count rate at an interval that contains
a cluster registered on December~28, 1998, see Table~2a.
The right plot shows the 5-second structure of the cluster
(bins No.~17--25).
}
\label{Fig:981228}
\end{figure}

\begin{itemize}
\item
	\textbf{24.08.98} and \textbf{24.12.98}:
	Both clusters were chosen for $T=10$~sec.
	They have close duration, consist of an equal number of EAS, and
     have a very small probability of appearance.
	The distribution of arrival times of EAS in these clusters
	is qualitatively the same: the highest count rate is observed
     at the first half of the events, though the amplitude of the
     highest peaks is different, see Fig.~\ref{Fig:980824,981224}.
     It is interesting to notice that if we slightly weaken the
     criterion of cluster search by taking $K=6$ instead of $K=7$ for
     $T=5$~sec, we find a shorter cluster inside the cluster selected
     on 24.08.98.
     For this cluster, 7~showers arrive in 4.06~sec with
     $\Tadj = 4.37$~sec and $\Padj = 3.0\times10^{-7}$.

\item
     \textbf{06.01.99(a)}:
	This cluster is approximately two times longer than two
	previous clusters, see Table~2a and Fig.~\ref{Fig:990106a}.
	It has an interesting ``interior structure" if we look at it
	using 4-second bins.
	Namely, one can see a ``hole" near the end of the cluster:
	there is a 4-second time interval, during which no EAS have arrived.
	We shall see below that such ``holes" are quite usual for
	sufficiently long clusters.

\item
     \textbf{28.12.98}:
	This event was chosen for $T=45$~sec with $K=17$ instead of $K=18$,
	see Table~1.
     We have added it to the list of events because of its possible
     correlation with some other astrophysical events, see
     Sec.~\ref{sec:disc}.
	The cluster has an interesting ``interior structure" if one
	considers 5-second time bins: there are two bins with zero
     count rate and two bins of equally high count rate equal to
     4~EAS per 5~sec, see Fig.~\ref{Fig:981228}.

\end{itemize}

     While it is obvious that our selection procedure could not lead
     to a discovery of a cluster embedded into a 5-second cluster,
     it may be a bit surprising that clusters which cover several minutes
     do not contain any shorter clusters inside (compare events
     registered on 05.10.97, 16.08.98, and 03.01.99 with
     those presented in Tables~2b and~2c).
     A natural question arises: Do arrival times of EAS that form
     these clusters have the uniform distribution?
     Our analysis reveals that the answer is negative.

\begin{figure}
\begin{center}
\includegraphics[width=9.0cm]{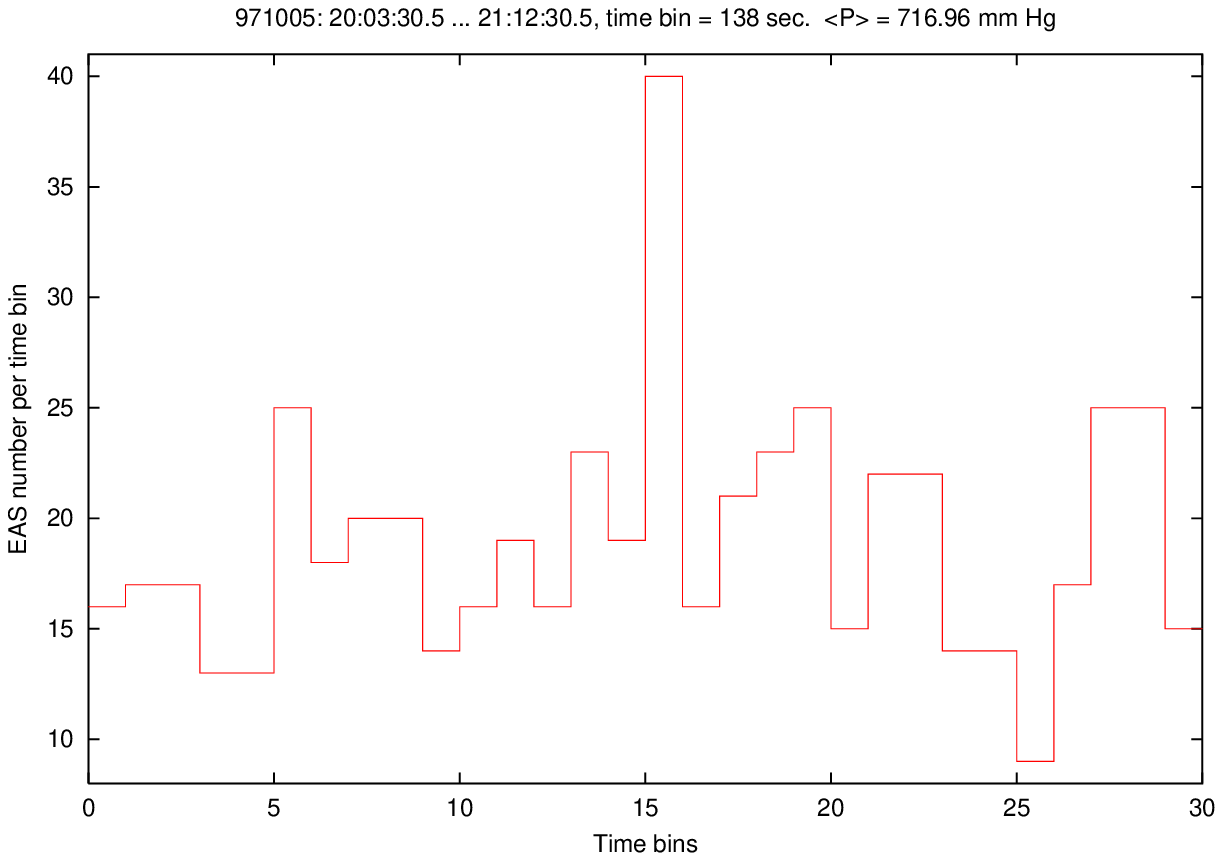}\\
\includegraphics[width=9.0cm]{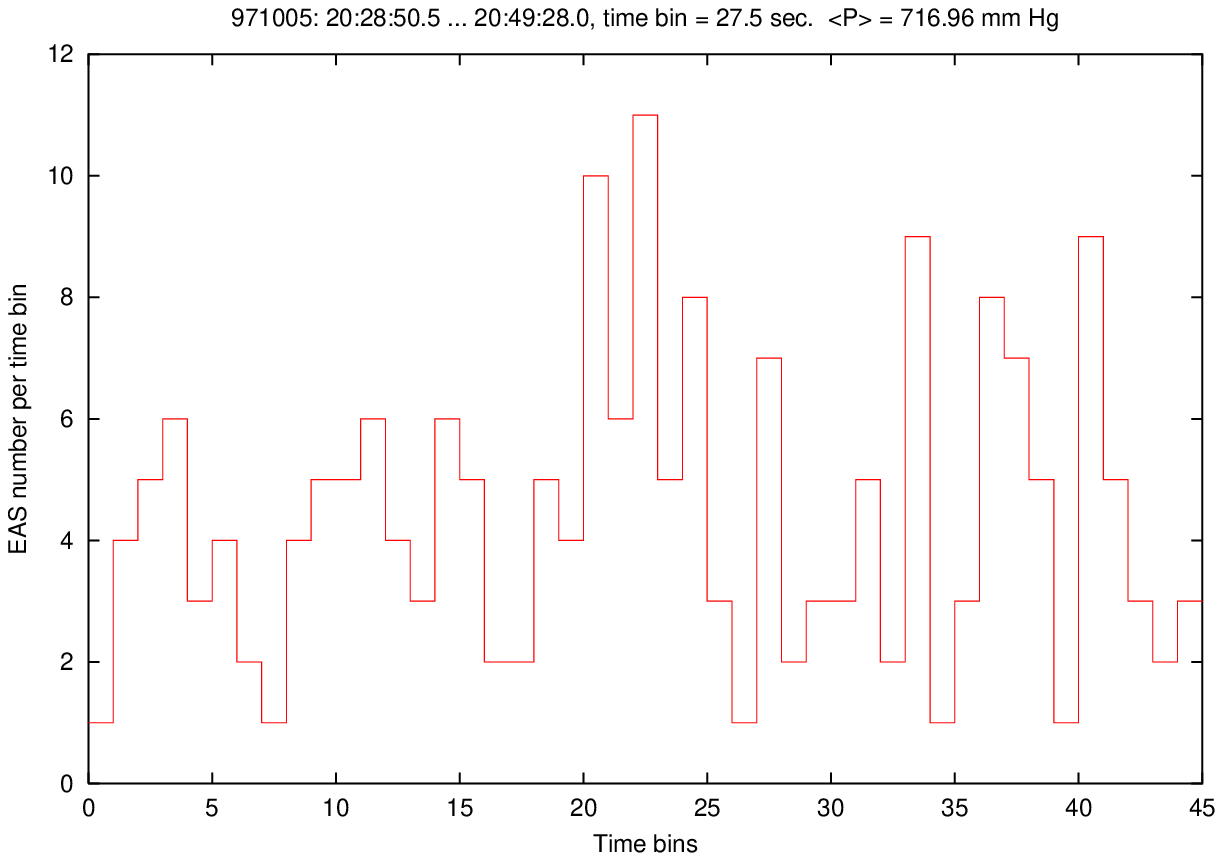}\\
\includegraphics[width=9.0cm]{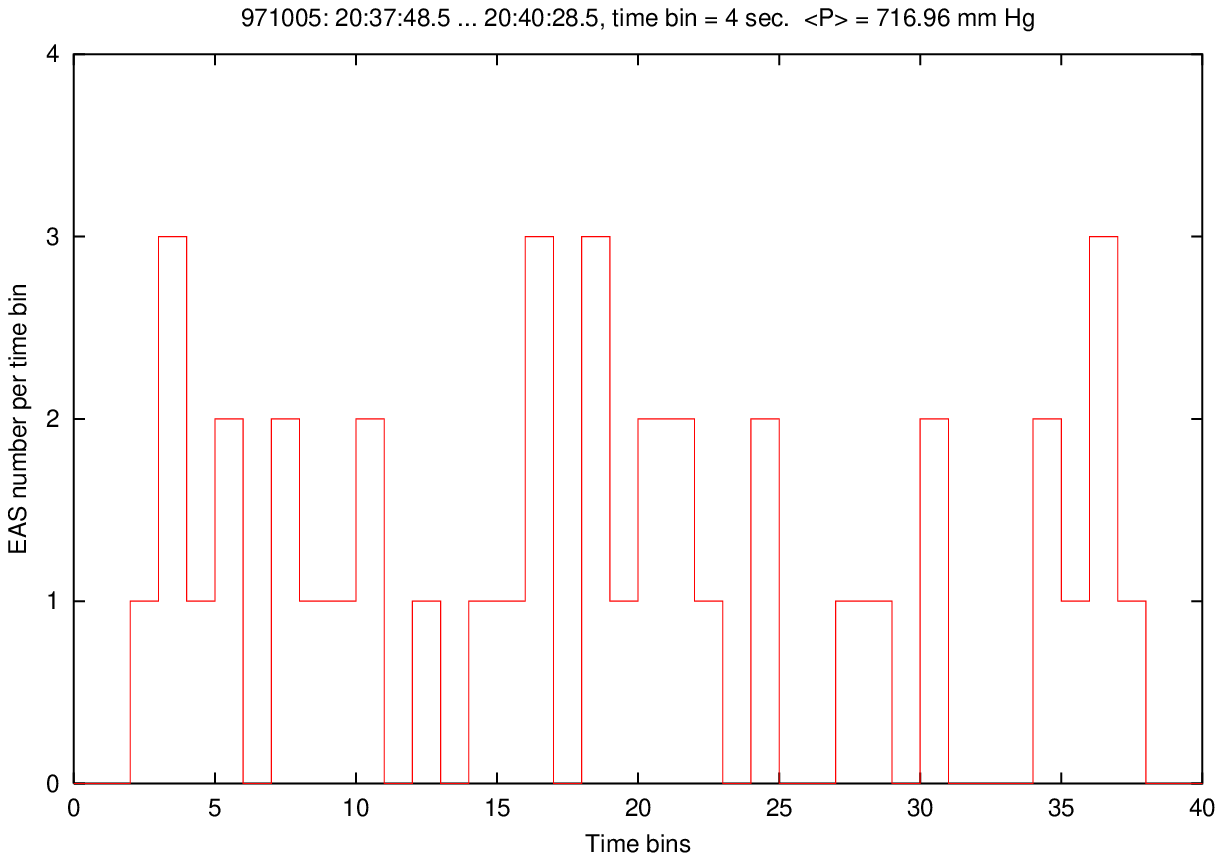}
\end{center}
\caption{The count rate during time intervals that contain
a cluster registered on October~5, 1997, see Table~2a.
Two bottom plots reveal the ``interior structure" of the cluster.
In the middle plot, the cluster occupies bins No.~21--25.
}
\label{Fig:971005}
\end{figure}

\begin{figure}
\begin{center}
\includegraphics[width=9.0cm]{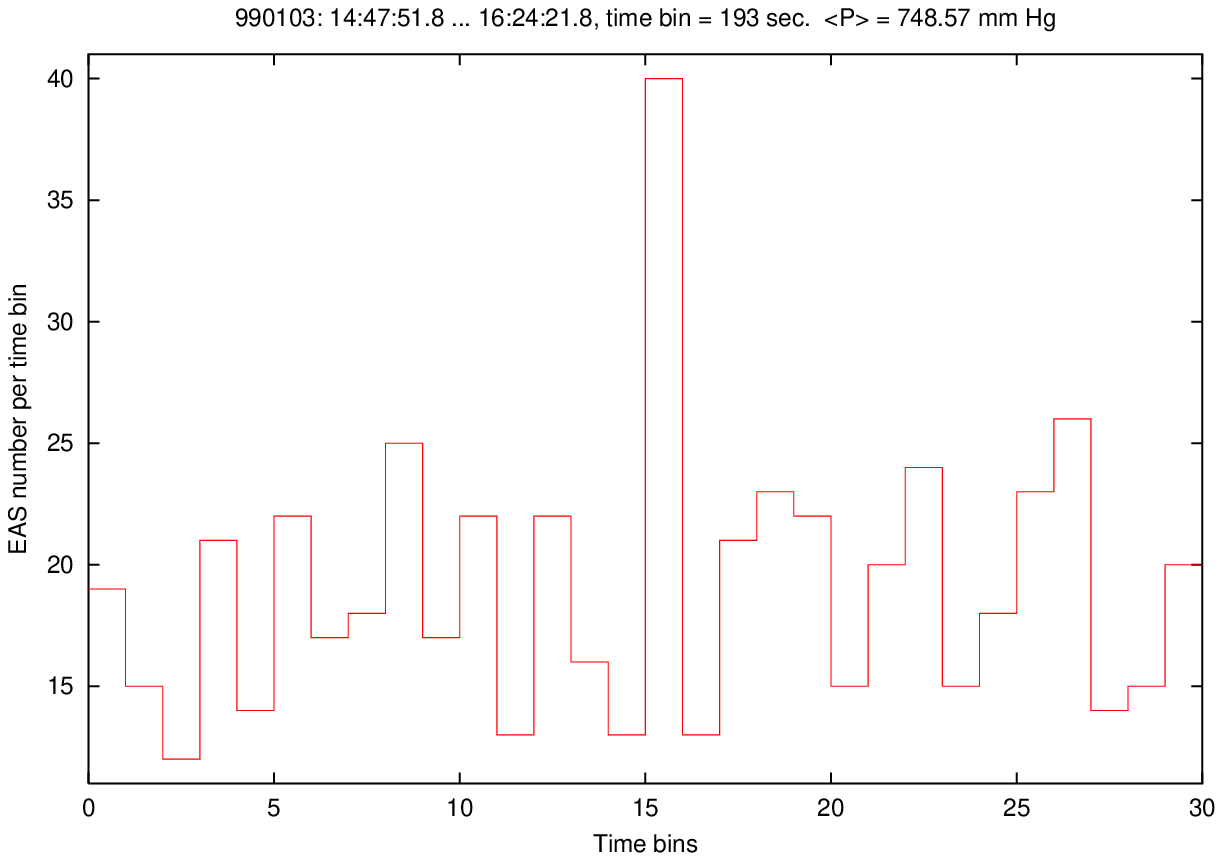}\\
\includegraphics[width=9.0cm]{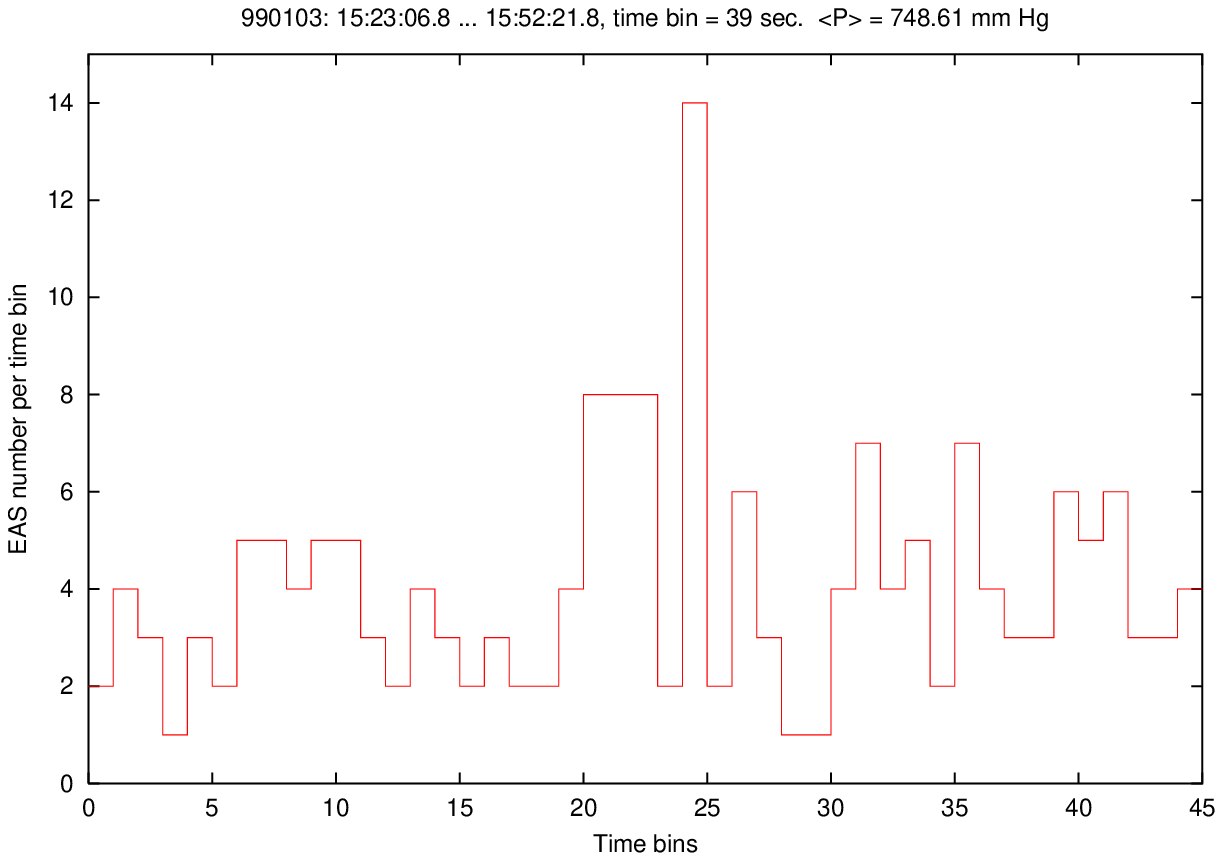}\\
\includegraphics[width=9.0cm]{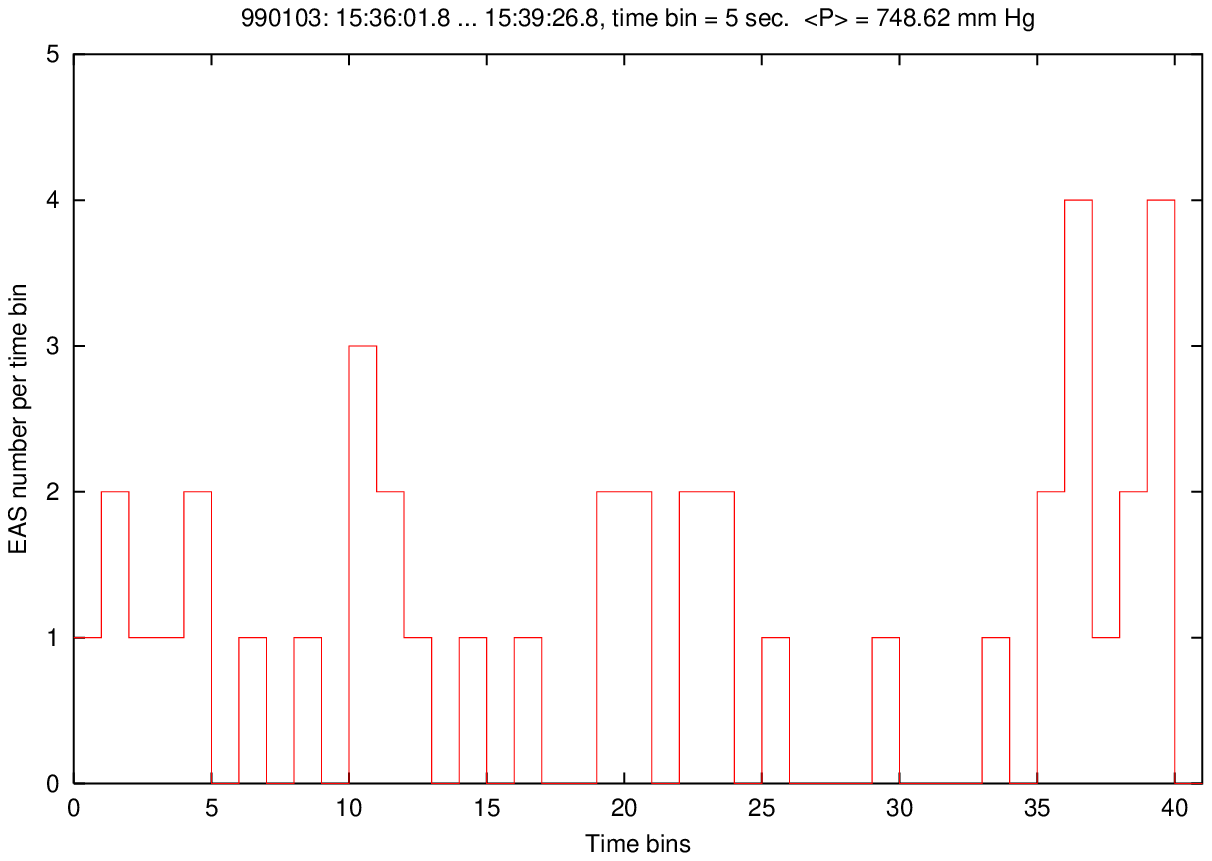}
\end{center}
\caption{The count rate during time intervals that contain
a cluster registered on January~3, 1999, see Table~2a.
Two bottom plots reveal the ``interior structure" of the cluster.
In the middle plot, the cluster occupies bins No.~21--25.
}
\label{Fig:990103}
\end{figure}

\begin{figure}
\begin{center}
\includegraphics[width=9.0cm]{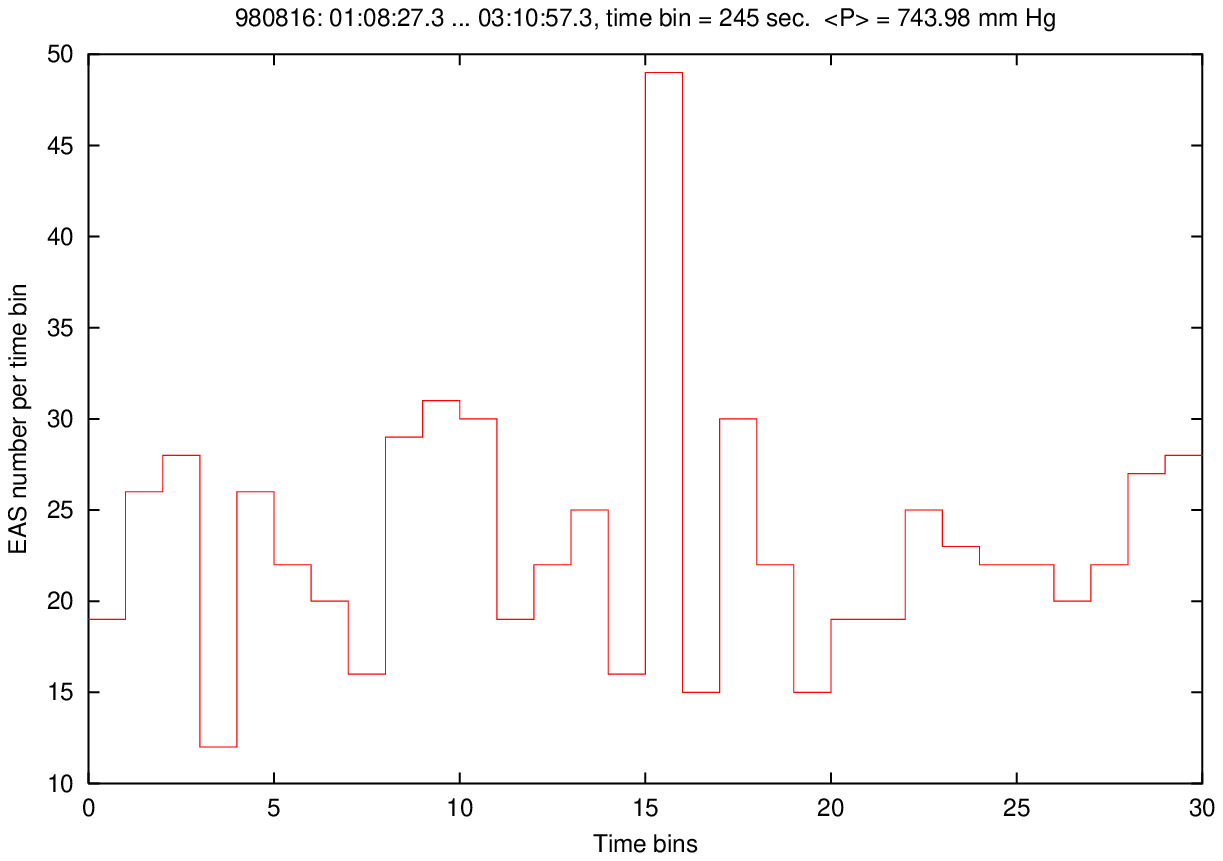}\\
\includegraphics[width=9.0cm]{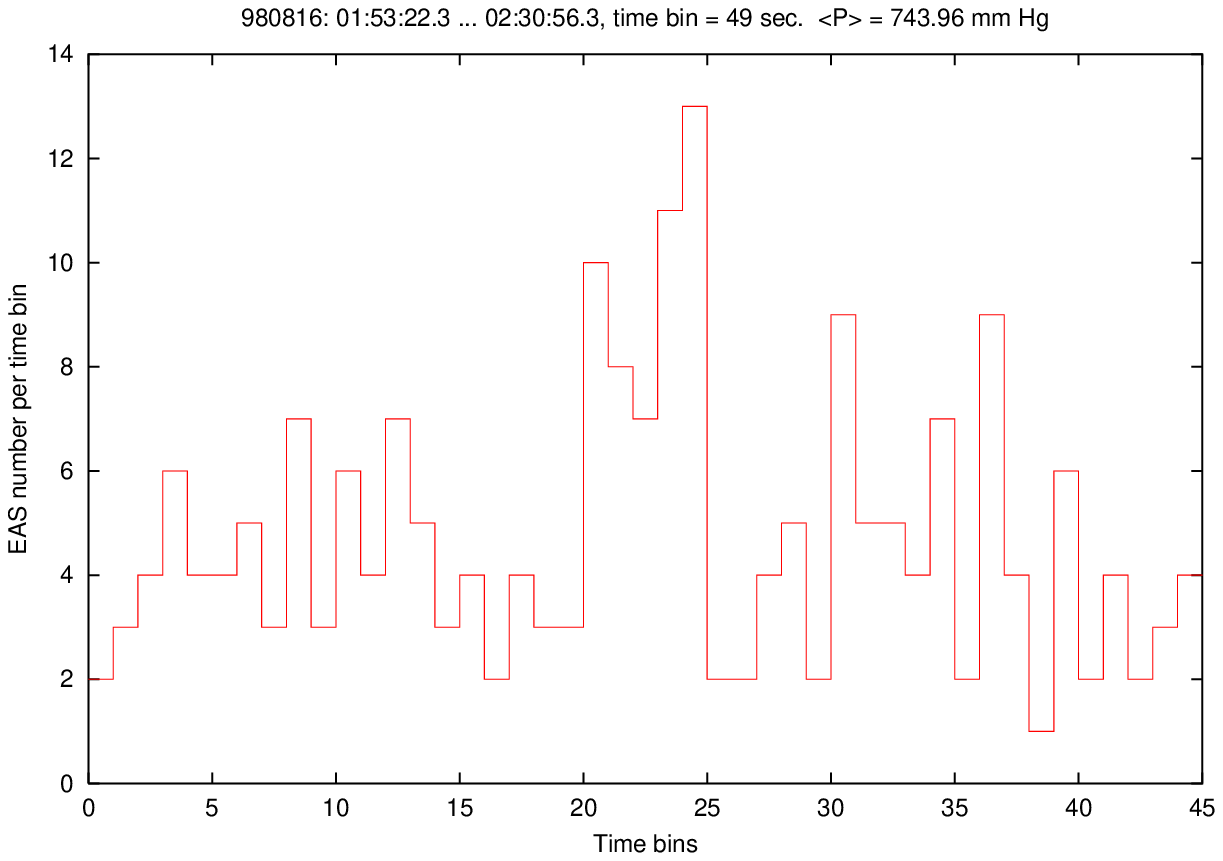}\\
\includegraphics[width=9.0cm]{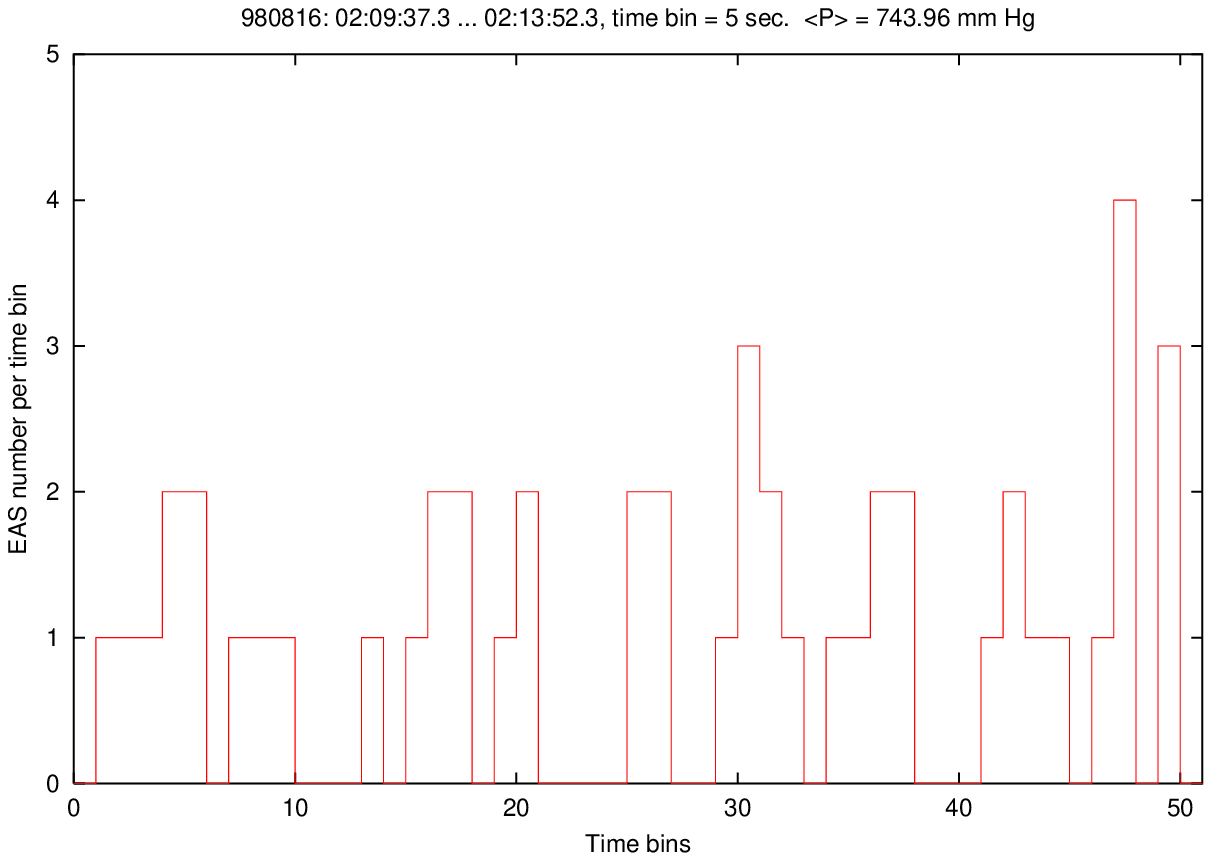}
\end{center}
\caption{The count rate during time intervals that contain
a cluster registered on August~16, 1998, see Table~2a.
Two bottom plots reveal the ``interior structure" of the cluster.
In the middle plot, the cluster occupies bins No.~21--25.
}
\label{Fig:980816}
\end{figure}

     Let us take a look at Figs.~\ref{Fig:971005}, \ref{Fig:990103},
     and~\ref{Fig:980816}.
     These figures depict the structure of the events
     registered on October~5, 1997, January~3, 1999, and August~16, 1998
     respectively.
     It is clearly seen that for all three events one can point out
     short time intervals within a cluster where the count rate
     is the highest.
     Let us consider these events in more details:
\begin{itemize}
\item
     \textbf{05.10.97} and \textbf{03.01.99}:
	Both events were selected for $T=3$~min, see Table~1.
	They contain equal number of showers, but have different~\Treal{}
	because of different atmospheric pressure at the moments of
	registration.
     It is clearly seen from the plots that for these two events,
     distributions of EAS arrival times are drastically different.
     For the first cluster, the distribution of EAS arrival times
	is highly non-uniform, and
     the highest count rate is observed in the middle and
     at the beginning of the event (see Fig.~\ref{Fig:971005}).
	For the second cluster, the situation is opposite:
     the highest count rate takes place at the end of the event
	(see Fig.~\ref{Fig:990103}).
     Both clusters contain high ``sub-peaks," but none of them is
	high enough to be selected as a separate cluster.

\item
     \textbf{16.08.98}:
	This cluster is the longest among single ones.  The count rate
	for this cluster is shown in Fig.~\ref{Fig:980816}.
	It is clearly seen that similar to the above two events,
	this cluster consists of a number of short bursts, though
	none of them is intensive enough to form a separate cluster.
     Fig.~\ref{Fig:980816} also demonstrates that the highest count
     rate for this event takes place in the end of the cluster.
\end{itemize}

     Thus the arrival times of EAS that form a cluster can have
     different distributions.
     More than this, single clusters are not single ``bursts" of the
     count rate with the uniform distribution of EAS arrival times,
     but consist of several ``sub-bursts."
     Hence the reason that we find single clusters is that
     the amplitude of ``sub-bursts" is not high enough to
     satisfy our selection criterion.
     This is true for all events in the first group.

\subsection{Embedded Clusters}

     Now let us turn to the second group of events.
     As one can see from Tables~2b and~2c, this group consists of
     twelve events.
     It is convenient to divide this group into two subgroups
     in accordance with the complexity of the events:

\bigskip
     1. Simple embedding, see Table~2b.

\begin{figure}
\centerline{
\includegraphics[width=8.4cm]{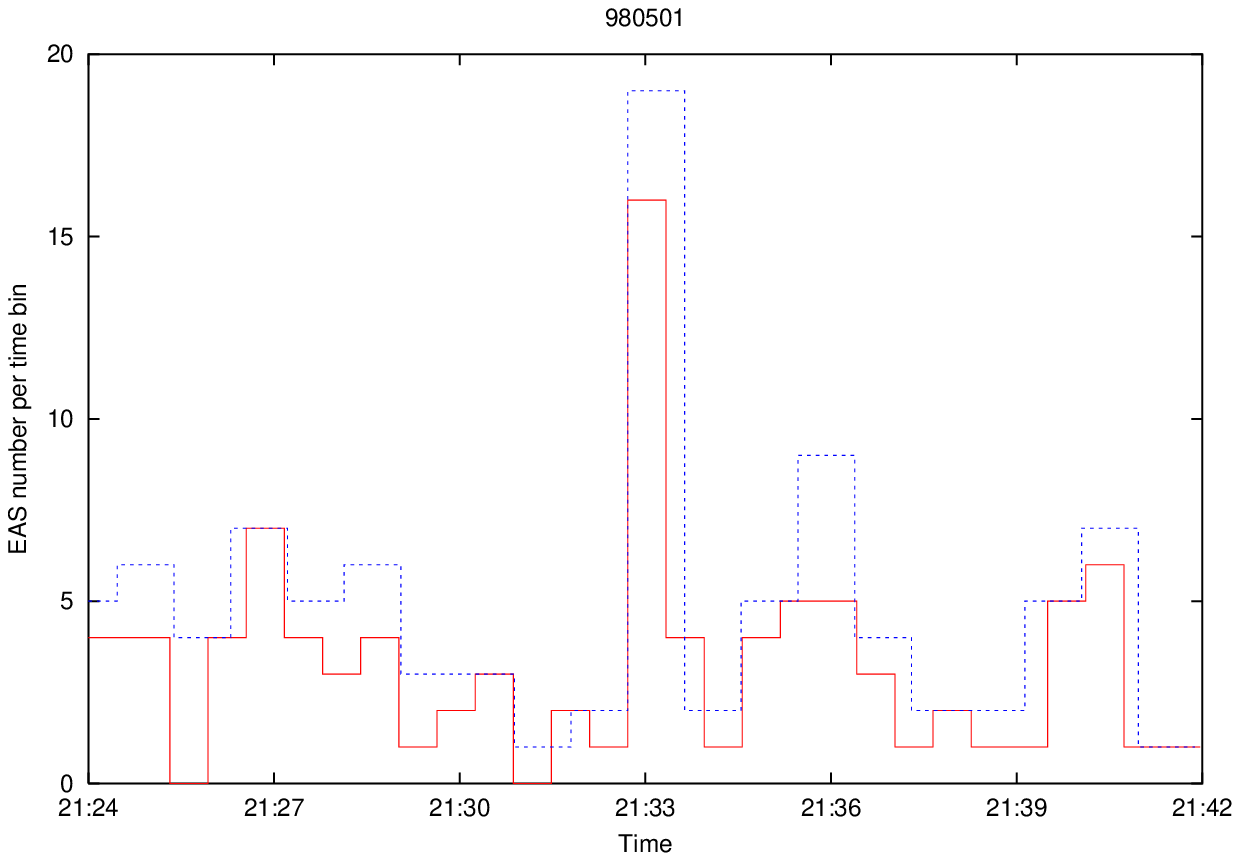}
\includegraphics[width=8.4cm]{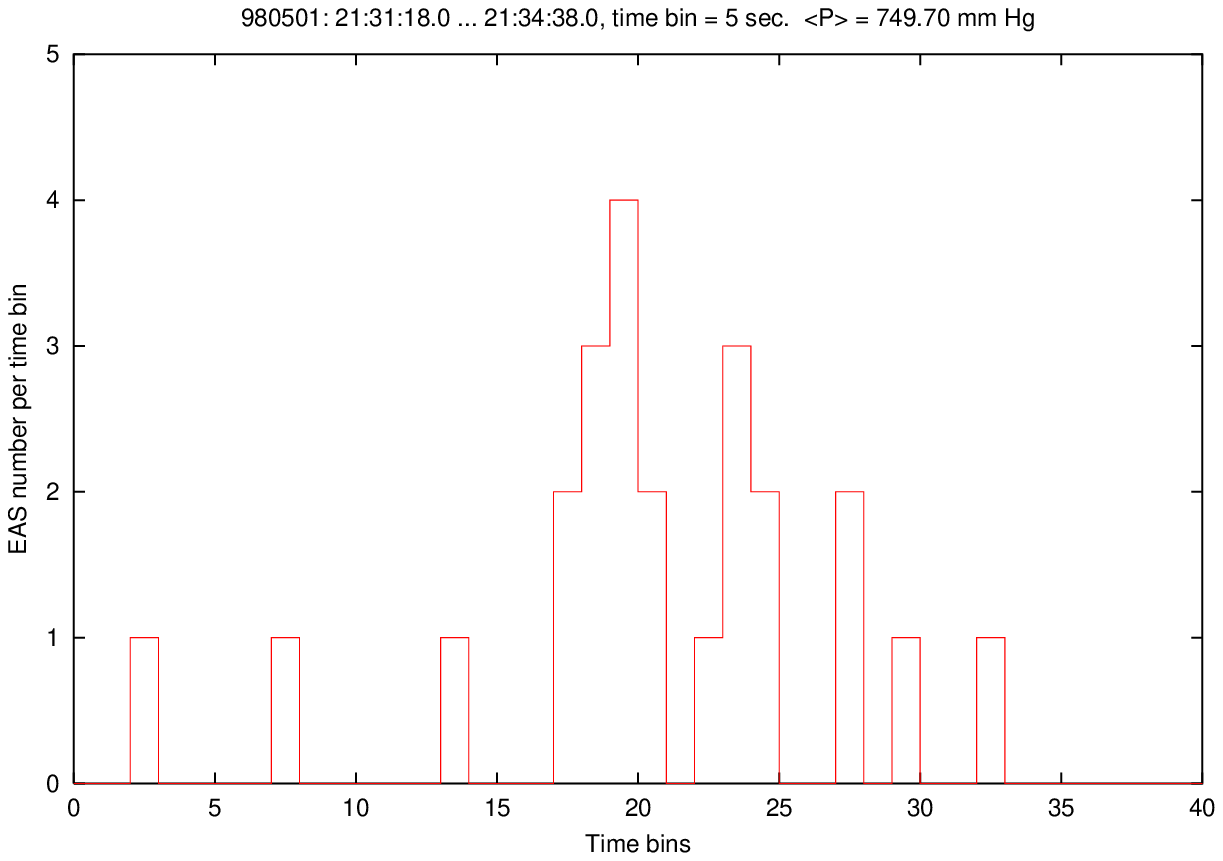}
}
\caption{
An event registered on May~1, 1998, see Table~2b.
The left plot: the innermost cluster (37-second bins)
embedded into the outer cluster (55-second bins).
The right plot: the 5-second ``substructure" of the event
(the outer cluster occupies bins No.~18--28).
}
\label{Fig:980501}
\end{figure}

\begin{figure}
\centerline{
\includegraphics[width=8.4cm]{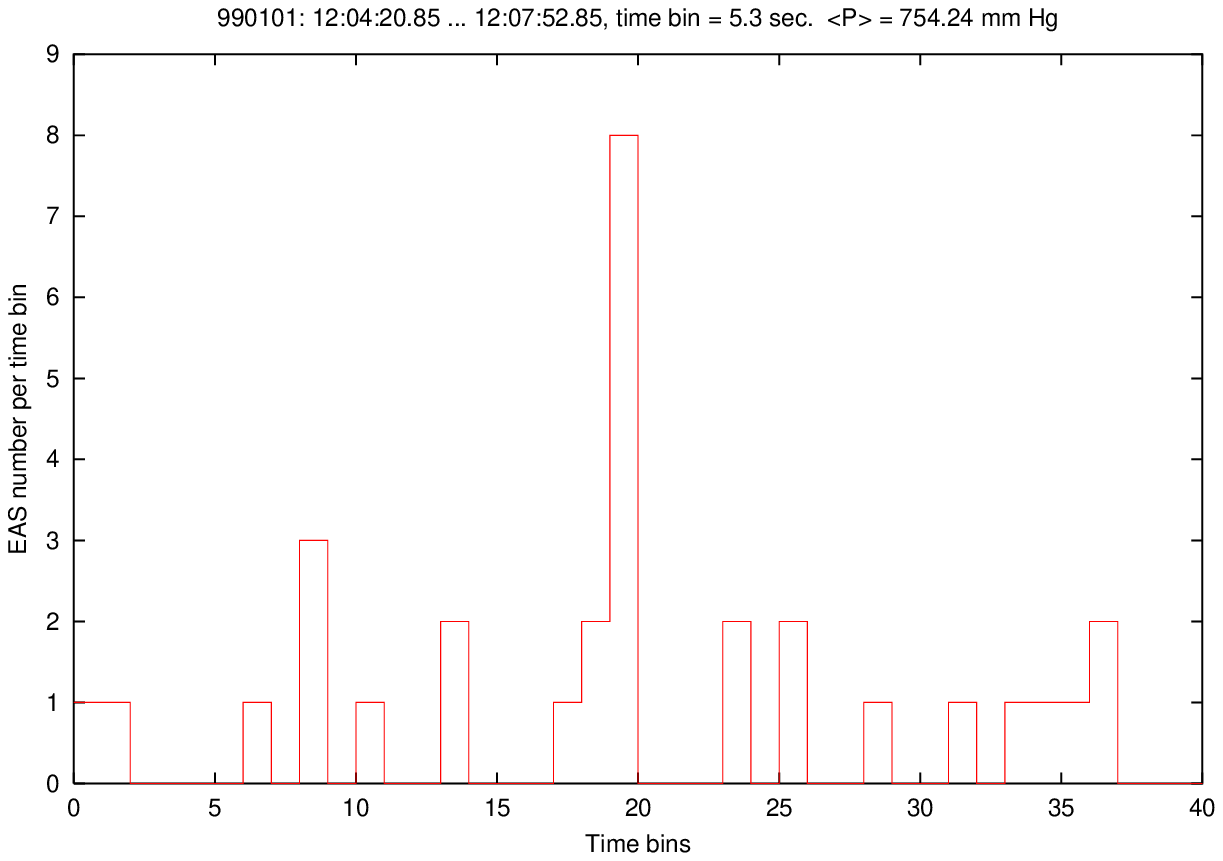}
\includegraphics[width=8.4cm]{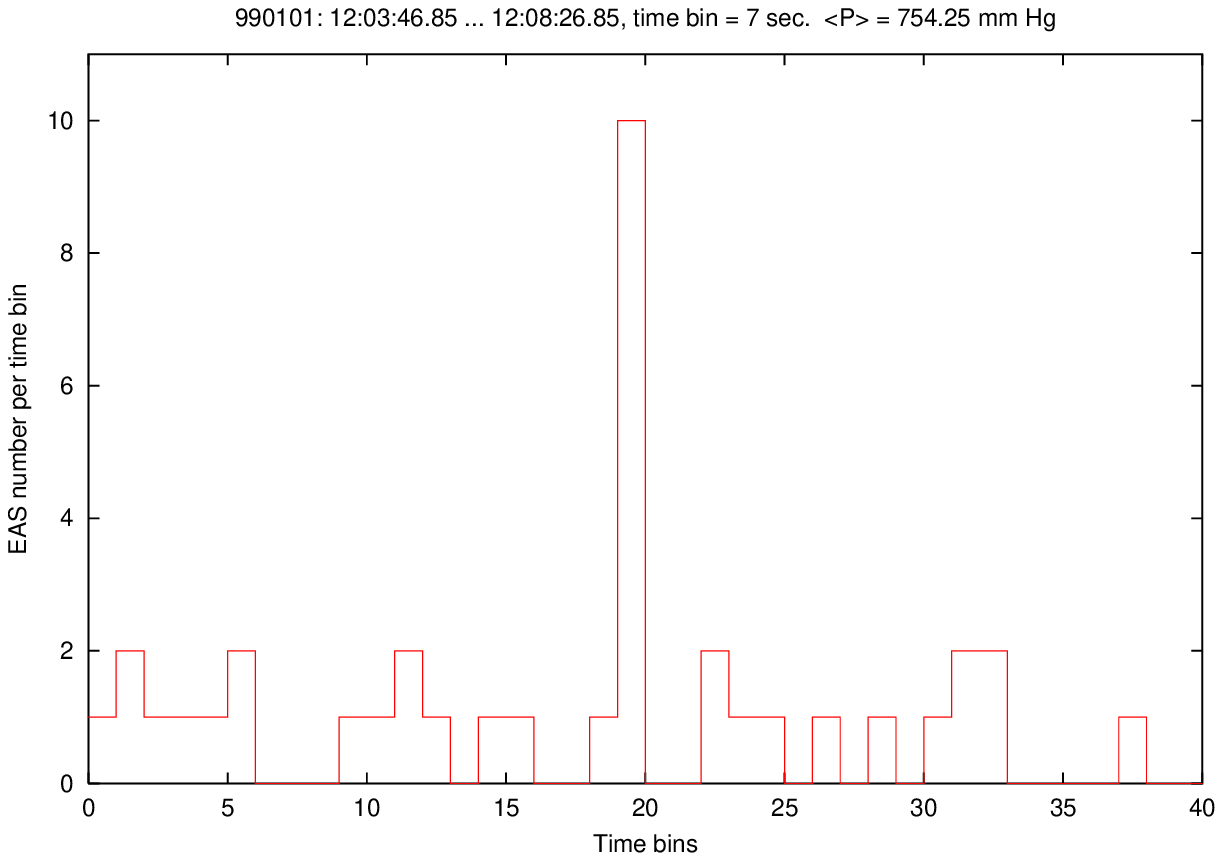}
}
\caption{An event registered on January~1, 1999, see Table~2b.
The left plot: the interior cluster.
The right plot: the outer cluster.
In both cases, the cluster ends at the center of the plot.
}
\label{Fig:990101}
\end{figure}

\begin{figure}
\centerline{
\includegraphics[width=8.4cm]{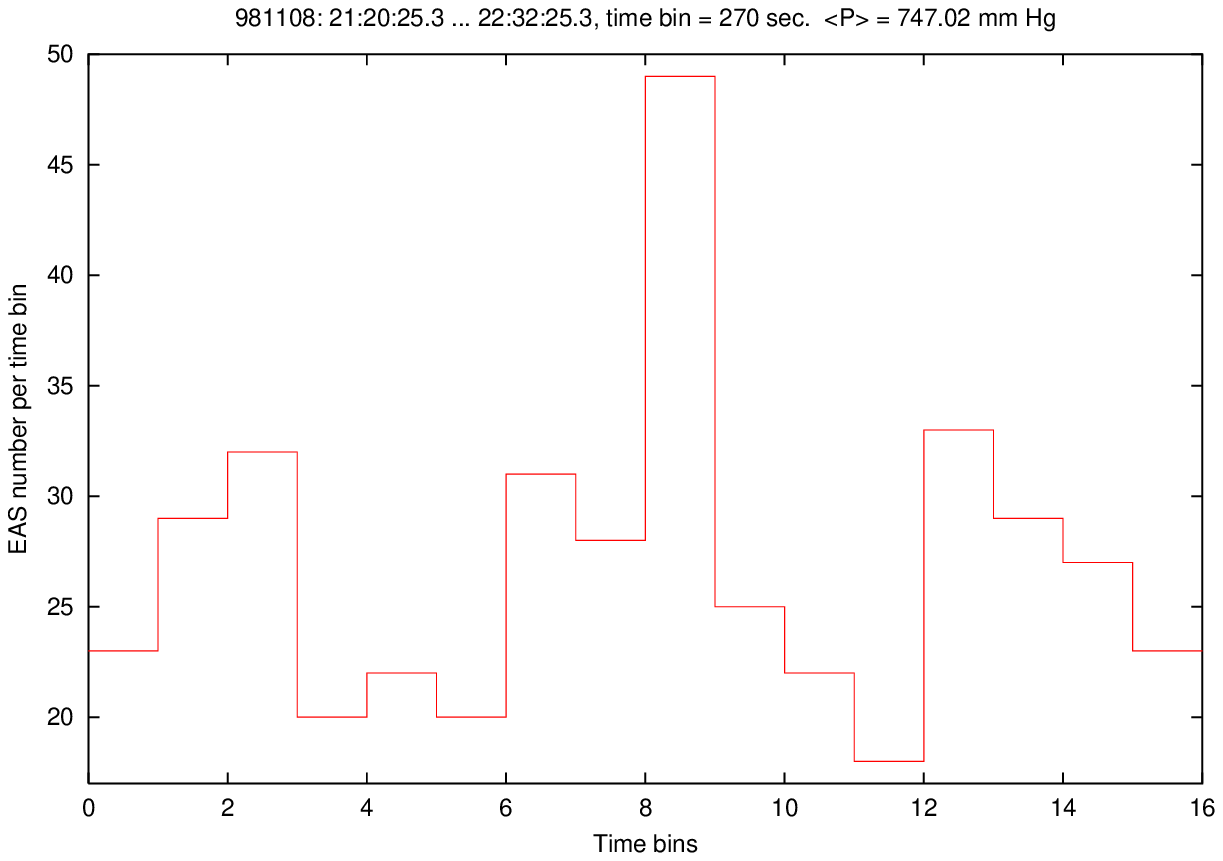}
\includegraphics[width=8.4cm]{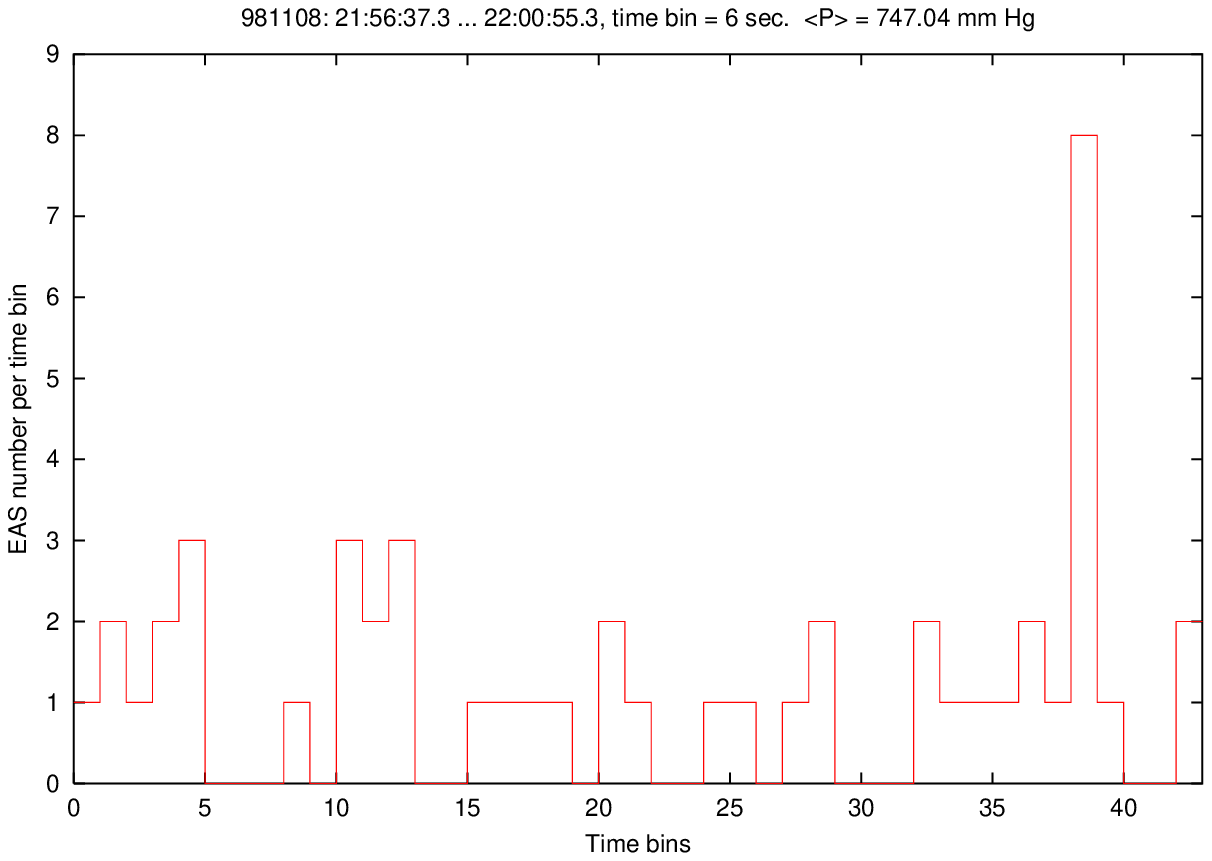}
}
\caption{An event registered on November~8, 1998: the outer cluster
(left) and its ``interior structure" (right).
Notice a short cluster (39th bin in the right plot) embedded into
the outer one, see Table~2b.
}
\label{Fig:981108}
\end{figure}

\begin{figure} 
\centerline{
\includegraphics[width=8.4cm]{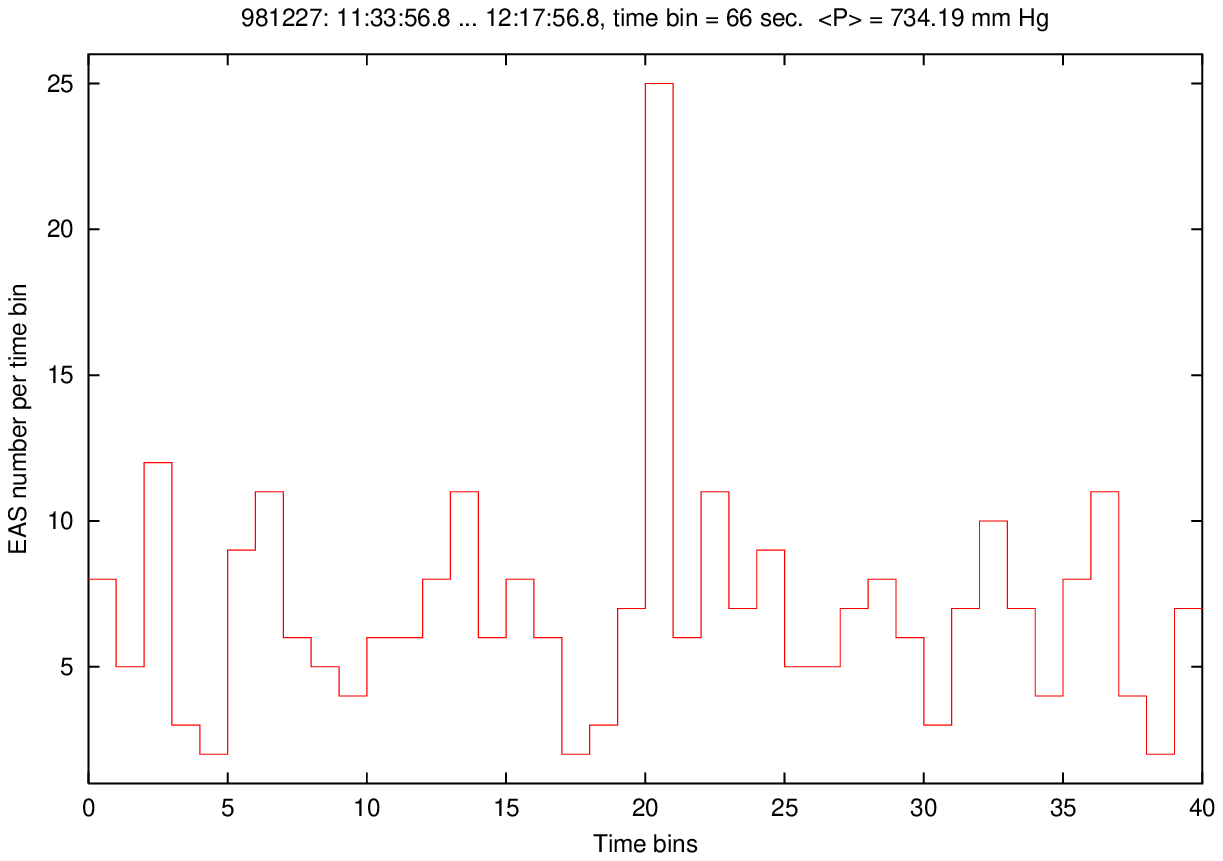}
\includegraphics[width=8.4cm]{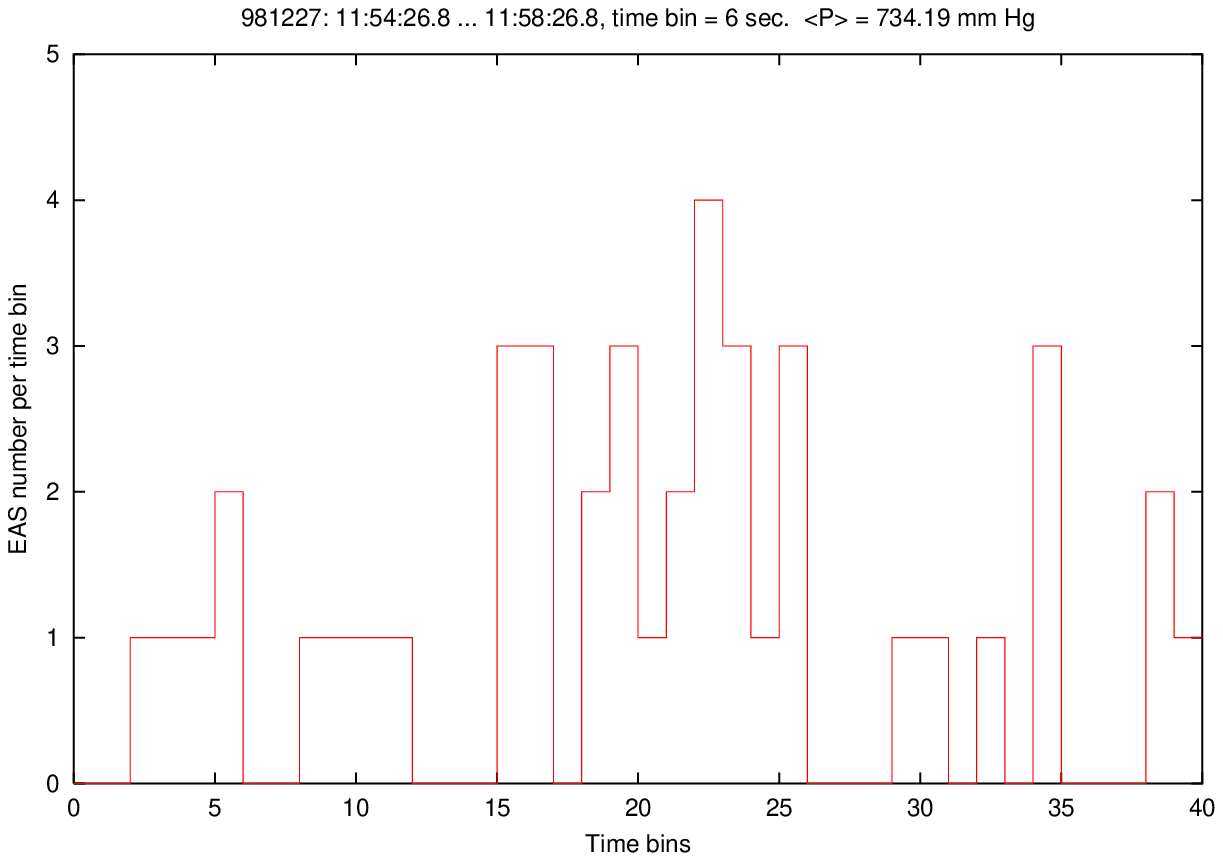}
}
\caption{An event registered on December~27, 1998
(left), see Table~2b, and its ``substructure" (right).
In the latter case, the cluster occupies bins No.~16--26.
}
\label{Fig:981227}
\end{figure}

\begin{itemize}
  \item
	\textbf{01.05.98}:
     For this event,
     two clusters are strictly embedded into the longest one.
     Notice that all clusters begin simultaneously.
	Fig.~\ref{Fig:980501} shows the innermost and the outer clusters
	and the 5-second substructure of this event.
	It is clearly seen how a short cluster appears inside a longer
	one.

	Notice that this event was selected for the values of~$K$ that
	are less by a unit than those given in Table~1.
	We have included this event in the list because of an interesting
	property of the arrival directions of EAS in it, see the discussion
	in Sec.~\ref{sec:AD}.

  \item
     \textbf{01.01.99}:
     This event is similar to the above one except that it consists
	of only two clusters, which do not begin, but end simultaneously.
     It is clearly seen in Fig.~\ref{Fig:990101} that the highest count
     rate is at the end of the event.
     Notice that the outer cluster has an extremely small~\Padj.

  \item
     \textbf{08.11.98}:
     An interesting situation takes place for this event, since a
     5-second cluster is embedded into a 4-minute cluster, and no
     clusters of intermediate duration are found, see
     Fig.~\ref{Fig:981108}.

  \item
     \textbf{27.12.98}:
     Two clusters that begin at consecutive showers are selected
     for $T=70$~sec.
     We assume that there is no phenomenological reason in the
     appearance of these two clusters, since they cover exactly the
     same interval of time as a cluster selected for $T=75$~sec.
     It is likely that a ``real" cluster is the outer one,
     and thus this event may be referred to the first group.
     This cluster and its ``substructure" are shown in
     Fig.~\ref{Fig:981227}.
\end{itemize}

\bigskip
\begin{center}
     Table 2b: Events that consist of several EAS clusters.
     See notation in Tables~1 and~2a.\\[4mm]
\begin{tabular}{|c|c|c|c|c|c|c|c|c|c|}
\hline
Date    & $P$,& $T$, &Beginning  &End        &\Treal,&\NAS&\Ncl&\Tadj,&\Padj           \\
dd.mm.yy&mm Hg&  sec &hh:mm:ss   &hh:mm:ss   &sec    &    &    &sec   &                \\
\hline
01.05.98&749.7&35$^*$&21:32:43.01&21:33:19.92& 36.91& 16 & 1  & 33.95&$3.0\times10^{-7}$\\
        &     &40$^*$&21:32:43.01&21:33:21.78& 38.77& 17 & 1  & 35.67&$1.1\times10^{-7}$\\
        &     &50$^*$&21:32:43.01&21:33:37.17& 54.16& 19 & 1  & 49.82&$5.8\times10^{-7}$\\
\hline
01.01.99&754.2&   5  &12:06:01.57&12:06:06.84&  5.27&  8 & 1  &  4.62&$2.4\times10^{-8}$\\
        &     &  10  &12:05:59.98&12:06:06.84&  6.86& 10 & 1  &  6.01&$6.4\times10^{-10}$\\
\hline
08.11.98&747.0&   5  &22:00:25.36&22:00:30.58& 5.22 &  8 & 1  &  4.94&$4.0\times10^{-8}$\\
        &     &  240 &21:56:41.54&22:00:54.63&253.09& 49 & 1  &239.65&$9.4\times10^{-7}$\\
\hline
27.12.98&734.2&  70  &11:55:56.85&11:57:02.59&      & 25 & 2  &      &                 \\
        &     &  75  &11:55:56.85&11:57:02.59& 65.74& 25 & 1  & 71.55&$5.7\times10^{-8}$\\
\hline
04.11.98&730.9&  30  &08:17:44.90&08:18:10.99& 26.09& 16 & 1  & 29.41&$4.7\times10^{-8}$\\
        &     &  35  &08:17:38.91&08:18:09.83& 30.92& 17 & 1  & 34.85&$8.3\times10^{-8}$\\
        &     &  40  &08:17:36.88&08:18:10.99&      & 19 & 2  &      &        \\
        &     &  45  &08:17:36.88&08:18:10.99& 34.11& 19 & 1  & 38.45&$1.3\times10^{-8}$\\
        &     &  50  &08:17:36.88&08:18:19.56& 42.68& 20 & 1  & 48.10&$8.1\times10^{-8}$\\
\hline
10.08.98&732.3&  50  &16:47:43.15&16:48:28.13&44.98 & 20 & 1  & 49.98&$1.5\times10^{-7}$\\
        &     &  240 &16:47:36.61&16:51:19.83&      & 53 & 5  &      &                  \\
\hline
11.11.98&752.3&  900 &01:21:17.47&01:38:02.27&      &136 & 3  &      &                 \\
\hline
06.01.99(b)&733.2& 180 &09:22:52.51&09:25:50.41&      & 43 & 4  &      &    \\
\hline
08.01.99(a)&726.9& 120 &03:56:45.66&03:58:28.26&      & 33 & 3  &      &                 \\
        &      & 180 &03:55:50.40&03:58:28.26&      & 44 & 5  &      &                 \\
        &      & 240 &03:55:05.31&03:58:28.26&      & 50 & 2  &      &                 \\
\hline
\end{tabular}
\end{center}

\bigskip

\begin{figure}
\centerline{
\includegraphics[width=8.4cm]{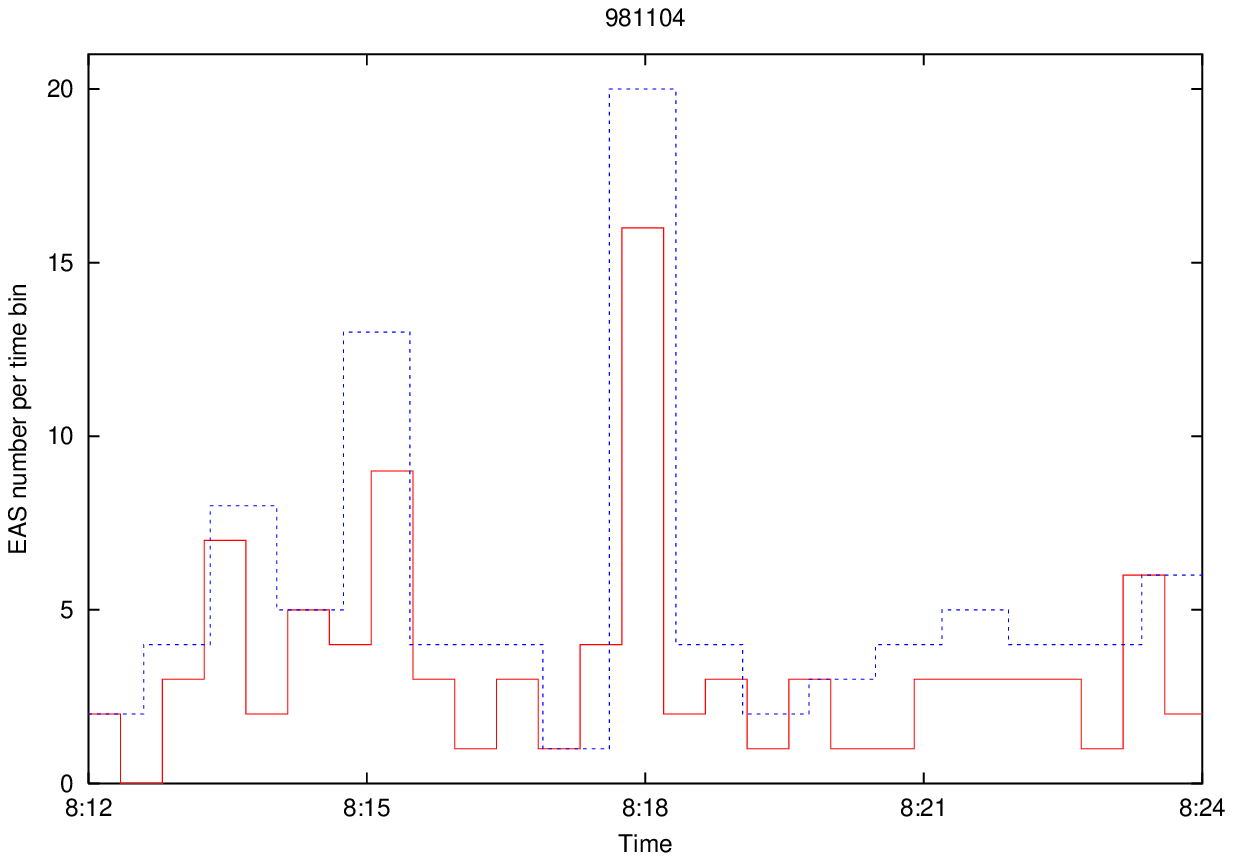}\\
\includegraphics[width=8.4cm]{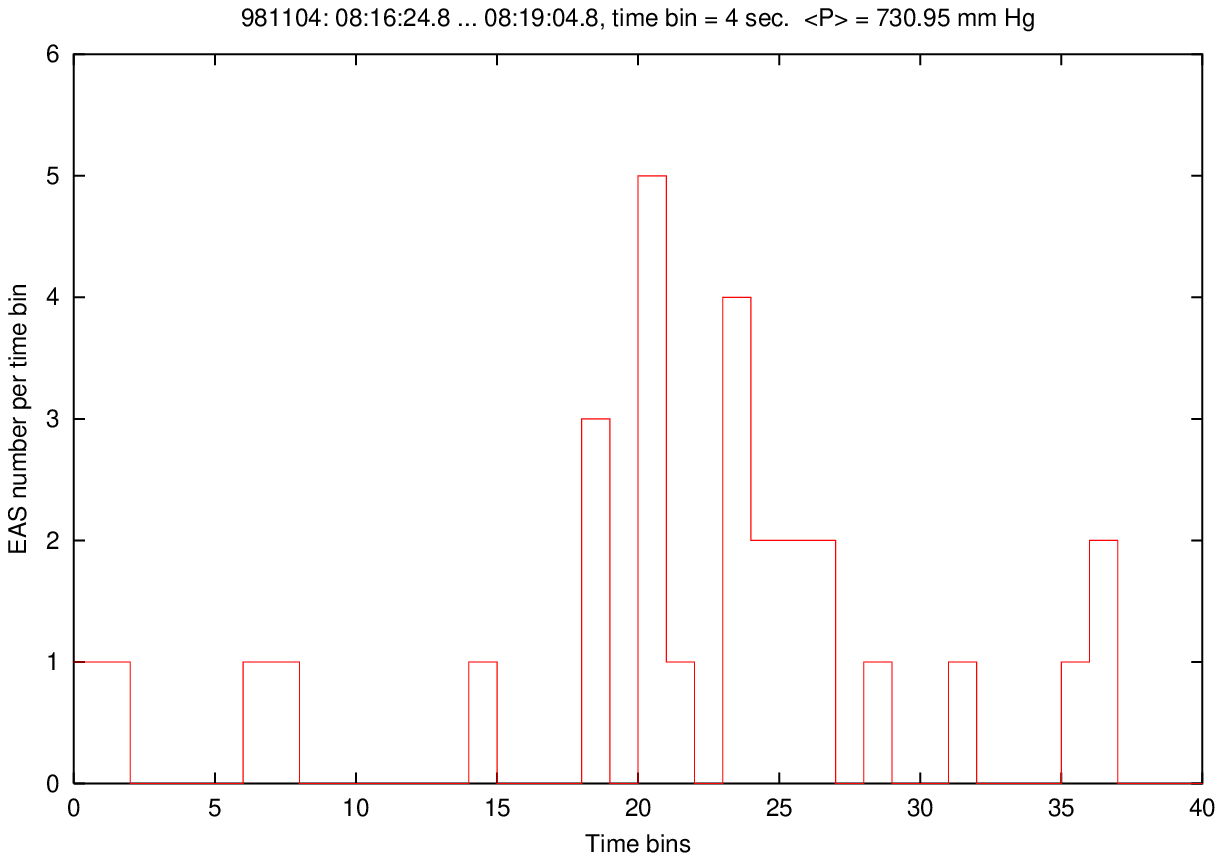}
}
\caption{
An event registered on November~4, 1998:
the innermost (27-second bins) and the outer (43-second bins)
clusters (left), see Table~2b, and the 4-second ``substructure" of the event
(right).
In the latter case, the outer cluster occupies bins No.~19--29,
the interior cluster occupies bins No.~21--27.
}
\label{Fig:981104}
\end{figure}

\begin{figure}
\begin{center}
\includegraphics[width=9.0cm]{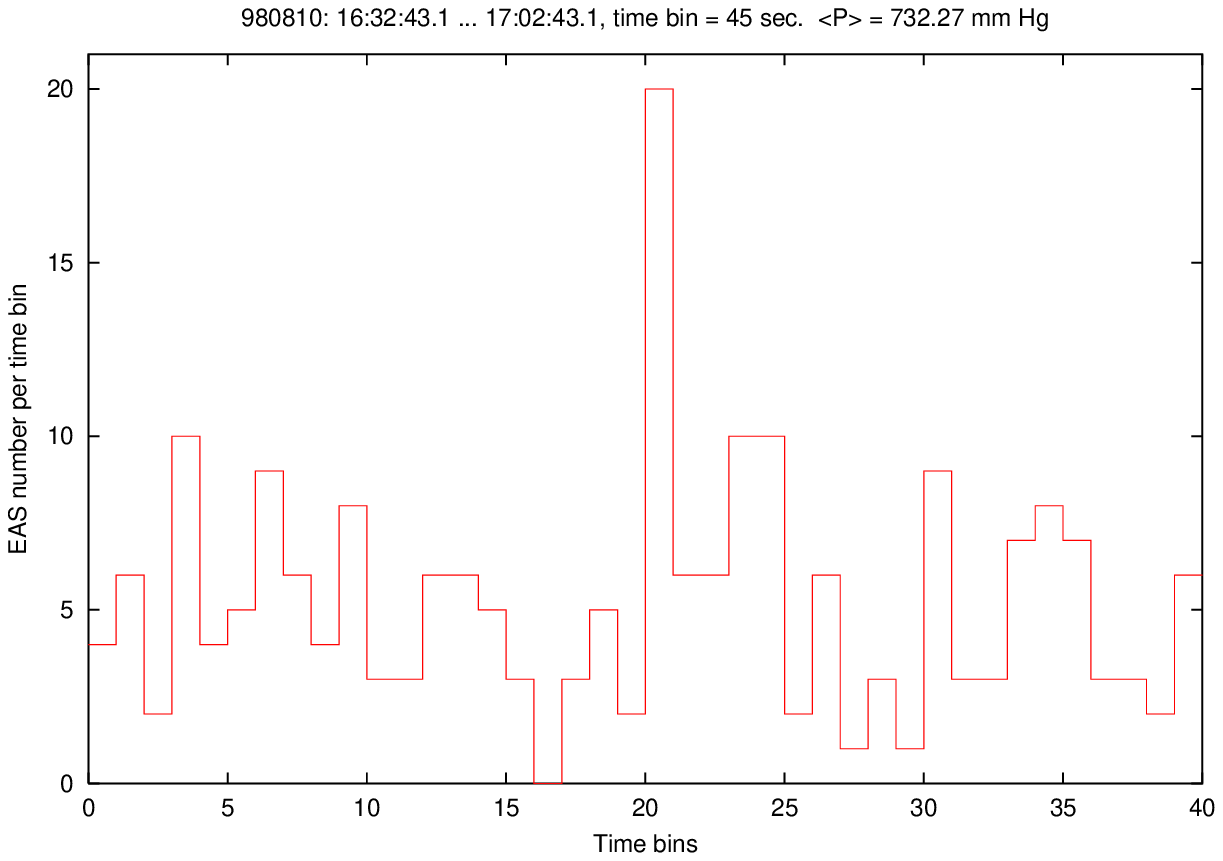}\\
\includegraphics[width=9.0cm]{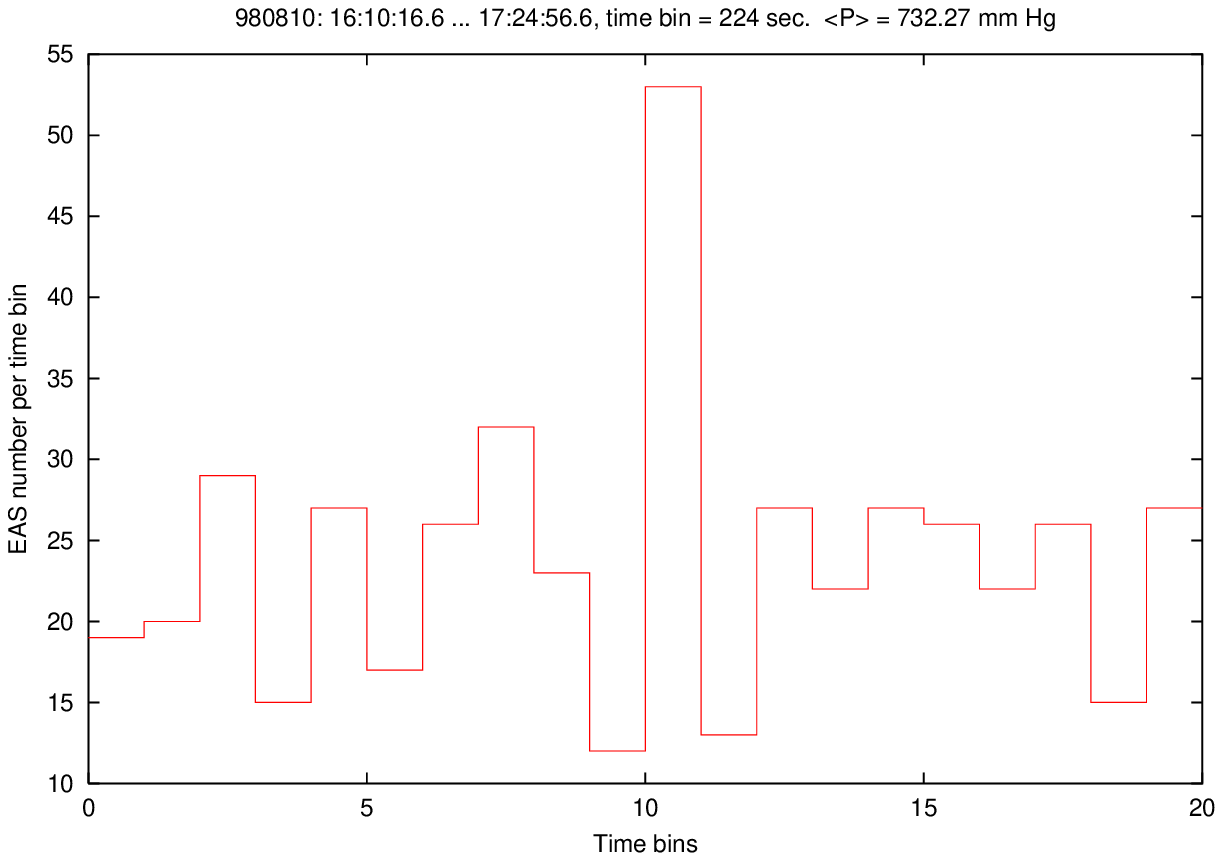}\\
\includegraphics[width=9.0cm]{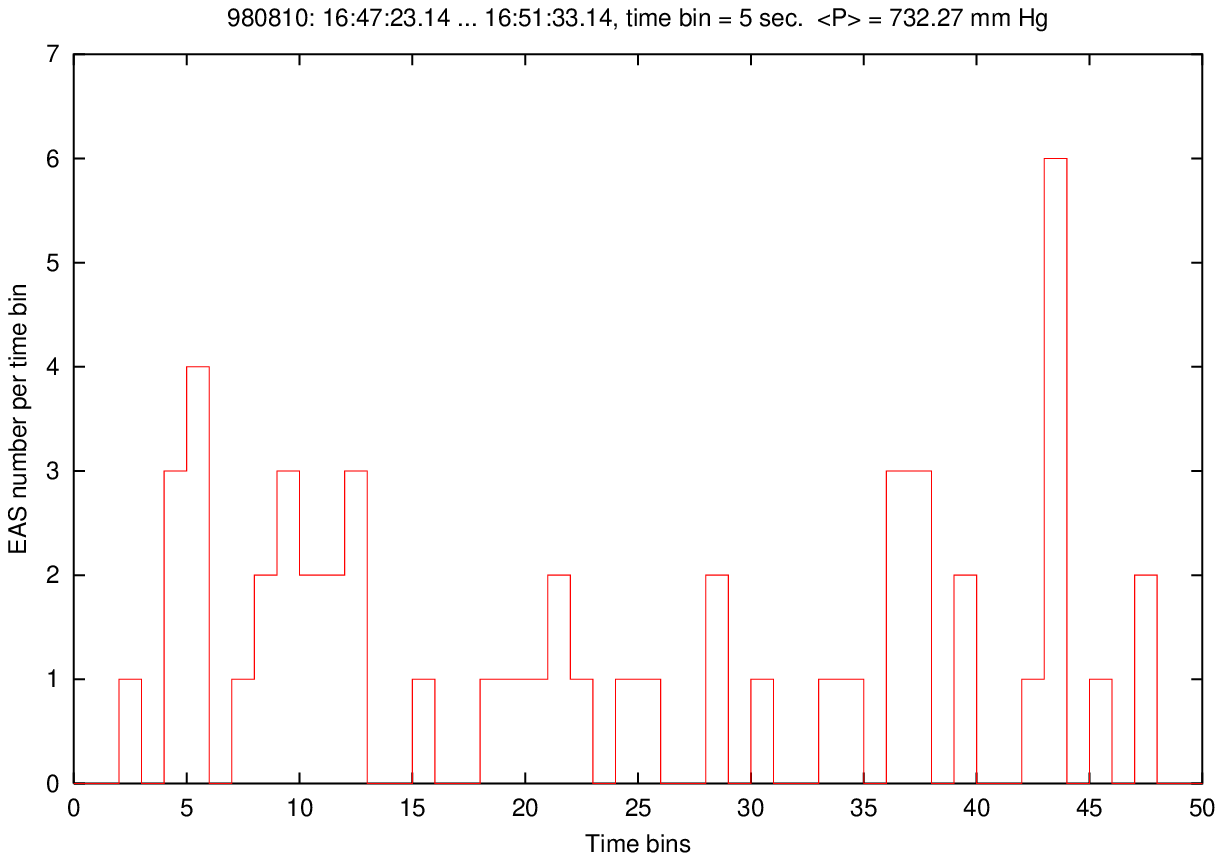}
\end{center}
\caption{An event registered on August~10, 1998, see Table~2b.
From top: the interior cluster, the group selected for $T=240$~sec,
and the 5-second ``substructure" of the event.
In the later case, the interior cluster occupies bins No.~5--13.
}
\label{Fig:980810}
\end{figure}

\begin{figure}
\begin{center}
\includegraphics[width=9.0cm]{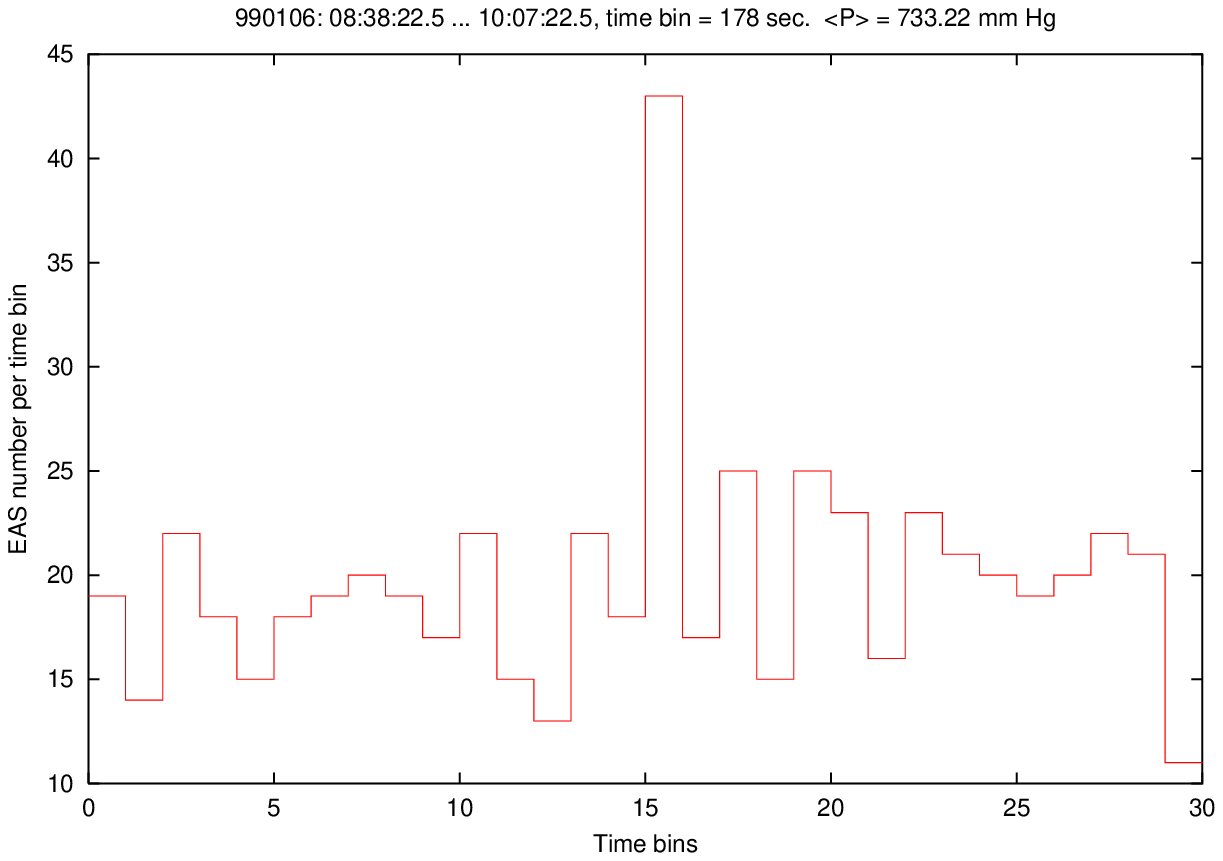}\\
\includegraphics[width=9.0cm]{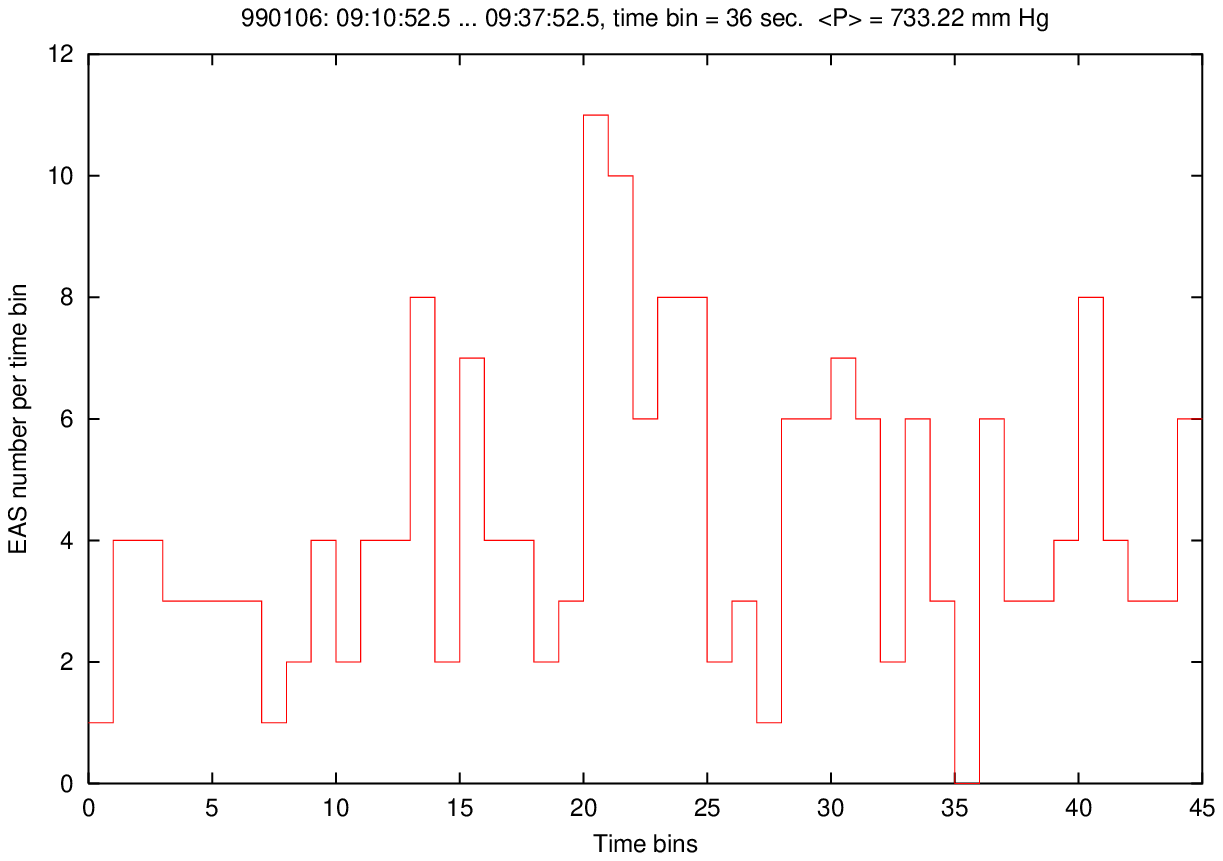}\\
\includegraphics[width=9.0cm]{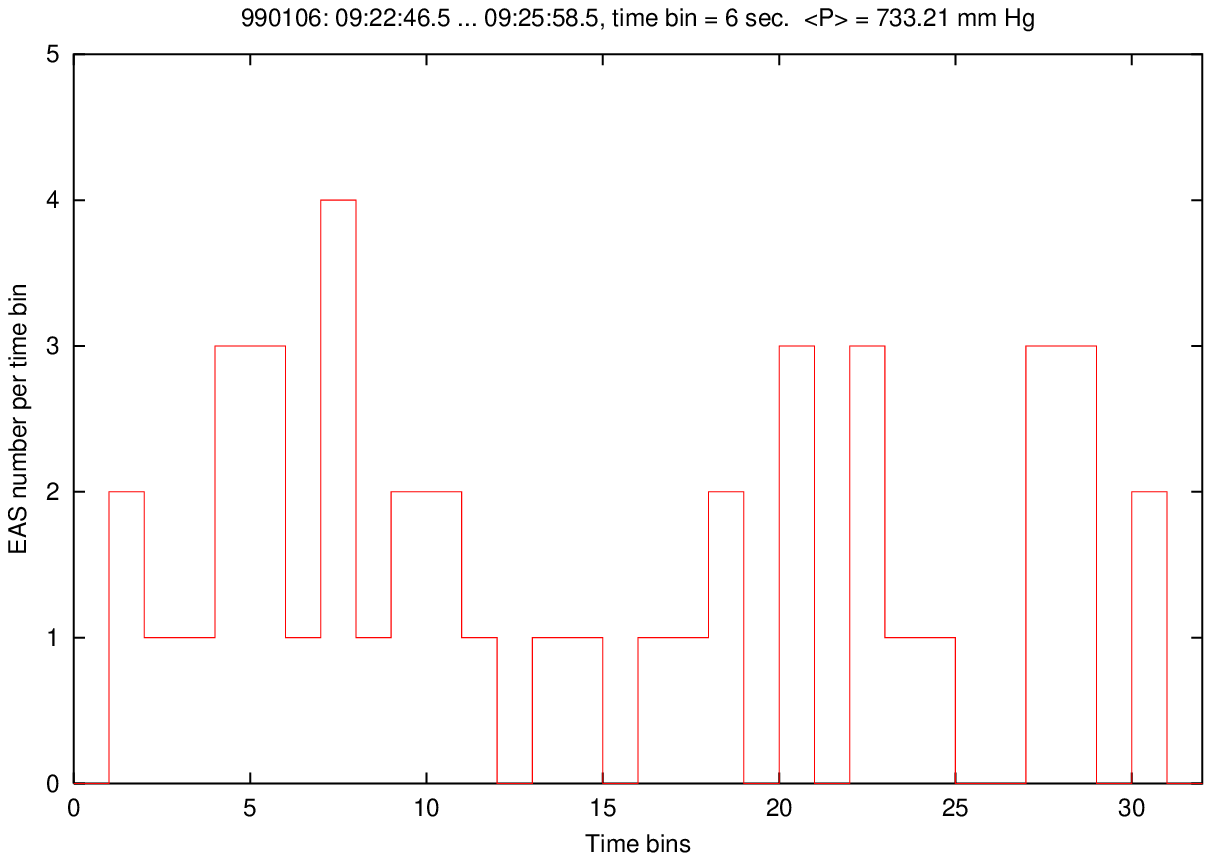}
\end{center}
\caption{An event registered on January~6, 1999(b) (top plot),
see Table~2b, and its ``substructure."
In the middle plot, the cluster occupies bins No.~21--25.
}
\label{Fig:990106b}
\end{figure}

\begin{figure}
\centerline{
\includegraphics[width=8.4cm]{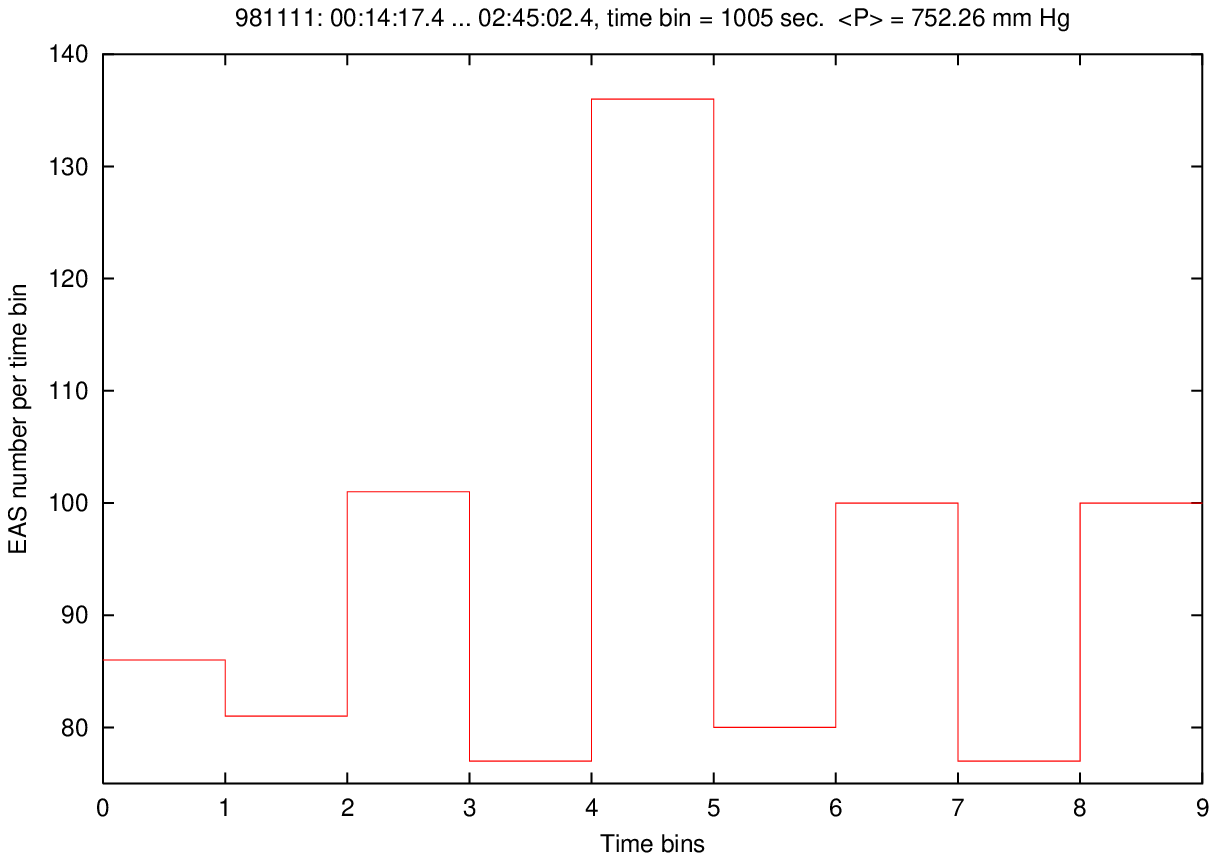}
\includegraphics[width=8.4cm]{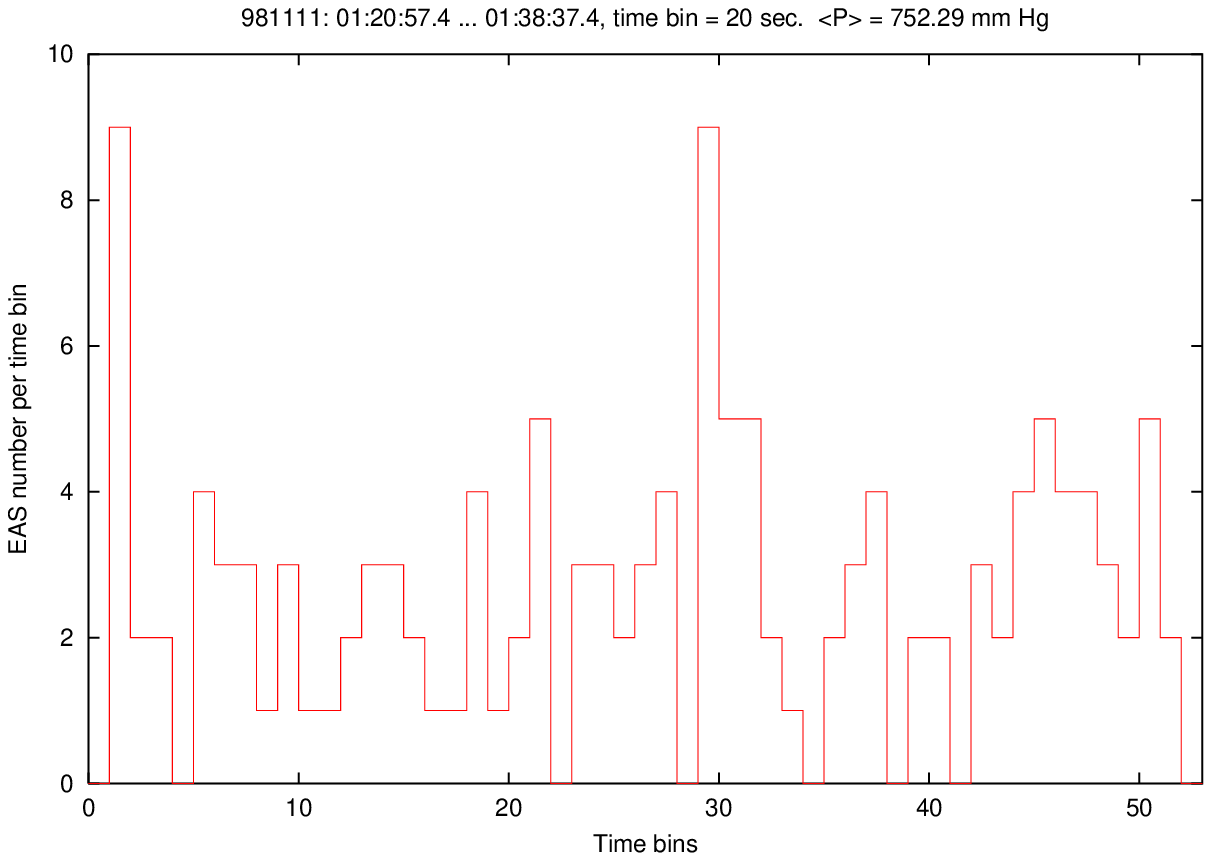}
}
\caption{An event registered on November~11, 1998, see Table~2b,
and its ``substructure."
}
\label{Fig:981111}
\end{figure}

\begin{figure}
\centerline{
\includegraphics[width=8.4cm]{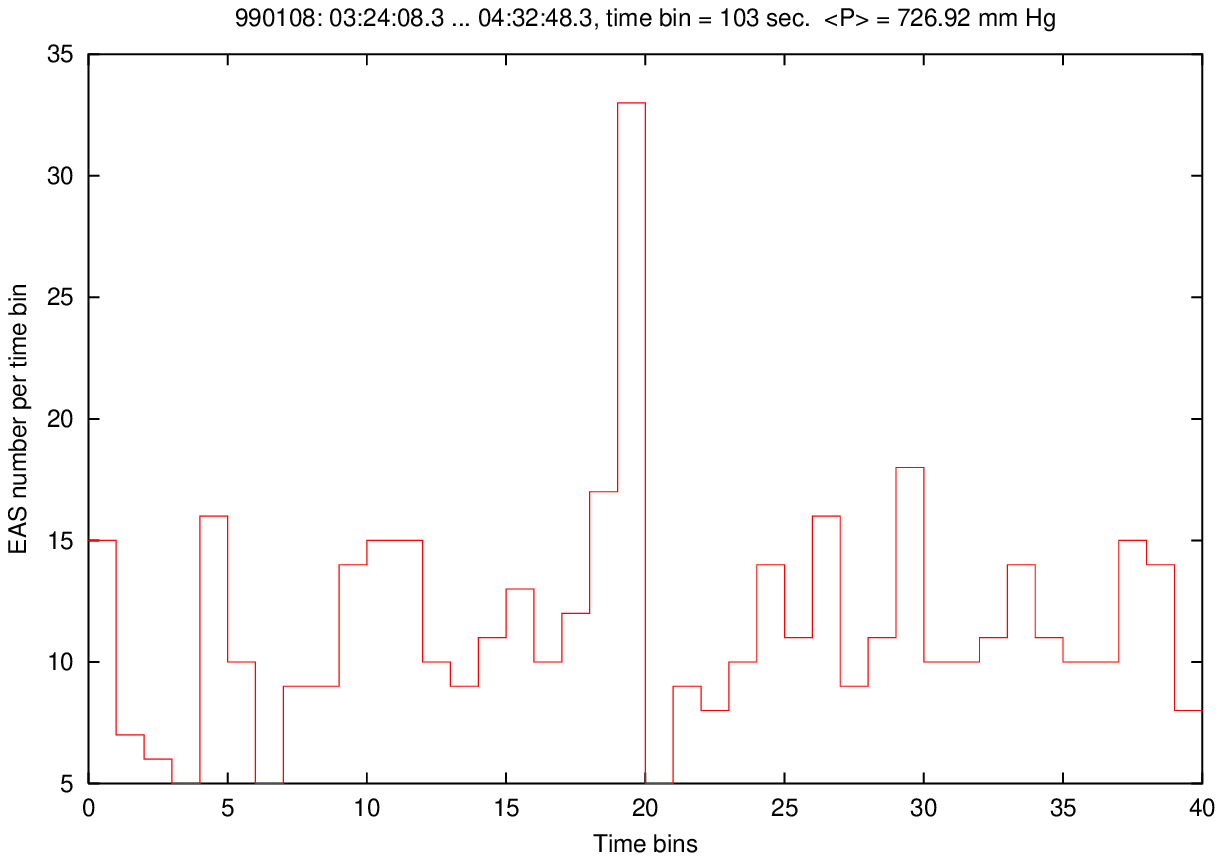}
\includegraphics[width=8.4cm]{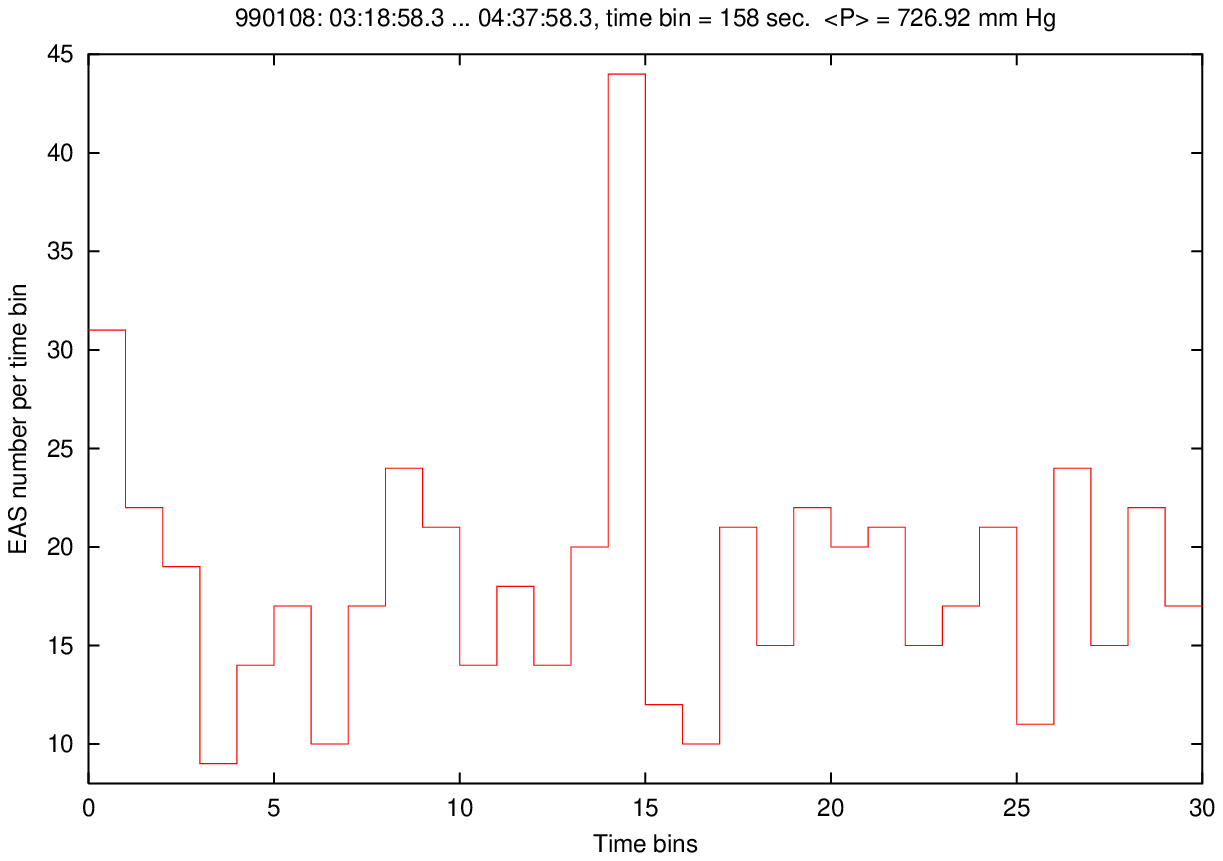}
}
\centerline{
\includegraphics[width=8.4cm]{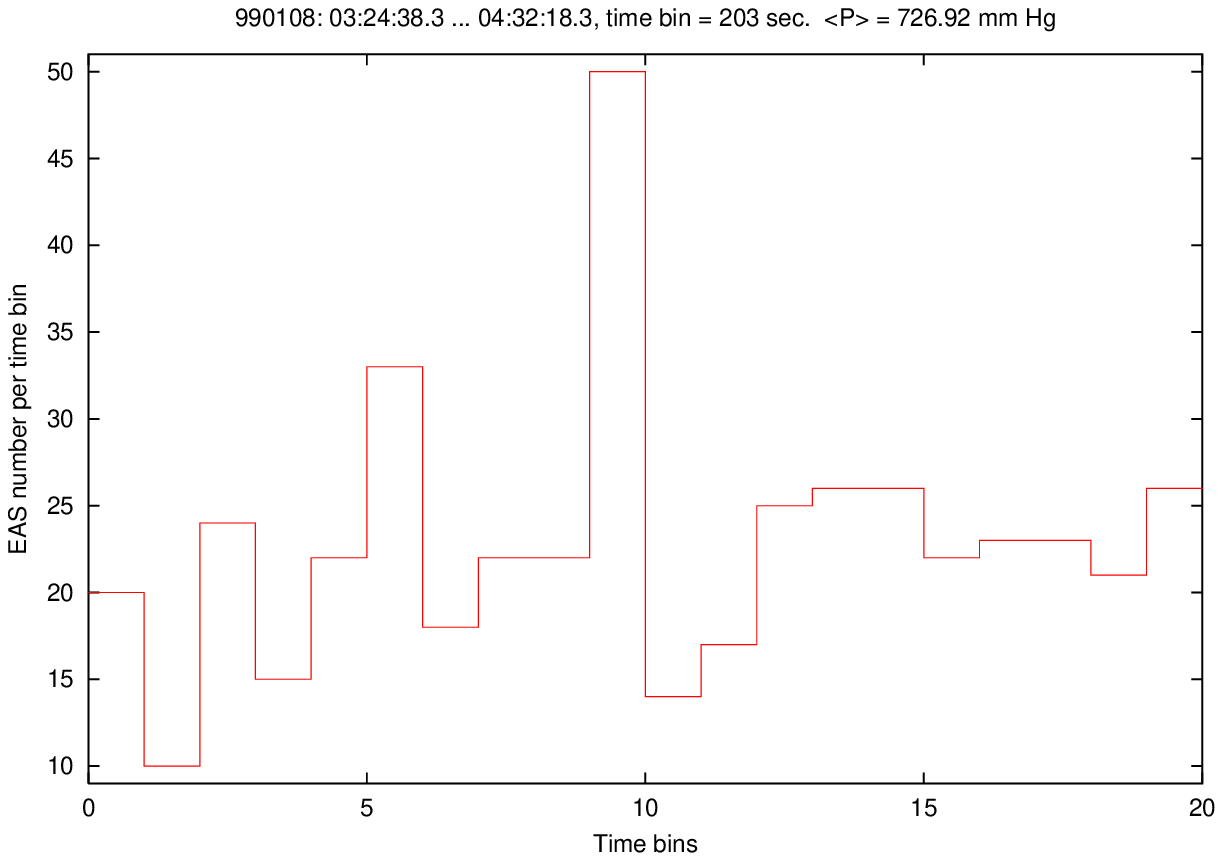}
\includegraphics[width=8.4cm]{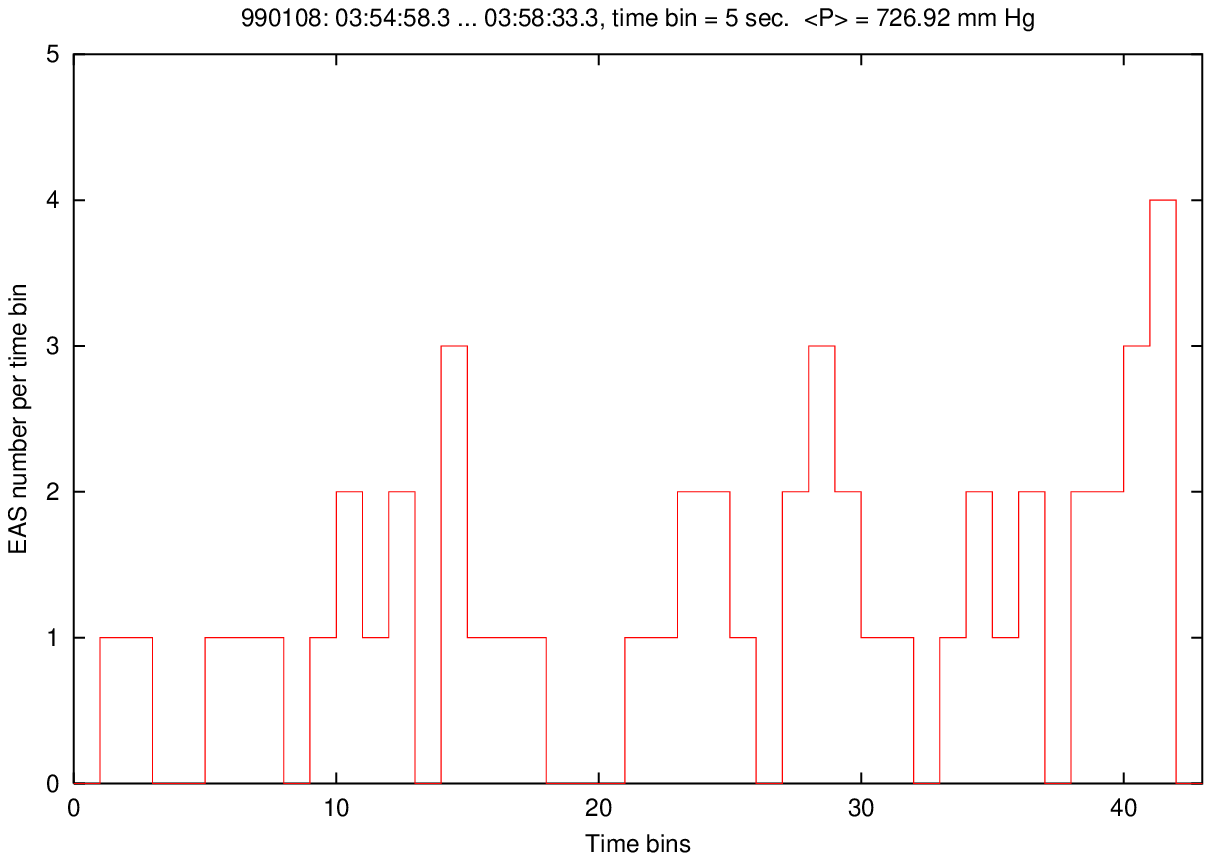}
}
\caption{An event registered on January~8, 1999(a), see Table~2b.
The embedding of clusters and the ``substructure"
of the event are shown.
Except for the right bottom plot, the clusters end in the center
of the plots.
}
\label{Fig:990108b}
\end{figure}

\begin{itemize}
  \item
     \textbf{04.11.98}:
     For this event, shorter clusters are almost exactly embedded
     into longer ones.
     Similar to the event registered on 27.12.98, two clusters
     selected for $T=40$~sec begin at consecutive EAS and cover
     exactly the same time interval as a cluster selected for
     $T=45$~sec.
     The innermost cluster points at the interval of the highest
     count rate, see Fig.~\ref{Fig:981104}.

	This event has another interesting feature, see the discussion
	in Sec.~\ref{sec:AD}.

  \item
     \textbf{10.08.98}:
     For this event, a bit more surprising situation takes place:
     a 45-second cluster is found within a sequence of five
	nearly 4-minute clusters that begin at consecutive arrival times.
     The situation is in a sense similar to that one with the event
     registered on 08.11.98.
     Notice that the short cluster and the sequence of long clusters begin
     nearly simultaneously.
     This event and its ``interior structure" are shown in
	Fig.~\ref{Fig:980810}.

  \item
     \textbf{11.11.98} and \textbf{06.01.99(b)}:
     For these events, a surprising situation takes place,
     since a sequence of long clusters (3 and 4 respectively) is
     found, and no other clusters are inside or outside.
     All clusters in these events begin at consecutive EAS, and thus
     it is likely that the only reason of such a situation is a
     comparatively big time step in~$T$ equal to one minute.
	As one can see from Figs.~\ref{Fig:990106b} and~\ref{Fig:981111},
	it is possible to find intervals of the maximum count rate for
	these events, too.
	Namely, the highest count rate is shifted to the beginning of
     the event registered on January~6, 1999, while for the other event
     one can find two 20-second bins with maximum count rate, located
     at the very beginning and soon after the middle of the event
     respectively.

  \item
     \textbf{08.01.99(a)}:
     An event that is similar to the two previous events, but
     three groups of clusters are found.
     Each group consists of clusters that begin at consecutive arrival
     times.
     The most likely reason of the fact that we do not find a separate
	cluster neither inside nor outside those given in Table~2b
	is again the discreteness of~$T$.
     Notice that all sequences end at the same moment of time,
     and the highest count rate is shifted to the second
	half of the event, see Fig.~\ref{Fig:990108b}.
	It is also interesting to mention that the first group of clusters
	(selected for $T=2$~min) is almost exactly two times shorter than
	the third group (selected for $T=4$~min).
\end{itemize}

     2. Complicated embedding, see Tables~2c and~3.

\begin{figure}
\centerline{
\includegraphics[width=8.4cm]{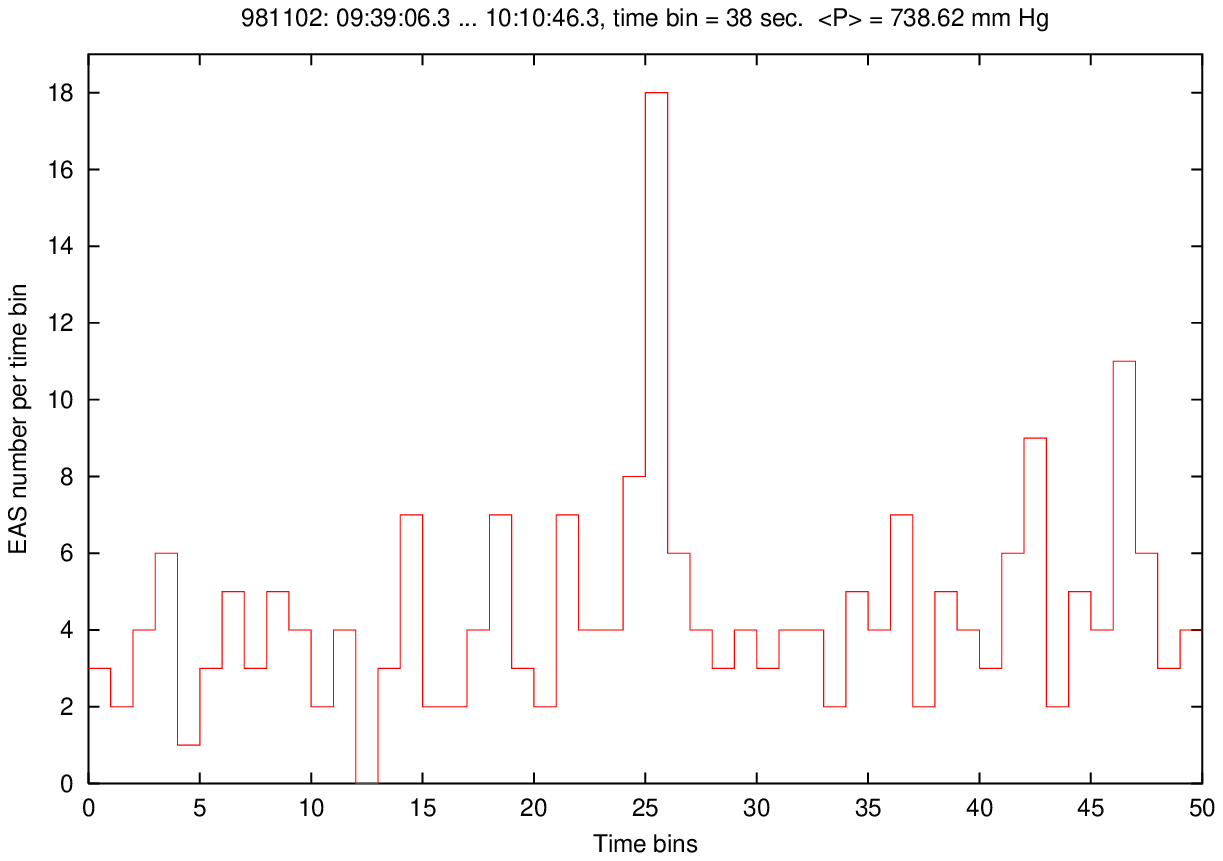}
\includegraphics[width=8.4cm]{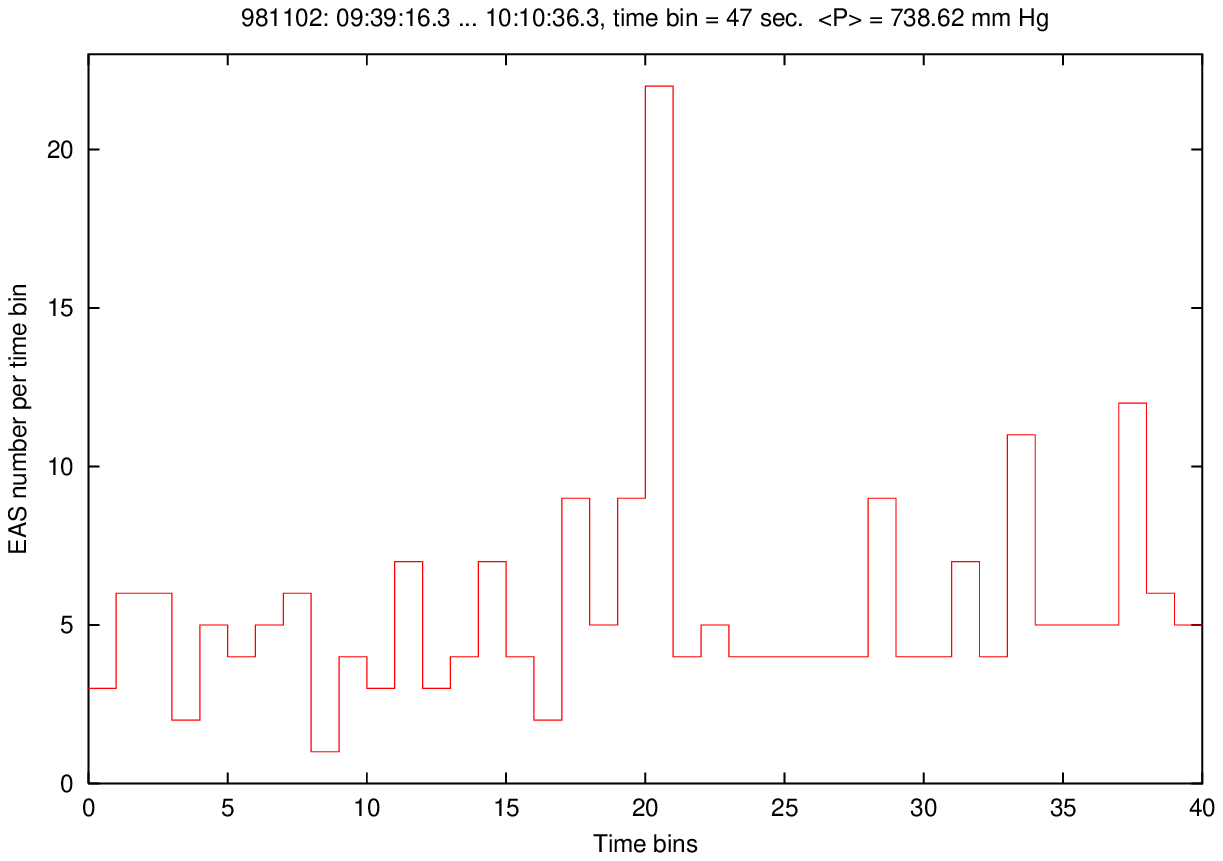}
}
\centerline{
\includegraphics[width=8.4cm]{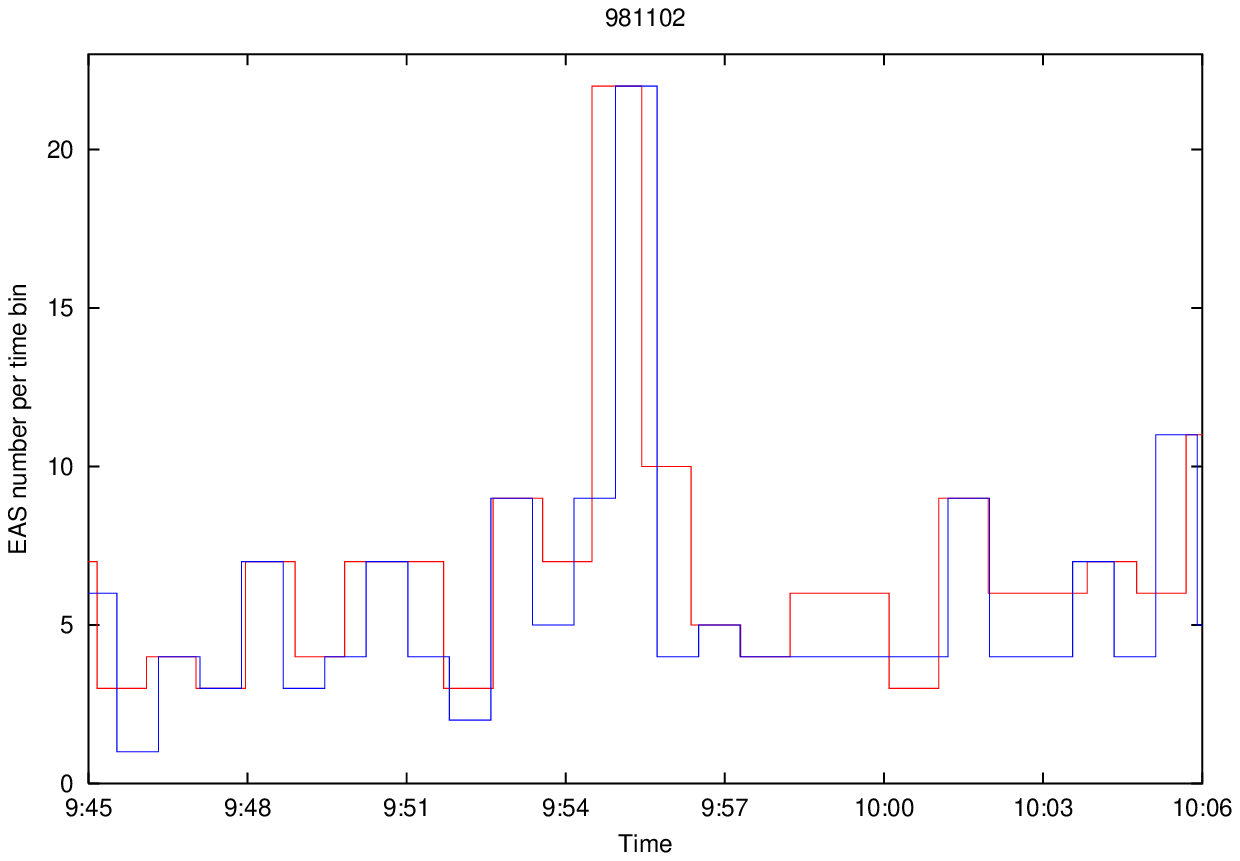}
\includegraphics[width=8.4cm]{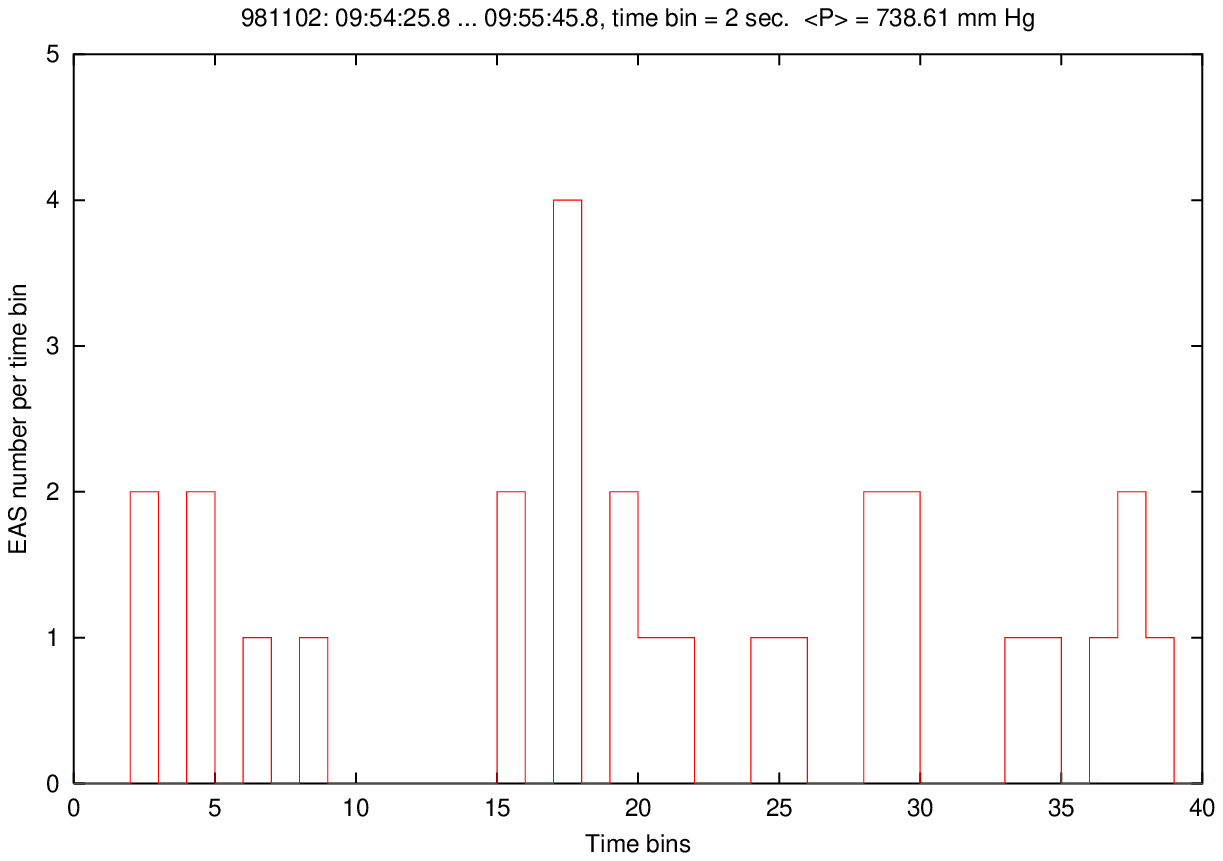}
}
\centerline{
\includegraphics[width=8.4cm]{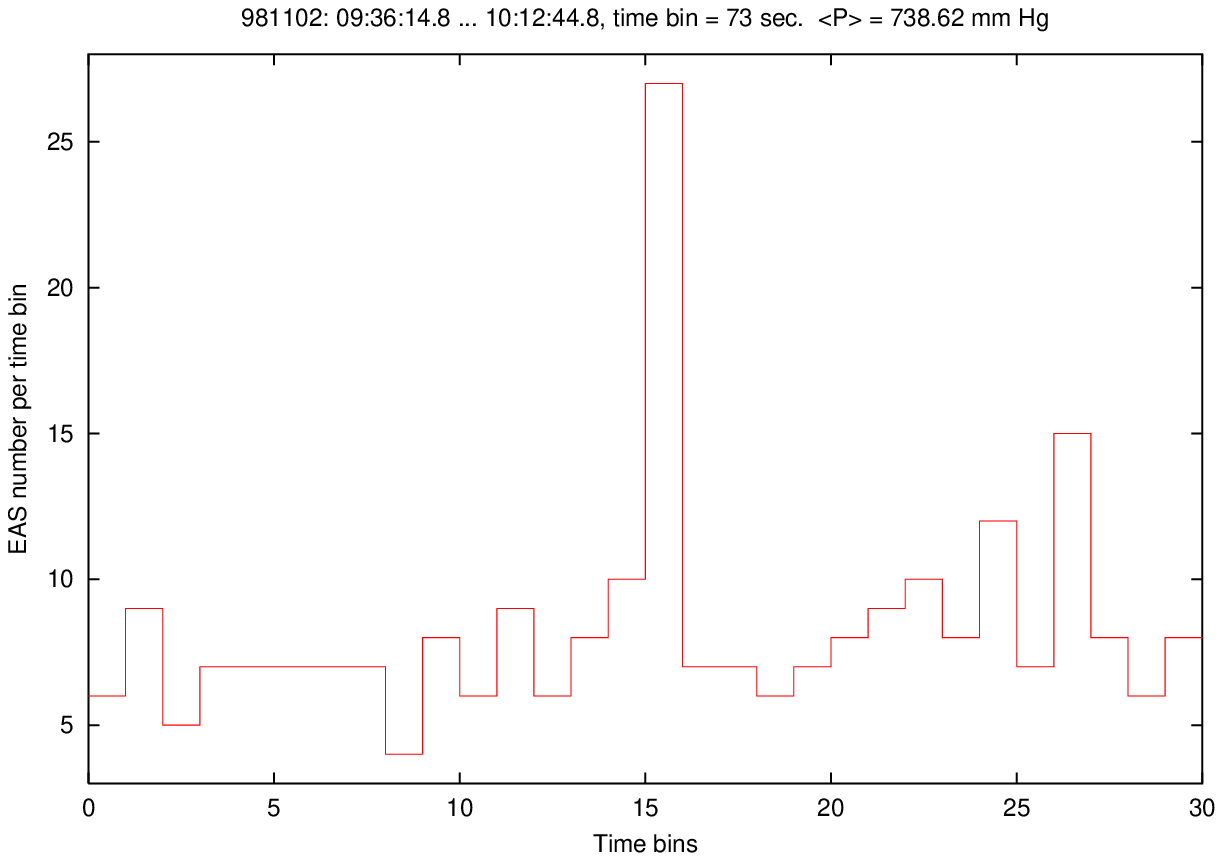}
\includegraphics[width=8.4cm]{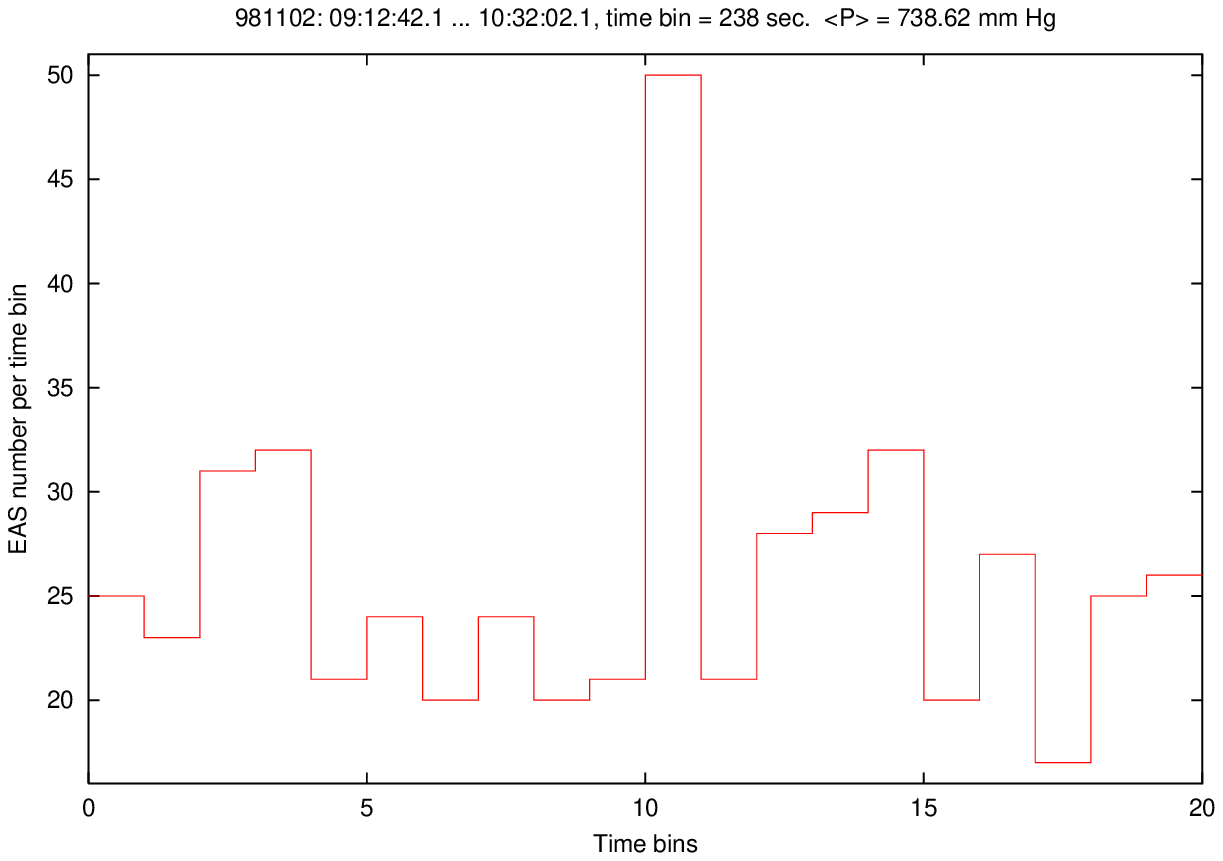}
}
\caption{An event registered on November~2, 1998, see Tables~2c and~3.
Top row: a cluster selected for $T=40$~sec (left), the first of the
clusters selected for $T=50$, 55~sec (right).
Middle row: two clusters found for $T=60$~sec (left) (56-second bins
are used for the first cluster, 47-second bins are used for the
second cluster); the ``interior structure" of this group (right): the
first cluster occupies bins No.~3--30, the second cluster occupies
bins No.~16--39.
Bottom row: one of the clusters selected for $T=80$~sec (left):
it consists of the showers that form two clusters found for $T=60$~sec,
see the text; the outer group of clusters (right).
}
\label{Fig:981102a}
\end{figure}

\begin{figure}
\centerline{
\includegraphics[width=8.4cm]{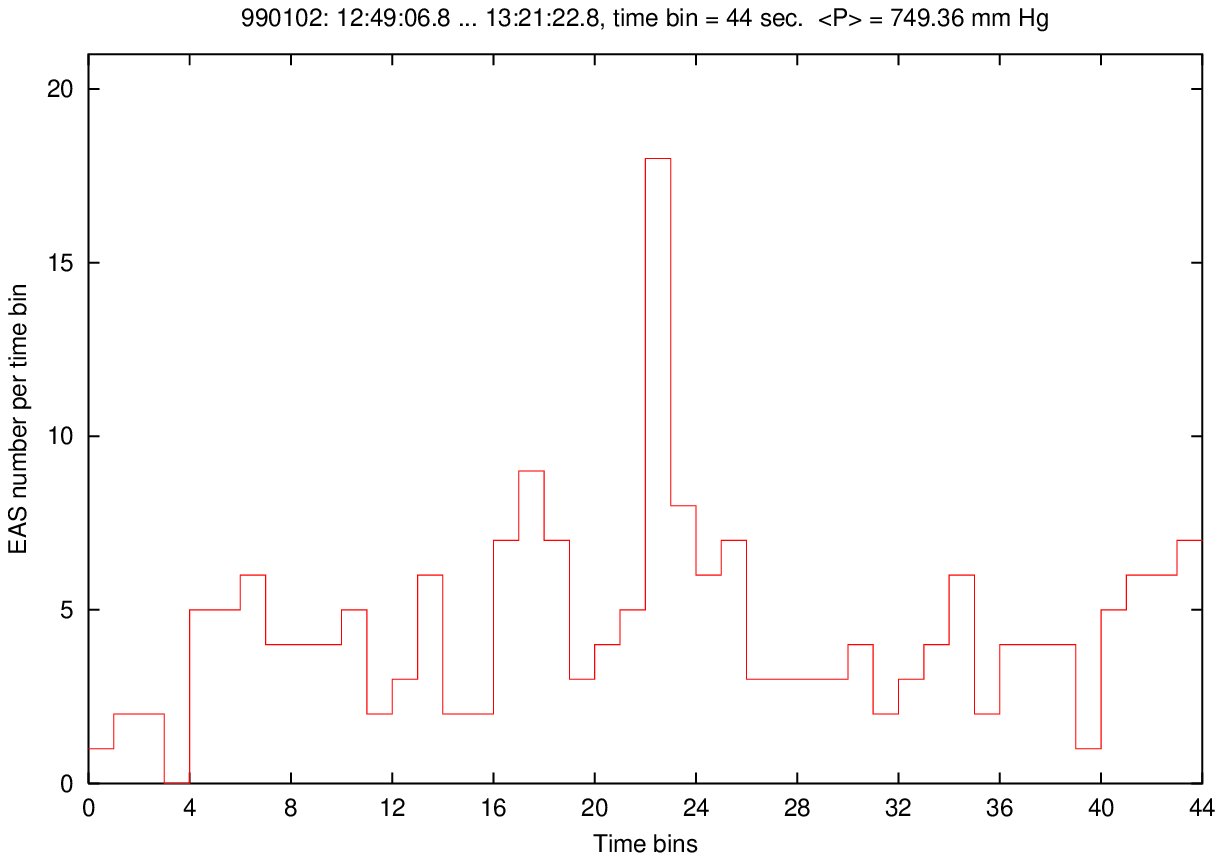}
\includegraphics[width=8.4cm]{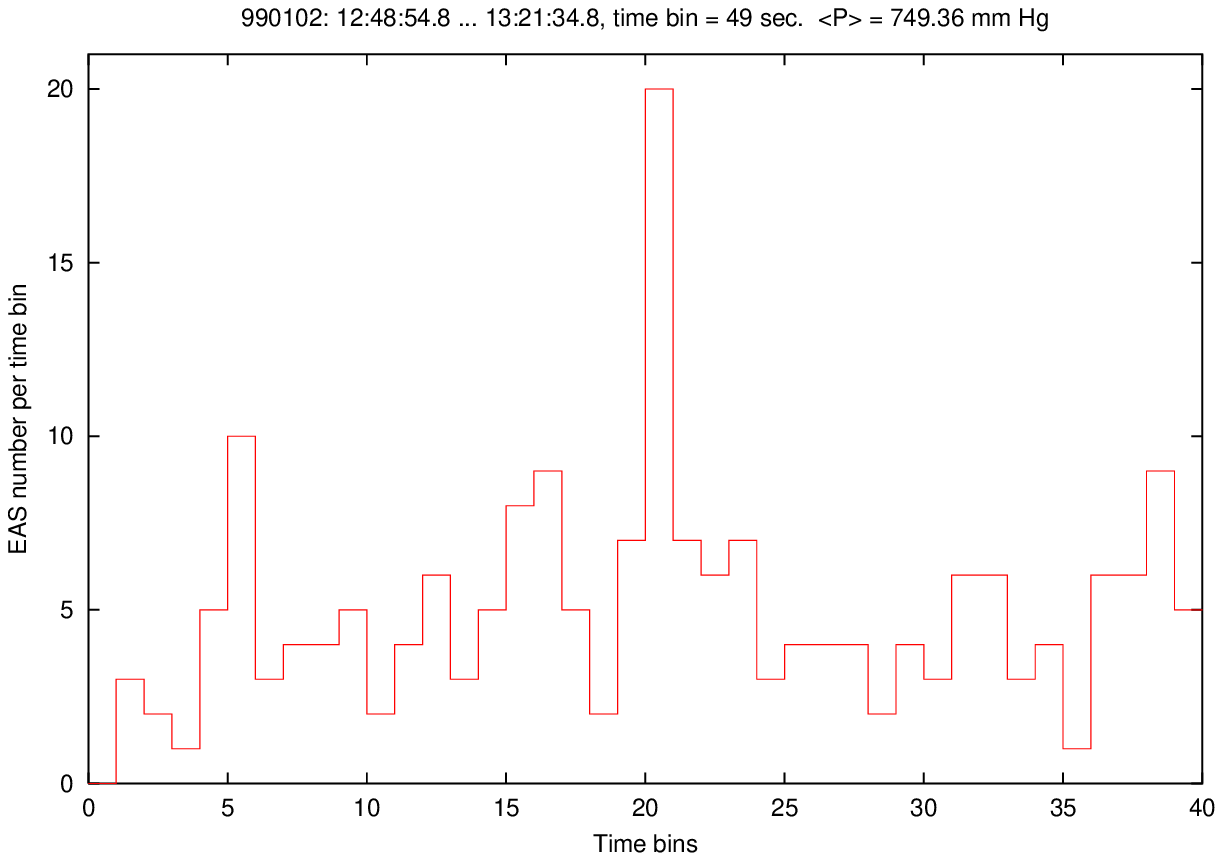}
}
\caption{
Two of the clusters found on January~2, 1999:
a single cluster selected for $T=40$~sec (left) and
the first of the clusters selected for $T=45$, 50~sec (right),
see Tables~2c, 3 and the text.
Notice that both clusters begin simultaneously.
}
\label{Fig:990102a}
\end{figure}

\begin{figure}
\centerline{
\includegraphics[width=8.4cm]{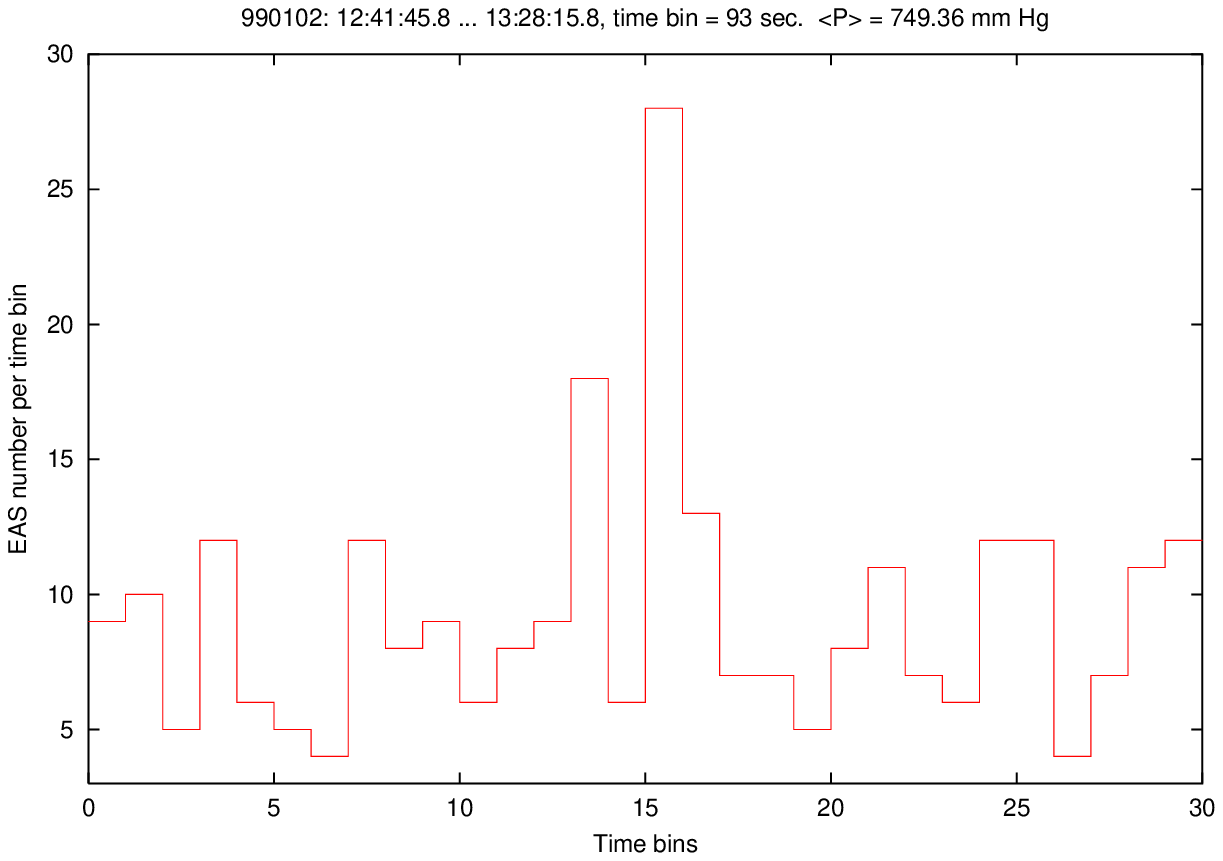}
\includegraphics[width=8.4cm]{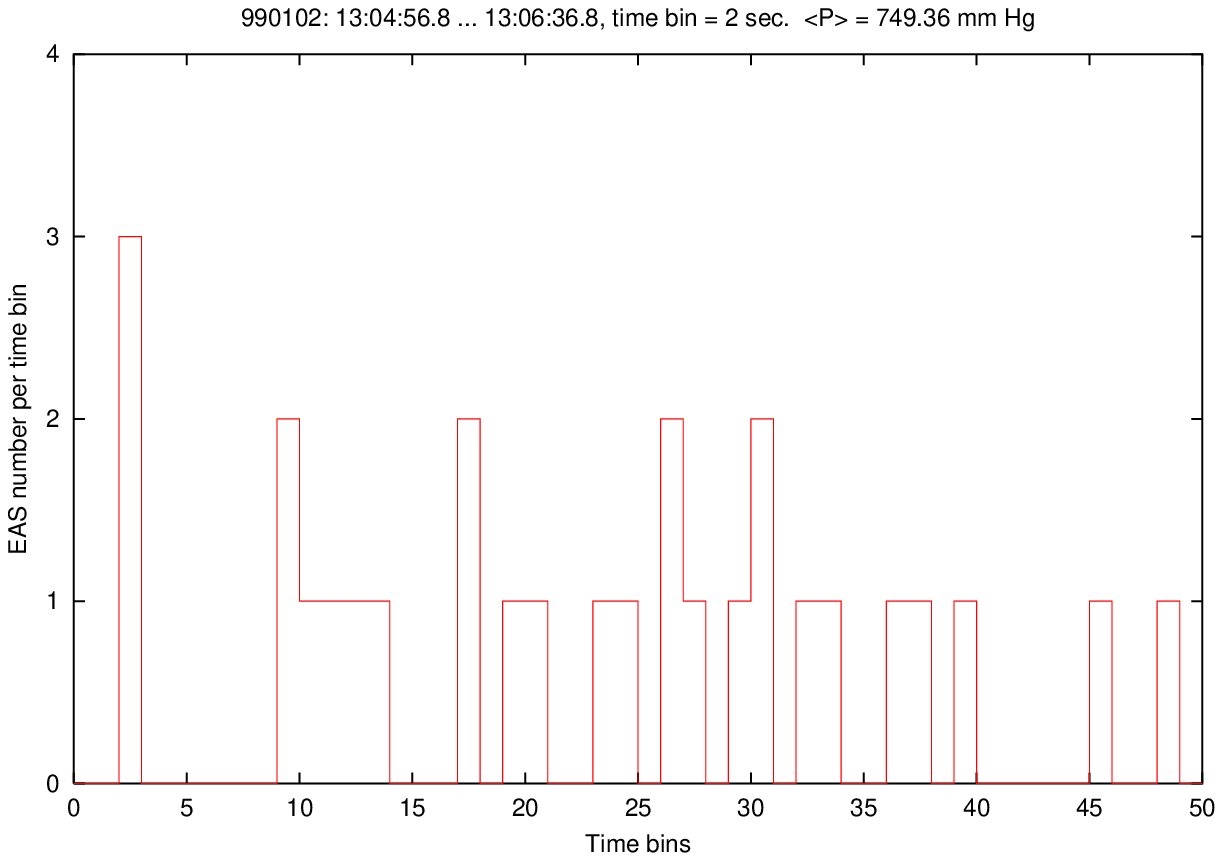}
}
\caption{
A single cluster found on January~2, 1999 for $T=90$~sec (left)
and its ``interior structure"  (right), see Tables 2b and~3.
Notice that the cluster selected for $T=40$~sec occupies bins
No.~9--31.
}
\label{Fig:990102b}
\end{figure}

\begin{figure}[!t]
\centerline{
\includegraphics[width=8.4cm]{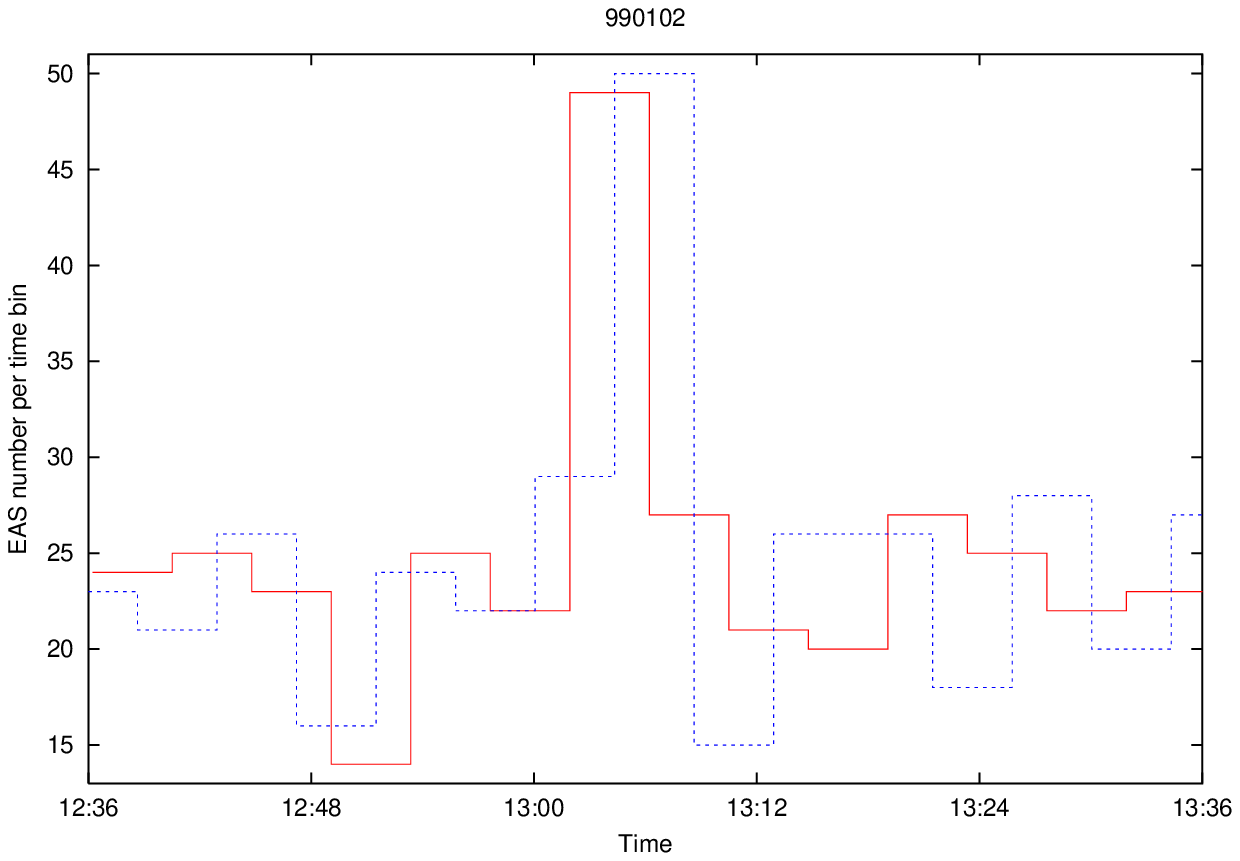}
\includegraphics[width=8.4cm]{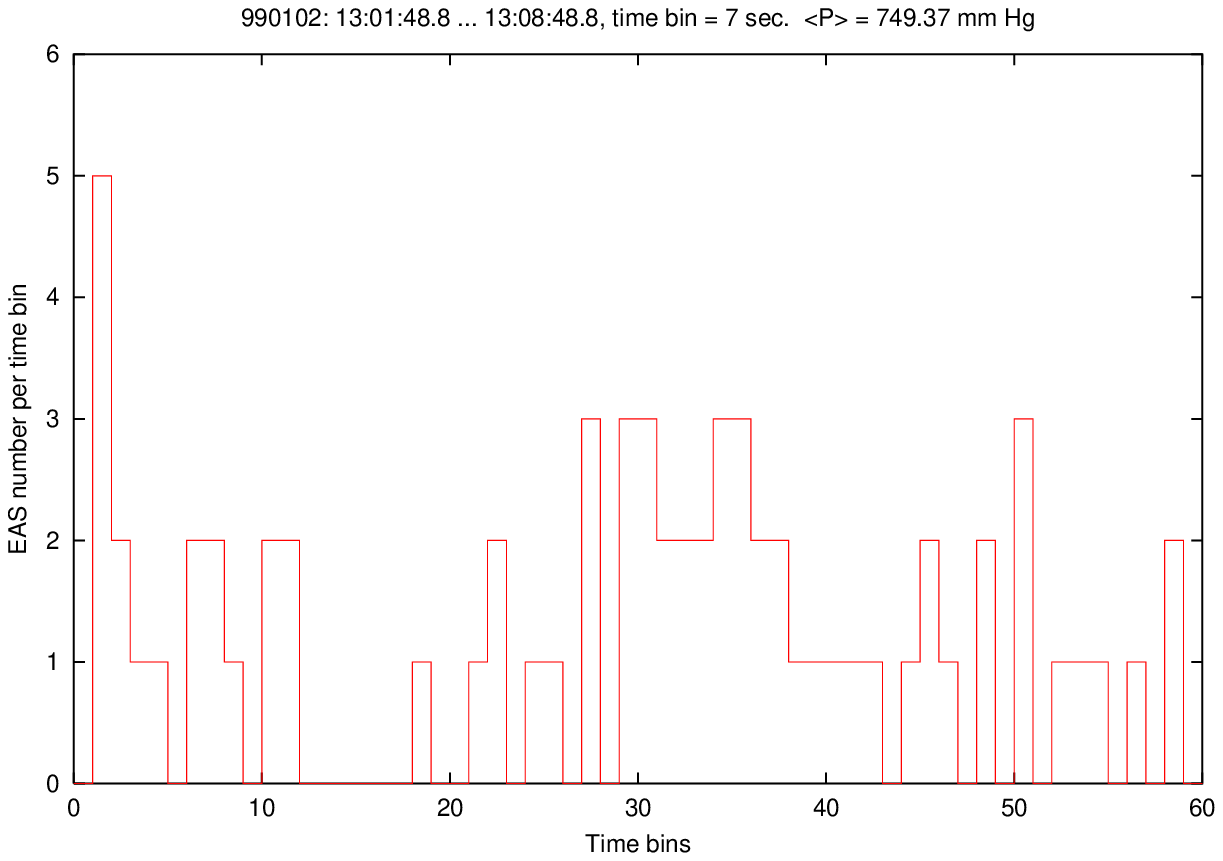}
}
\caption{
Left plot:
two ``shifted" clusters selected for $T=4$~min on January~2, 1999,
see Tables~2c, 3, and the text.
Time bins are equal to 257~sec.
Right plot: the ``interior structure" of these clusters.
The first cluster occupies bins No.~2--38, the second cluster
occupies bins No.~22--59.
Notice that the clusters shown in Fig.~\ref{Fig:990102a} occupy
bins No.~30--36 and 30--37 respectively.
}
\label{Fig:990102c}
\end{figure}

\begin{itemize}
  \item
     \textbf{02.11.98}:
	First of all, there is a single cluster found for $T=40$~sec,
     see the top left plot in Fig.~\ref{Fig:981102a}.
	Next, for $T=45, 50$, and $55$~sec there are sequences of~4,
	3, and~2 consecutive clusters, which begin simultaneously
	with the 40-second cluster.
	All three sequences cover the same range of EAS and contain
	the 40-second cluster.
     It is remarkable that the first of the clusters selected for $T=50$
     and $55$~sec coincide and cover the complete range of EAS arrival
     times, containing the following clusters inside.
     Thus this one is the most pronounced of the clusters.
     It has very small $\Padj = 4.0\times10^{-9}$:
     22 showers arrive in 46.58~sec with $\Tadj=48.32$~sec,
     see the top right plot in Fig.~\ref{Fig:981102a}.

	For $T=60$~sec, the situation changes: there are two clusters,
     but to the contrary to the above sequences, they do not begin
	at consecutive arrival times.
	The first cluster consists of 22 showers arriving in 55.75~sec
	($\Tadj = 57.84$~sec, $\Padj = 8.4\times10^{-8}$).
     The second cluster appears 26.4 seconds later and is more
	intensive: it also contains 22 EAS, but they arrive in 46.58~sec
     ($\Tadj=48.32$~sec, $\Padj = 4.0\times10^{-9}$),
     see Fig.~\ref{Fig:981102a}, the middle row.

     Next, for $T=65\dots90$~sec the range of EAS becomes a bit wider.
	It is interesting that for $T=80$~sec there appears a cluster
	that covers EAS No.~3717--3744 thus containing the 60-second
	group (28 showers), see the left bottom plot in
	Fig.~\ref{Fig:981102a}.
     For this cluster, $\Treal=72.99$~sec, $\Tadj=75.73$~sec, and
     $\Padj = 3.1\times10^{-9}$.
     The same cluster is identified for $T=85$~sec.
	This makes us conclude that two ``shifted" clusters found for
     $T=60$~sec do not present different astrophysical events but
     are two ``bursts" within one event.

     Finally, for $T=2$, 3, and 4~min the range of EAS covered by
	clusters becomes more wide, though the right end shifts very little:
	the first event in $T=4$~min group begins 124.12~sec earlier than
     the 40-second cluster, but ends only 45.75~sec later, see Table~2c.
     Thus the highest count rate for this event is shifted
     toward its end.

	The group of clusters selected for $T=4$~min is shown in the
	right bottom plot of Fig.~\ref{Fig:981102a}.

  \item
     \textbf{02.01.99}:
	Similar to the above event, the shortest cluster in this event
     is a single one.
     This cluster is shown in Fig.~\ref{Fig:990102a}.
     As~$T$ grows, the number of clusters increases.
     As one can see from Table~3, for $T=45$ and 50~sec we have found
     groups that contain two and three consecutive clusters
     respectively.
	It is remarkable that for each of these two groups,
	the first cluster appears simultaneously with the shortest one,
	while the right boundary of each sequence slightly moves.
     For both groups, the first cluster covers the range of
	EAS No.~4231--4250 thus containing 20 showers arriving
	in 48.72~sec, see Fig.~\ref{Fig:990102a}.
     For this cluster, $\Tadj=44.98$~sec, $\Padj = 2.8\times10^{-8}$.

     Next, for $T=55\dots90$~sec all events begin at shower No.~4228
     thus appearing nearly 14~sec earlier than the shortest cluster.
     It is interesting that for $T=80$ and 85~sec the first cluster
	begins at EAS No.~4228, while the second one is ``shifted" to the EAS
     No.~4231, which is the first shower
	for the shortest cluster in this event.
     For $T=80$~sec, the first cluster consists of 26 EAS arriving
     in 75.85~sec with $\Tadj=70.03$~sec and $\Padj=1.0\times10^{-8}$.
     The second cluster is a bit less intensive:
     it also consists of 26 EAS, but they occupy
     85.96~sec with $\Tadj=79.37$~sec and $\Padj=1.1\times10^{-7}$.
     No other clusters are identified between them.
     Finally, for $T=90$~sec a single cluster is found, see Table~2c
	and Fig.~\ref{Fig:990102b}.
     Notice that it consists of exactly the same EAS that are covered
     by three clusters selected for $T=75$~sec.
     Notice also that the right boundary of this cluster has moved
     by 35.92~sec in comparison with the shortest cluster, while
     the left boundary has shifted by 13.95~sec only.
     Thus the peak count rate is closer to the beginning of the event.
	This is clearly seen in the figure.

     For $T=2$ and 3~min, besides other clusters, there are clusters that
	begin at EAS No.~4228 and~4231.
	For $T=2$~min, they cover the range of EAS No.~4228--4260 and
	4231--4261 and contain 33 and 31 showers with
     $\Padj = 1.0\times10^{-7}$ and
     $\Padj = 2.9\times10^{-7}$ respectively.
	For $T=3$~min, they cover the range of EAS No.~4228--4269 and
	4231--4270 and contain 42 and 40 showers with
     $\Padj = 1.0\times10^{-7}$ and
     $\Padj = 7.3\times10^{-7}$ respectively.

	An interesting situation takes place for $T=4$~min.
	There are only three clusters in the group, and the first one
	covers the range of EAS No.~4204--4252.
	It consists of 49 showers arrived in 256.28~sec.
	The other two clusters cover the range of EAS No.~4223,
	4224--4272 thus containing 50 and 49 showers respectively.
     The second cluster begins 144.95~sec later than the first one and
	ends 145.66~sec later (so that their durations are nearly equal),
	see Fig.~\ref{Fig:990102c}.
     This allows us to assume that these two clusters are found not
     because of the search method, but may represent an astrophysical
     process that consists of two consecutive bursts.

     For $T=5$~min, the beginning of the group of clusters moves more
     to the left, so that four clusters in this group cover the range
     of EAS No.~4196--4253, 4197--4254, and 4203--4260, 4204--4261
	respectively, thus splitting into two subgroups.
	Notice that the last cluster begins simultaneously with the first
	4-minute cluster.

	For $T=6$~min, the group again splits into two subgroups
	of consecutive clusters.
	The first one begins at EAS No.~4195 (at 13:00:44.42)
	and consists of four clusters,
	the last of them ending at EAS No.~4263 (at 13:07:28.11).
	The second subgroup consists of five clusters, the first of them
	beginning at EAS No.~4210 (at 13:01:20.17) and the last one
	ending at EAS No.~4270 (at 13:08:25.23).

	Finally, for $T=7$~min there are only five consecutive clusters.
	They cover the widest range of EAS arrival times that occupies nearly
     eight minutes, see Tables~2c, 3, and the left plot in
	Fig.~\ref{Fig:990102d}.
	An analysis of the ``interior structure" of this group of clusters
	clearly demonstrates how two subgroups of clusters appear in this event,
	see the right plot in Fig.~\ref{Fig:990102d}.
\end{itemize}
\newpage
{\narrower\noindent
     Table 2c: Events that have a complicated structure of embedded
     clusters.
     See notation in Tables~1 and~2a.

}
\begin{center}
\begin{tabular}{|c|c|c|c|c|c|c|c|c|c|}
\hline
Date    & $P$,& $T$, &Beginning  &End        &\Treal,&\NAS&\Ncl&\Tadj,&\Padj           \\
dd.mm.yy&mm Hg&  sec &hh:mm:ss   &hh:mm:ss   &sec    &    &    &sec   &                \\
\hline
02.11.98&738.6&  40  &09:54:56.30&09:55:34.04& 37.74& 18 & 1  & 39.15&$8.3\times10^{-8}$\\
        &     &$\cdots$&         &           &      &    &    &      &                 \\
        &     &  60  &09:54:29.89&09:55:42.88&      & 28 & 2  &      & \\
        &     &$\cdots$&         &           &      &    &    &      &                 \\
        &     &  90  &09:54:19.67&09:55:56.61&      & 31 & 4  &      & \\
        &     &  120 &09:53:39.25&09:56:19.79&      & 39 & 9  &      & \\
        &     &$\cdots$&         &           &      &    &    &      &                 \\
        &     &  240 &09:52:22.18&09:56:19.79&      & 50 & 2  &      &                 \\
\hline
02.01.99&749.4&  40  &13:05:14.81&13:05:57.82& 43.01& 18 & 1  & 39.71&$1.0\times10^{-7}$\\
        &     &$\cdots$&         &           &      &    &    &      &                 \\
        &     &  65  &13:05:00.86&13:06:16.71&      & 26 & 4  &      & \\
        &     &$\cdots$&         &           &      &    &    &      &                 \\
        &     &  90  &13:05:00.86&13:06:33.74& 92.88& 28 & 1  & 85.75&$3.9\times10^{-8}$\\
        &     & 120  &13:04:20.77&13:07:28.11&      & 41 & 10 &      & \\
        &     &$\cdots$&         &           &      &    &    &      &                 \\
        &     & 240  &13:01:55.82&13:08:37.76&      & 69 & 3  &      & \\
        &     &$\cdots$&         &           &      &    &    &      &                 \\
        &     & 420  &13:00:44.42&13:08:37.76&      & 78 & 5  &      &                 \\
\hline
08.01.99(b)&726.7& 55  &00:20:50.36&00:21:36.94&46.58 & 21 & 1  & 54.95&$1.5\times10^{-7}$\\
        &      & 60  &00:20:50.36&00:21:40.29&49.93 & 22 & 1  & 58.90&$1.1\times10^{-7}$\\
        &      & 65  &00:20:22.51&00:21:33.15&      & 30 & 5  &      & \\
        &      &$\cdots$&        &           &      &    &    &      &                 \\
        &      & 90  &00:20:11.97&00:21:40.29&      & 34 & 7  &      & \\
	   &      & 120 &00:19:32.48&00:21:40.29&      & 43 & 13 &      & \\
	   &      & 180 &00:18:59.36&00:23:08.77&      & 58 & 13 &      & \\
        &      &$\cdots$&        &           &      &    &    &      &                 \\
        &      & 420 &00:19:32.48&00:25:49.64&      & 79 & 5  &      & \\
        &      &$\cdots$&        &           &      &    &    &      &                 \\
        &      & 660 &00:19:32.48&00:28:39.25&546.77&105 & 1  &645.07&$1.8\times10^{-7}$\\
        &      &$\cdots$&        &           &      &    &    &      &                 \\
        &      & 900 &00:19:47.03&00:32:26.31&759.28&134 & 1  &895.75&$4.0\times10^{-7}$\\
\hline
\end{tabular}
\end{center}
\bigskip

\begin{figure}
\centerline{
\includegraphics[width=8.4cm]{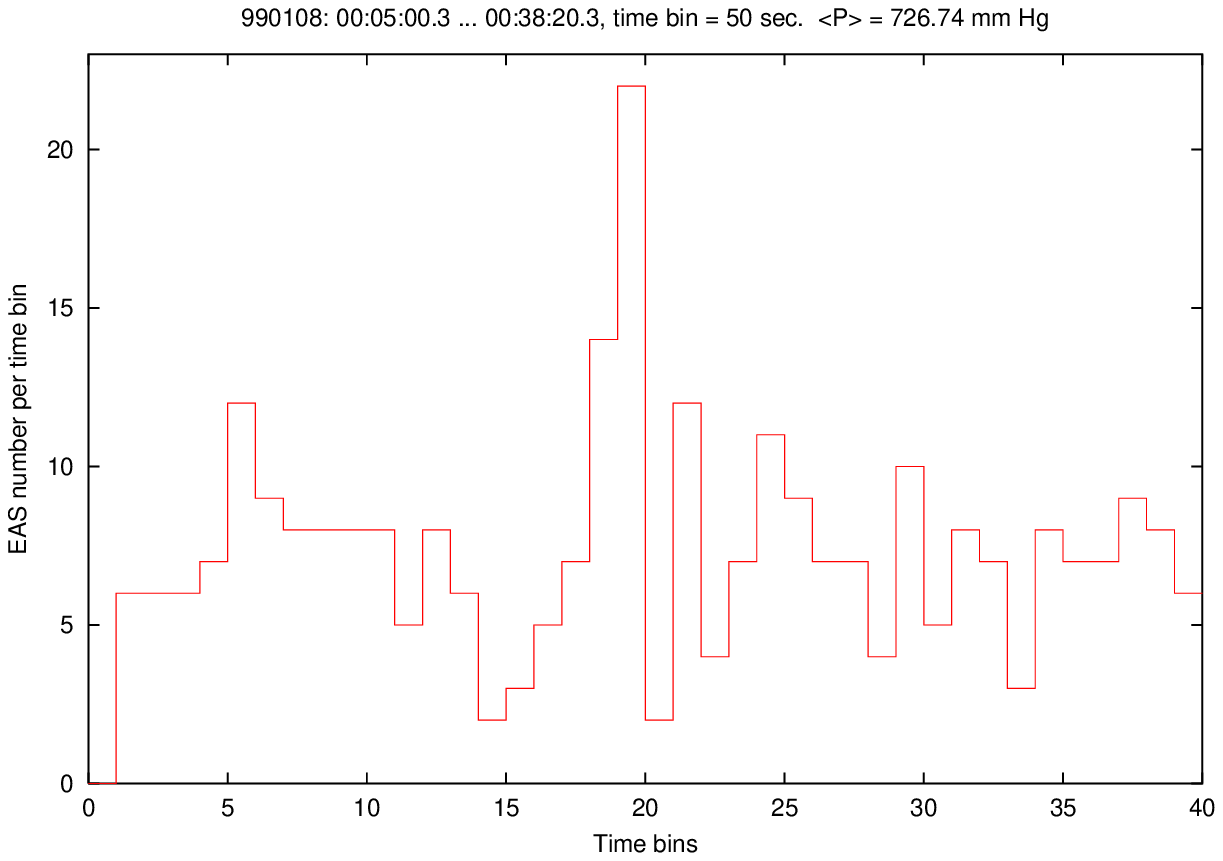}
\includegraphics[width=8.4cm]{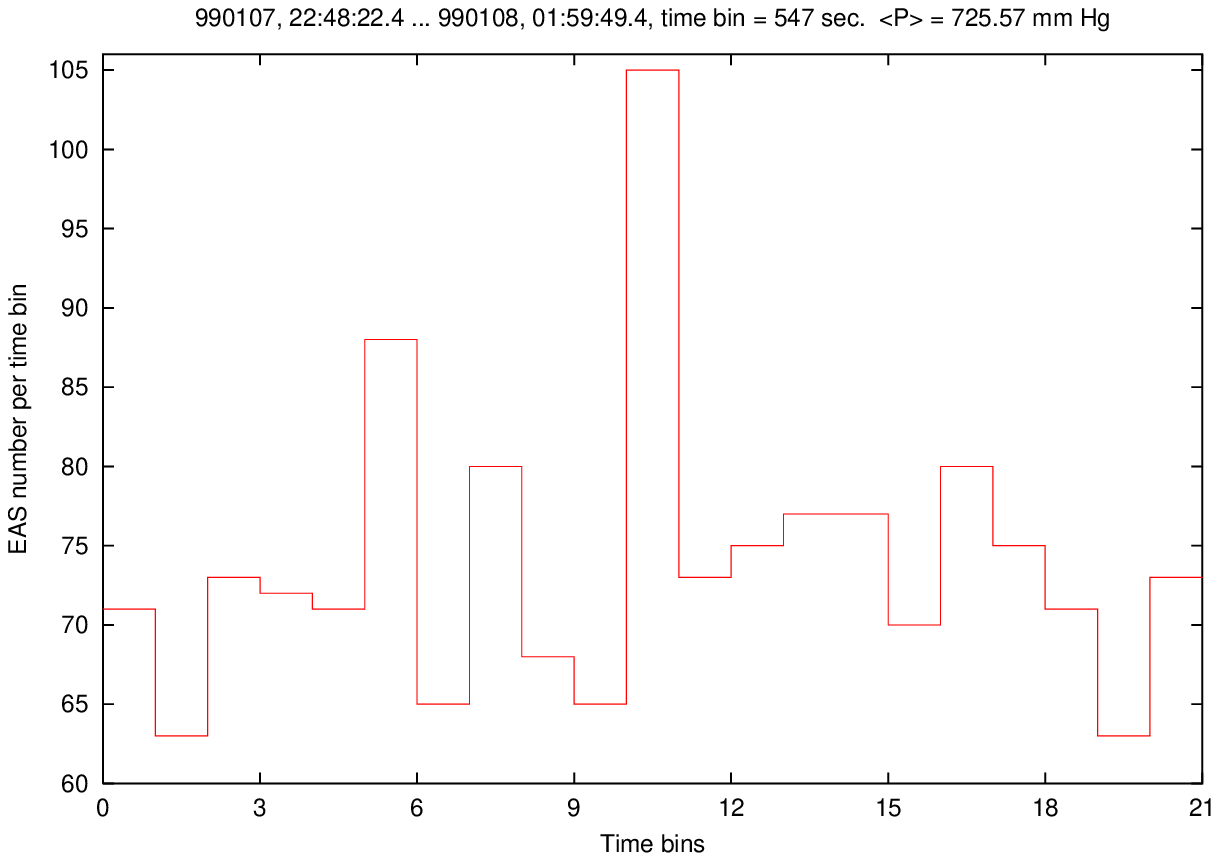}
}
\centerline{
\includegraphics[width=8.4cm]{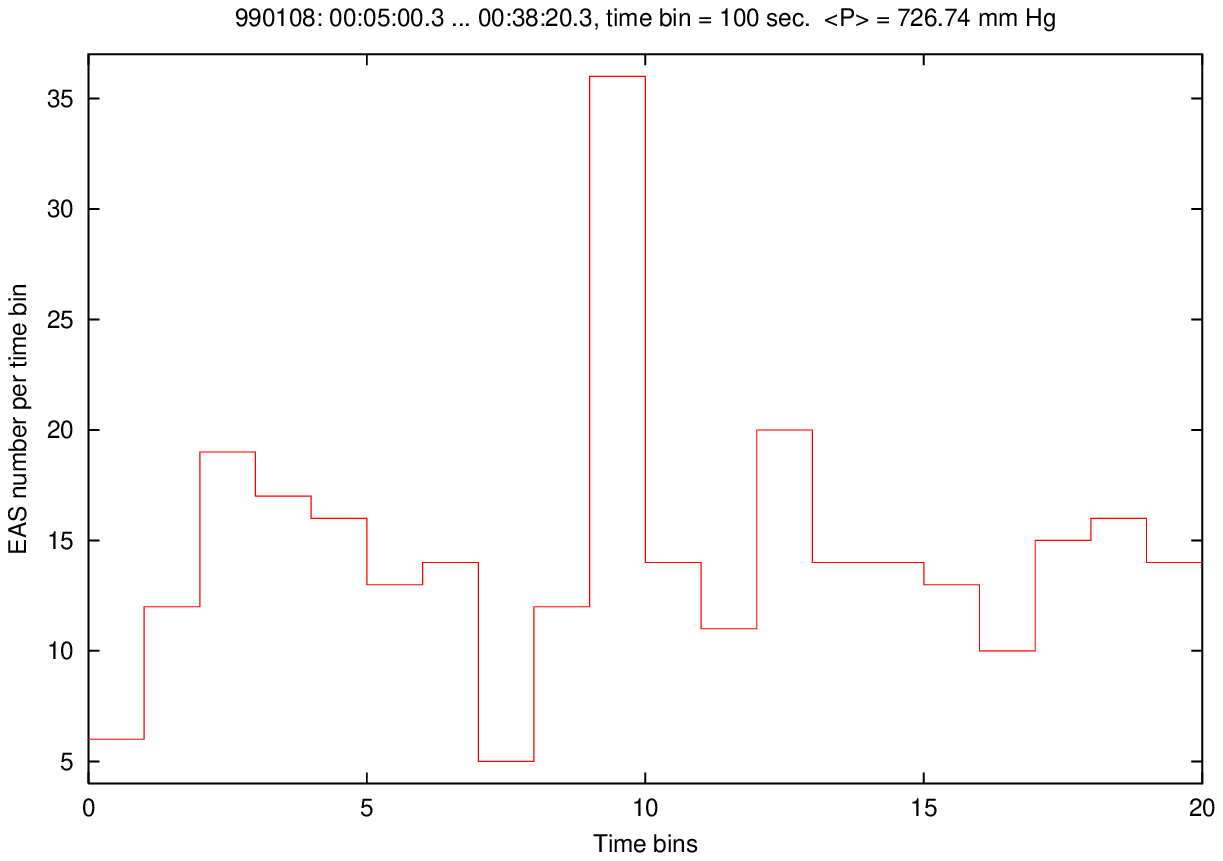}
\includegraphics[width=8.4cm]{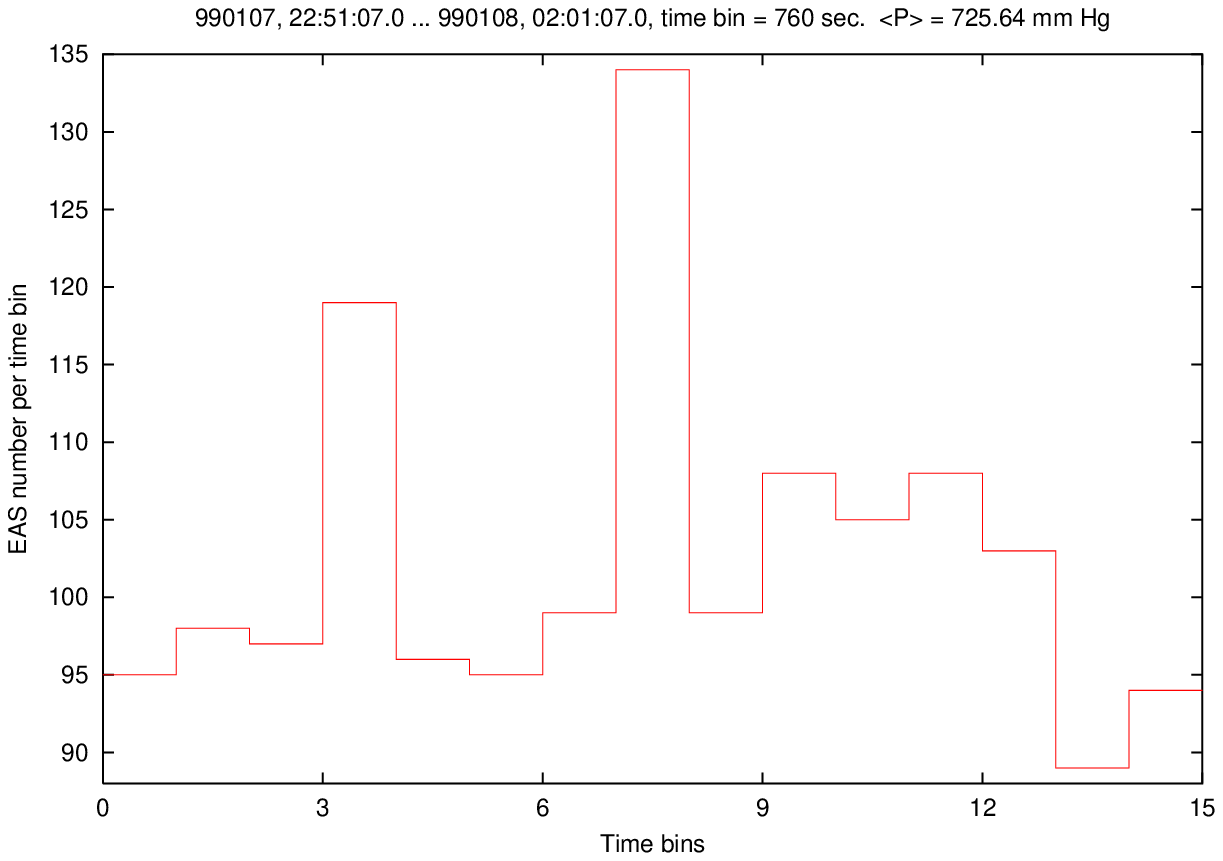}
}
\caption{Some of the clusters that constitute an event
registered on January~8, 1999(b), see Tables~2c and~3.
Top row: single clusters found for $T=60$~sec and $T=11$~min.
Bottom row: the eighth of the clusters selected for $T=2$~min
and a single cluster found for $T=15$~min.
Notice that clusters in the left column end simultaneously.
}
\label{Fig:990108a}
\end{figure}

\begin{figure}
\centerline{
\includegraphics[width=12cm]{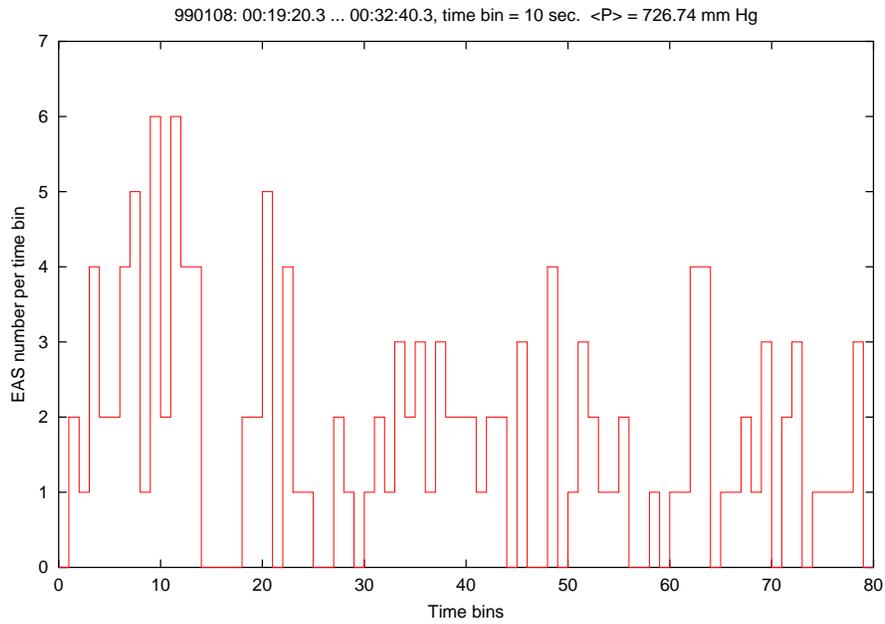}
}
\caption{
The ``interior structure" of the outer cluster found for the event
registered on January~8, 1999(b).
Bins No.~10--14 represent the single cluster selected for $T=60$~sec,
see Table~2c and Fig.~\ref{Fig:990108a}.
Bins No.~5--14 represent the eighth of the clusters selected for
$T=2$~min; the whole group occupies bins No.~2--14.
Bins No.~39 and~56 are the last bins for the clusters found for
$T=7$ and~11~min respectively.
}
\label{Fig:990108sub}
\end{figure}

\begin{itemize}
  \item
     \textbf{08.01.99(b)}:
     This is an event with the biggest number of clusters.
     As one can see from Table~3, separate clusters are found for $T=55$
     and 60~sec and for $T=11$ and 15~min.
     The last three clusters are shown in Fig.~\ref{Fig:990108a}.
     Surprisingly enough, but no clusters were found for $T=6$ and
     13~min.

     If we compare the clusters selected for $T=55$ and 60~sec with
     longer ones then we find that for $T=65\dots90$~sec, the left
     boundary of the sequences of clusters is shifted from EAS No.~164 to
     EAS No.~152--154, while the right boundary remains practically
     unchanged, see Tables~2c and~3.
     Beginning with $T=2$~min, the left boundary of the selected groups
     jumps further to EAS No.~141--145.
     Then at $T=3$~min, the right boundary begins to move
     considerably to the right.
     It is interesting to notice that while the right boundary
     moves monotonically as~$T$ grows, the left boundary fluctuates
     around EAS No.~143, so that the longest group ($T=15$~min)
     appears later than the majority of the groups selected for
     $T=2\dots14$~min.
\end{itemize}
\bigskip
{\narrower
\noindent
     Table 3: Three events with a complicated structure of embedded
	clusters.
     For each event, the first number gives the number of clusters
     for a given~$T$; letter~``c" means that the corresponding clusters
     begin at consecutive arrival times.
     Then the range of EAS numbers for each event follows.

}
\begin{center}
\begin{tabular}{|rl|rc|rc|rc|}
\hline
   \multicolumn{2}{|c}{$T$} &
   \multicolumn{2}{|c}{02.11.98}&
   \multicolumn{2}{|c}{02.01.99}&
   \multicolumn{2}{|c|}{08.01.99(b)}       \\
\hline
   40 &sec&  1 :& 3723--3740   &   1 :& 4231--4248   &      &             \\
   45 &   &  4c:& 3723--3744   &   2c:& 4231--4250   &      &             \\
   50 &   &  3c:& 3723--3744   &   3c:& 4231--4252   &      &             \\
   55 &   &  2c:& 3723--3744   &   4 :& 4228--4253   &   1 :& 164--184    \\
   60 &   &  2 :& 3717--3744   &   4 :& 4228--4253   &   1 :& 164--185    \\
   65 &   &  4 :& 3717--3745   &   4c:& 4228--4253   &   5 :& 154--183    \\
   70 &   &  5 :& 3715--3744   &   4c:& 4228--4254   &   6 :& 153--184    \\
   75 &   &  5c:& 3716--3744   &   3 :& 4228--4255   &   8 :& 153--185    \\
   80 &   &  5c:& 3715--3744   &   2 :& 4228--4256   &   9c:& 152--185    \\
   85 &   &  4c:& 3715--3744   &   2 :& 4228--4257   &   7c:& 153--185    \\
   90 &   &  4c:& 3715--3745   &   1 :& 4228--4255   &   7c:& 152--185    \\
    2 &min&  9c:& 3710--3748   &  10c:& 4223--4263   &  13c:& 143--185    \\
    3 &   &  7 :& 3701--3748   &   8 :& 4223--4270   &  13 :& 141--198    \\
    4 &   &  2c:& 3699--3748   &   3 :& 4204--4272   &  10c:& 143--200    \\
    5 &   &     &              &   4 :& 4196--4261   &   5 :& 141--203    \\
    6 &   &     &              &   9 :& 4195--4270   &      &             \\
    7 &   &     &              &   5c:& 4195--4272   &   5 :& 143--221    \\
    8 &   &     &              &      &              &   7c:& 143--230    \\
    9 &   &     &              &      &              &   5 :& 143--237    \\
   10 &   &     &              &      &              &   7c:& 143--245    \\
   11 &   &     &              &      &              &   1 :& 143--247    \\
   12 &   &     &              &      &              &   3c:& 145--258    \\
   13 &   &     &              &      &              &      &             \\
   14 &   &     &              &      &              &   3c:& 143--271    \\
   15 &   &     &              &      &              &   1 :& 145--278    \\
\hline
\end{tabular}
\end{center}
\bigskip
\begin{itemize}
\item[]
     Some of the clusters that form the groups found for $T=2$ and
     3~min are really remarkable.
     In particular, the fourth of the clusters selected for $T=2$~min
     begins at EAS No.~146 (at 00:19:52.63) and consists of 38
     showers arriving in 100.51~sec with
     $\Tadj=118.58$~sec and $\Padj=3.1\times10^{-10}$;
	the cluster ends simultaneously with a group found for $T=65$~sec
	and contains it inside itself.
     There is another cluster inside this one: it begins at EAS
     No.~147 and ends simultaneously with the outer cluster.
     For this cluster, $\Tadj=116.19$~sec, $\Padj=6.1\times10^{-10}$.
     The eighth cluster is also worth mentioning:  it begins at EAS
     No.~150 and consists of 36~showers arriving in 99.19~sec with
     $\Tadj=117.03$~sec and $\Padj=2.4\times10^{-9}$.
	This cluster ends simultaneously with the single cluster
     selected for $T=60$~sec, see the left bottom plot in
     Fig.~\ref{Fig:990108a}.

     For $T=3$~min, there is a cluster with an extremely small
     $\Padj=2.3\times10^{-10}$.
     It covers the range of EAS No.~145--185 and consists of
     41~showers arriving in 113.25~sec with $\Tadj=133.62$~sec.
     It is remarkable that this cluster begins simultaneously with the
     cluster selected for $T=15$~min and ends simultaneously with the one
     found for $T=60$~sec.

     The groups selected for $T>3$~min do not contain any
     outstanding clusters.
     Thus we assume that this event is presented by the cluster
     selected for $T=15$~min with the peak count rate located
     at the most intensive interior clusters, see
     Fig.~\ref{Fig:990108sub} for the ``substructure" of this
     cluster.
\end{itemize}

     Now we see that though the presented clusters have been selected
     by the same formal procedure, they are not identical but differ in
     a considerable number of properties such that the duration, the
     interior structure, the location of the peak count rate, etc.
     It seems to be a great challenge to try to figure out which
     astrophysical events are responsible for the observed EAS clusters
     and the variety of their properties.


\section{Arrival Directions of EAS in Clusters} \label{sec:AD}

\begin{figure}
\begin{center}
\includegraphics[width=9.0cm]{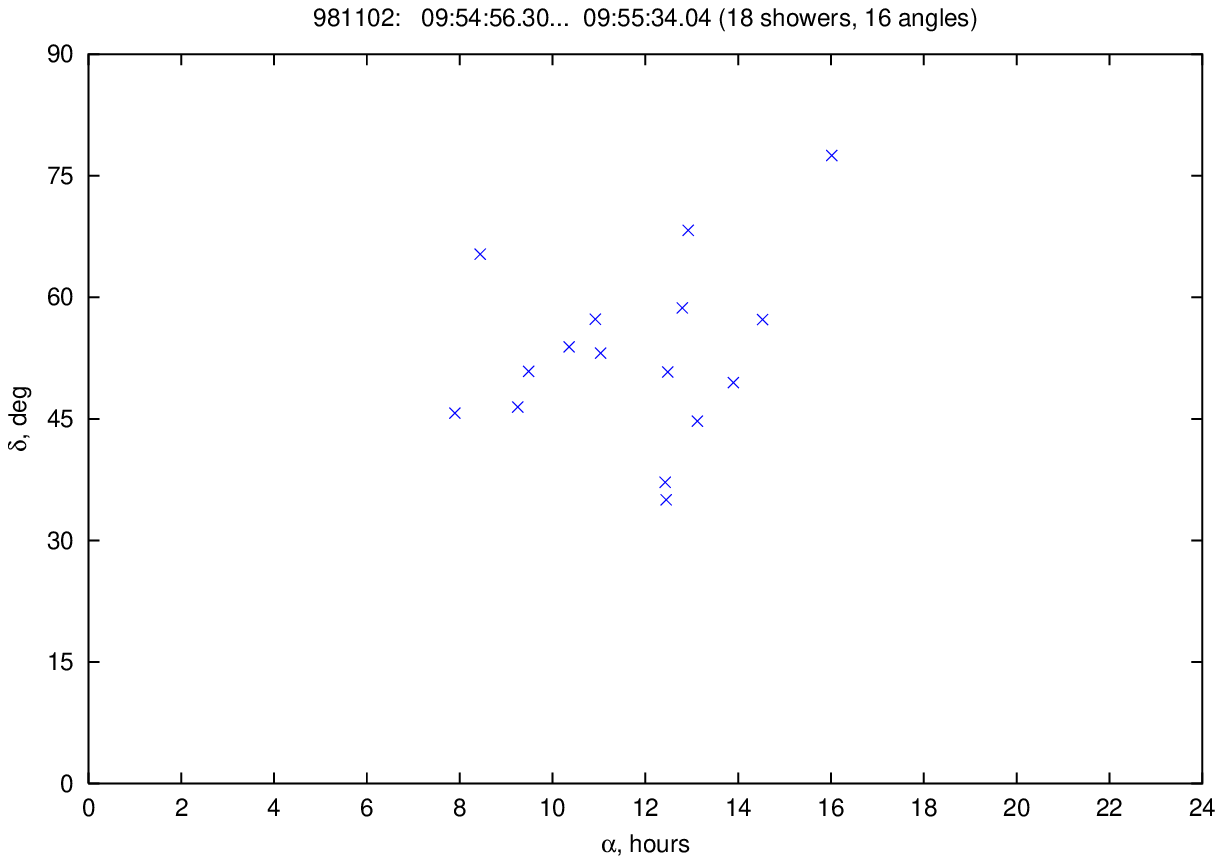}\\
\includegraphics[width=9.0cm]{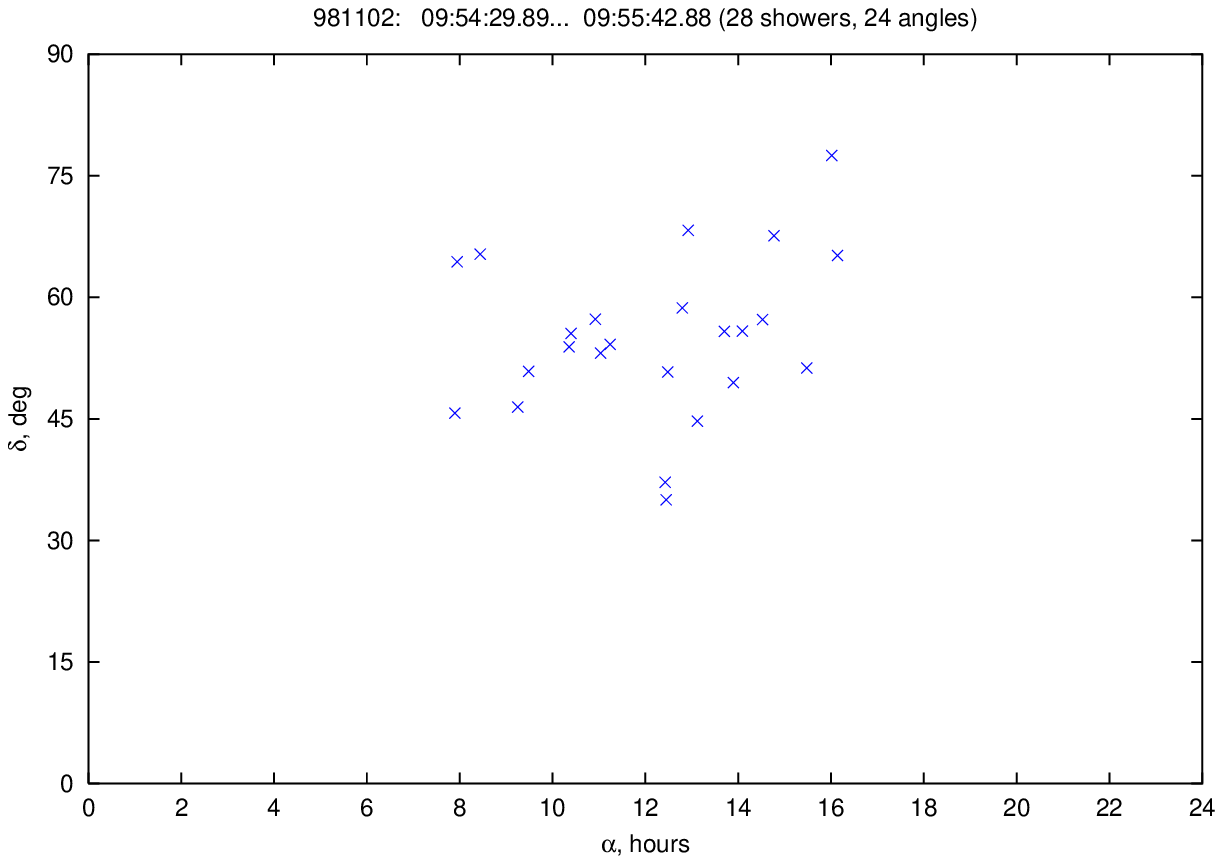}\\
\includegraphics[width=9.0cm]{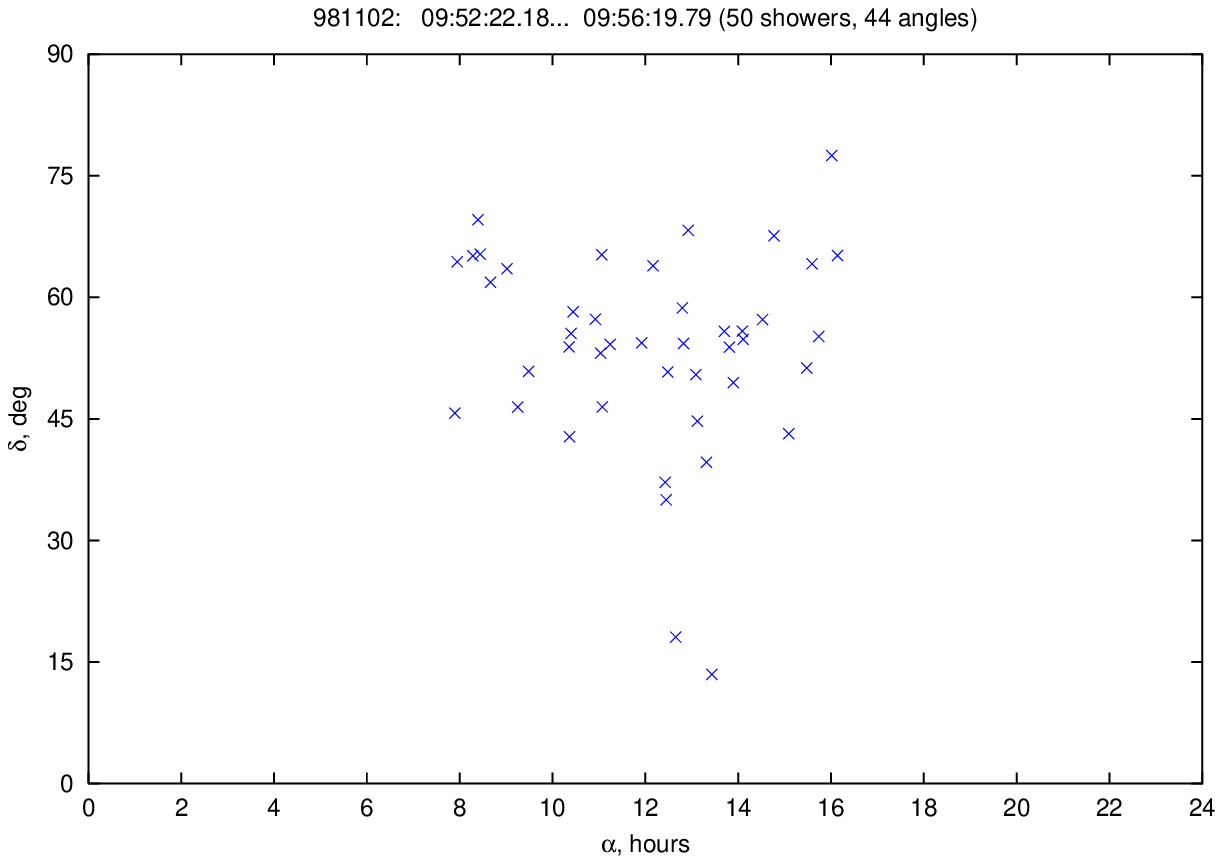}
\end{center}
\caption{Arrival directions of EAS in the cluster event registered on
November~2,1998.
One can see how showers with close arrival directions appear
if an interior cluster is compared with an outer one, see Tables~2b
and~2c.
The titles of the plots contain the total number of showers in
a cluster and the number of arrival directions (``angles")
determined.
}
\label{Fig:dirsA}
\end{figure}

\begin{figure}
\begin{center}
\includegraphics[width=12cm]{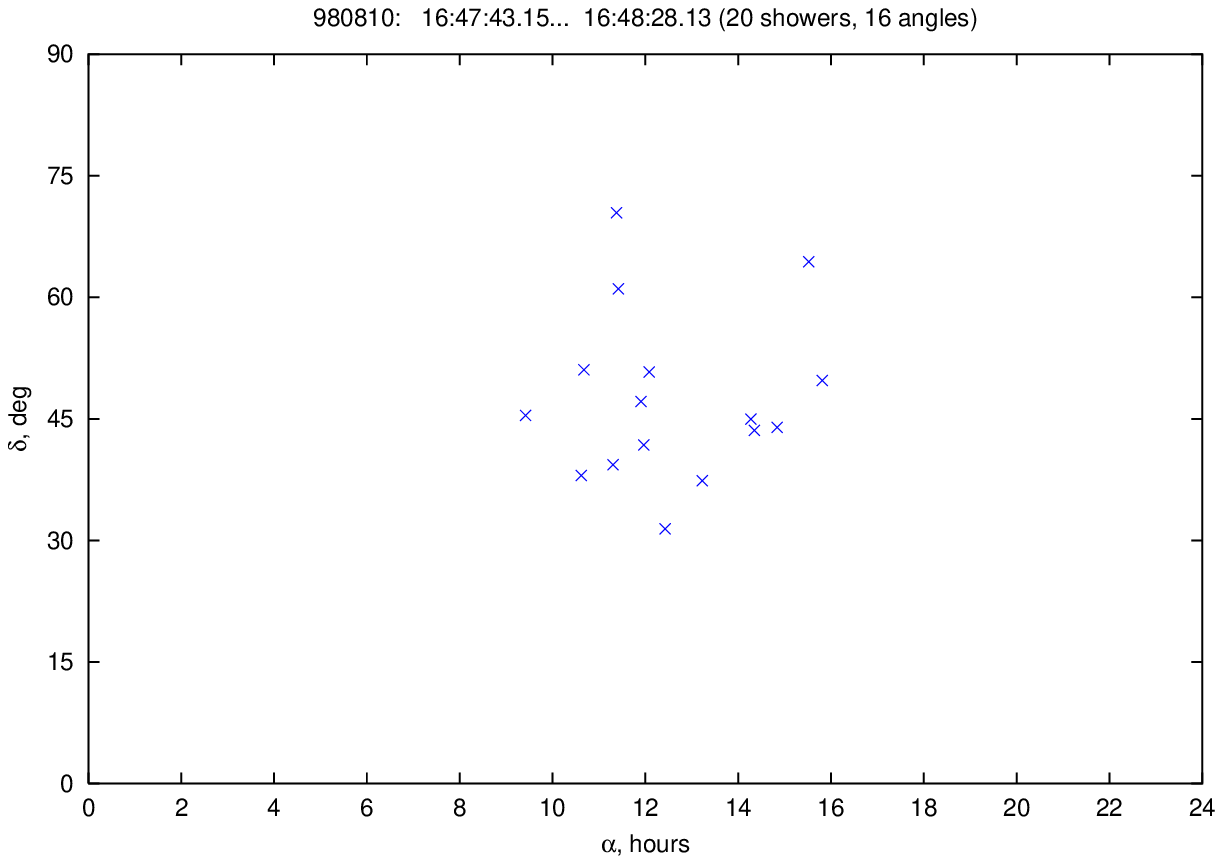}\\
\includegraphics[width=12cm]{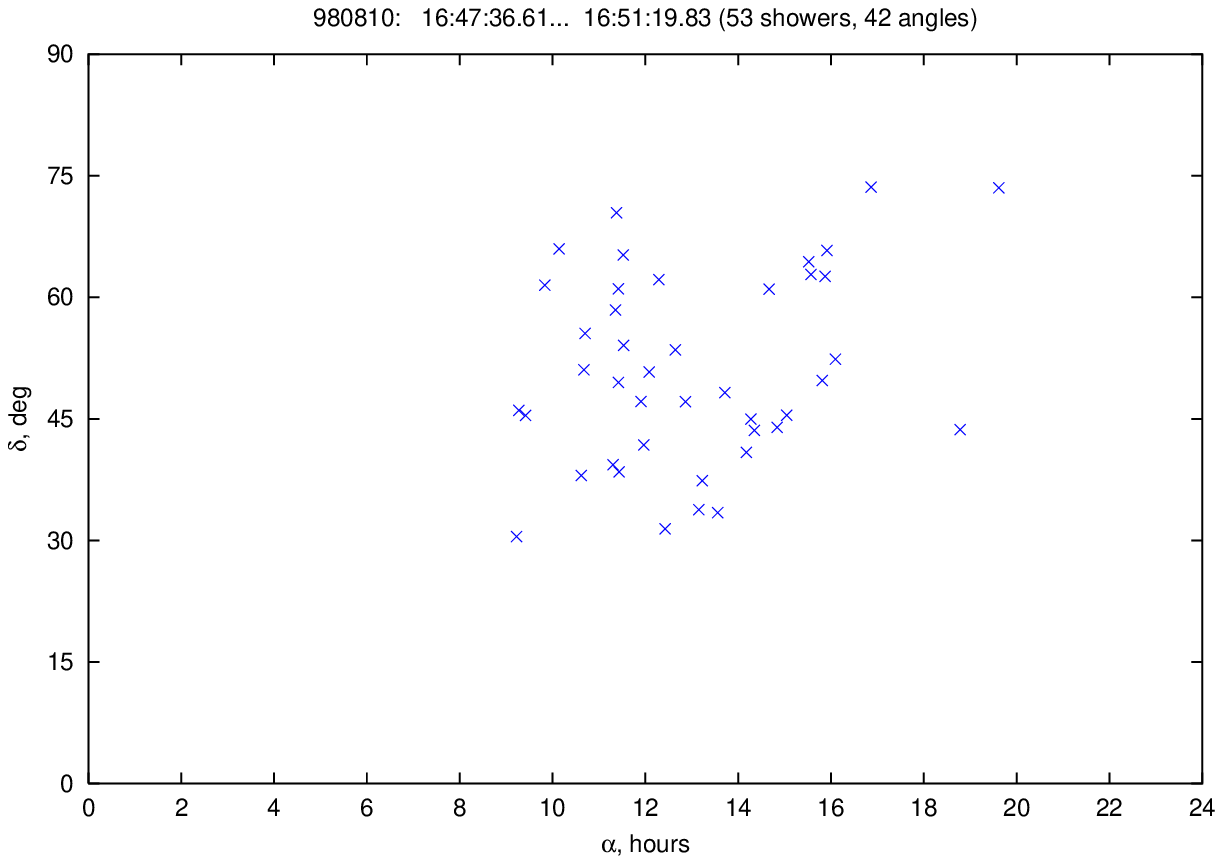}
\end{center}
\caption{Arrival directions of EAS in the cluster event
registered on August~10, 1998. }
\label{Fig:dirsC}
\end{figure}

\begin{figure}
\begin{center}
\includegraphics[width=12cm]{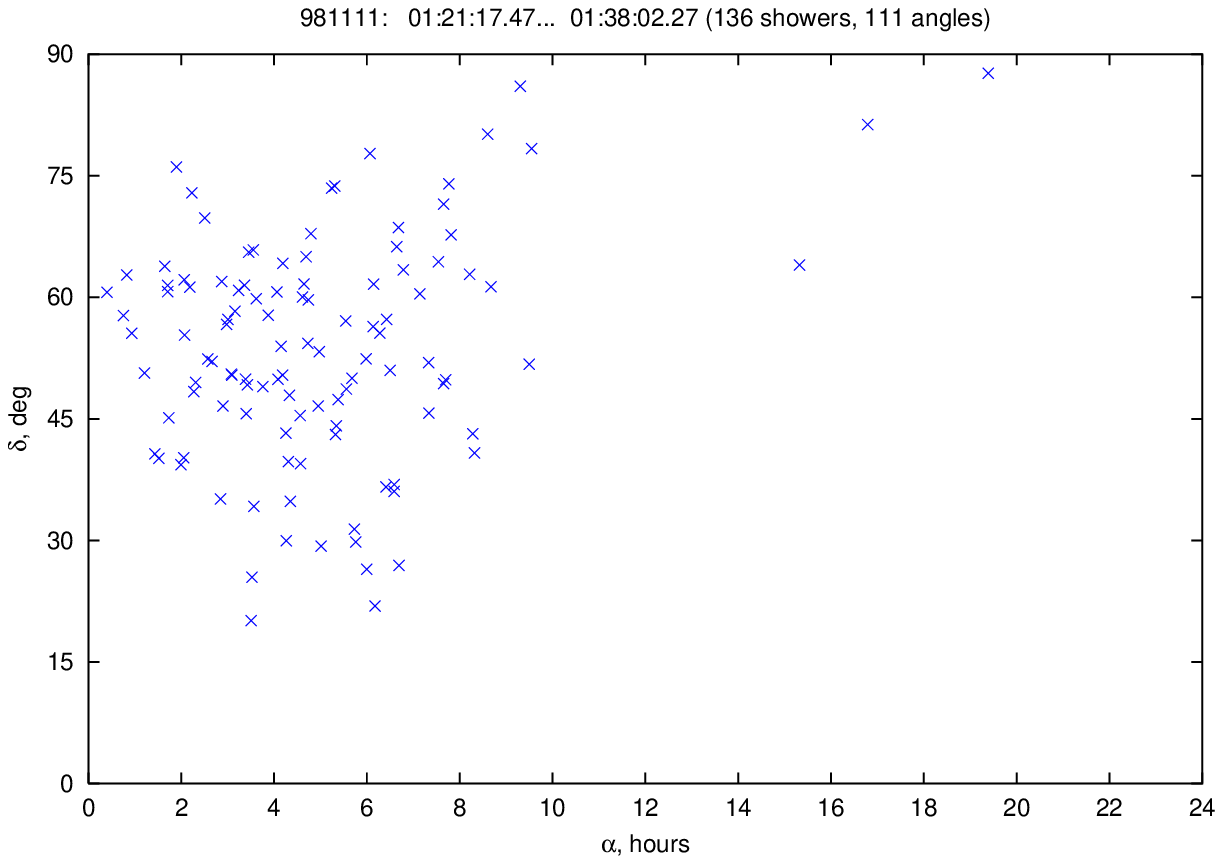} \\[1cm]
\includegraphics[width=12cm]{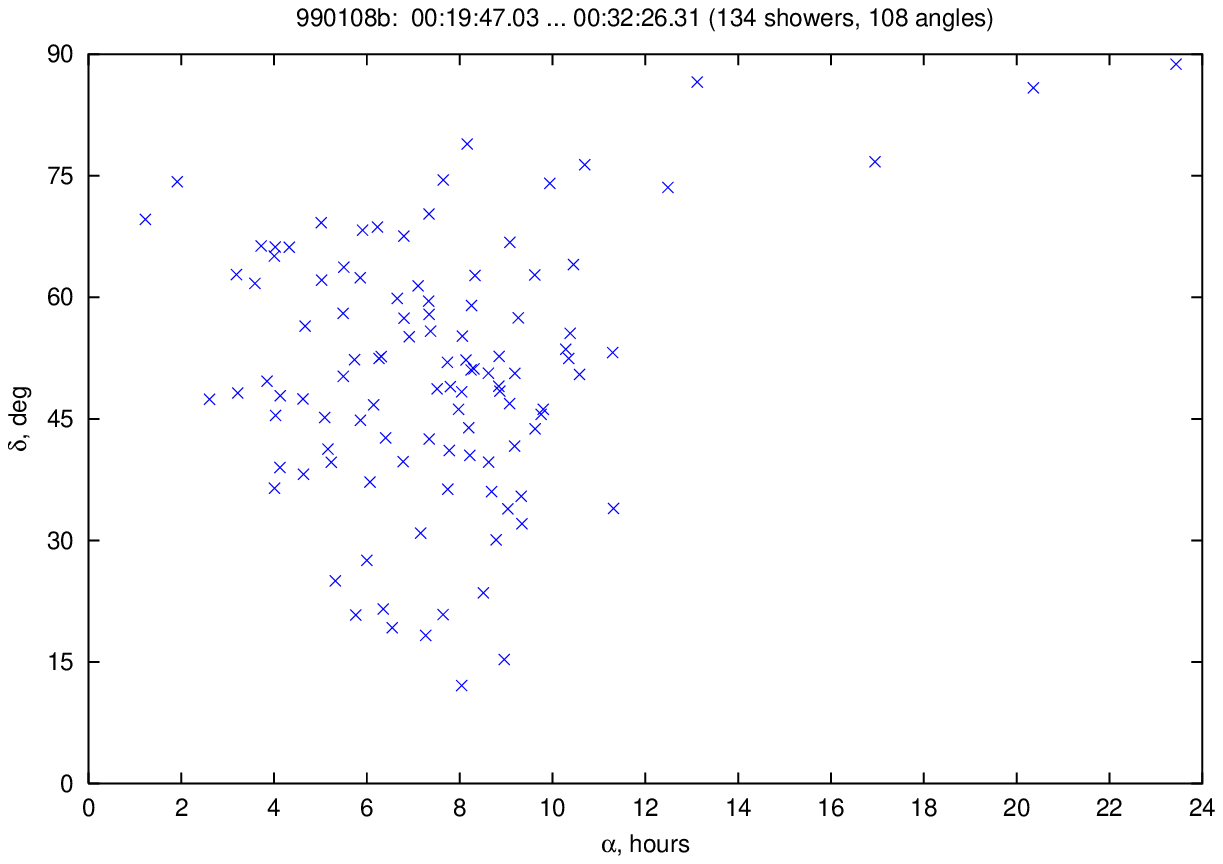}
\end{center}
\caption{Arrival directions of EAS in two longest events
(11.11.98 and 08.01.99(b), see Tables~2b and~2c).
One can see pairs and triplets of showers with almost
coincident arrival directions.
}
\label{Fig:dirsB}
\end{figure}

     As soon as one finds an EAS cluster, a natural question arises:
     Do any of the showers that form a cluster have the same
     astrophysical source, i.e., do they have the same arrival
     direction?
     As one can see from Figs.~\ref{Fig:dirsA}, \ref{Fig:dirsC},
     and~\ref{Fig:dirsB},
     the majority of showers in a cluster do not have a joint source.
     (We recall that Japanese researchers have come to a similar
     conclusion~\cite{Jap91,Jap93,Jap95}.)
     On the other hand, one can often find pairs and
     sometimes even triplets of showers that have very close or
     coincident arrival directions.%
\footnote{
     It is interesting to note that the list of AGASA events
     with the primary particle energy $>4\times10^{19}$~eV
     also contains several pairs and a triplet of EAS with close
     arrival directions~\cite{AGASA}.
}
     The number of such groups grows for clusters that
     consist of a sufficiently large number of showers.
     Figs.~\ref{Fig:dirsA} and~\ref{Fig:dirsC} show how the number of
     EAS with close arrival directions grows if an interior cluster is
     compared with an exterior one.
     This leads us to the conjecture that a part of showers in a
     cluster may have common sources.
     Nevertheless, we must remark that a similar situation sometimes
     takes place for samples that do not form a cluster.
     Thus this is an open question to be clarified in a future
     investigation: Do pairs and triplets of showers with close
     arrival directions found in some clusters have purely
     statistical origin or not?

\begin{figure}
\begin{center}
\includegraphics[width=12cm]{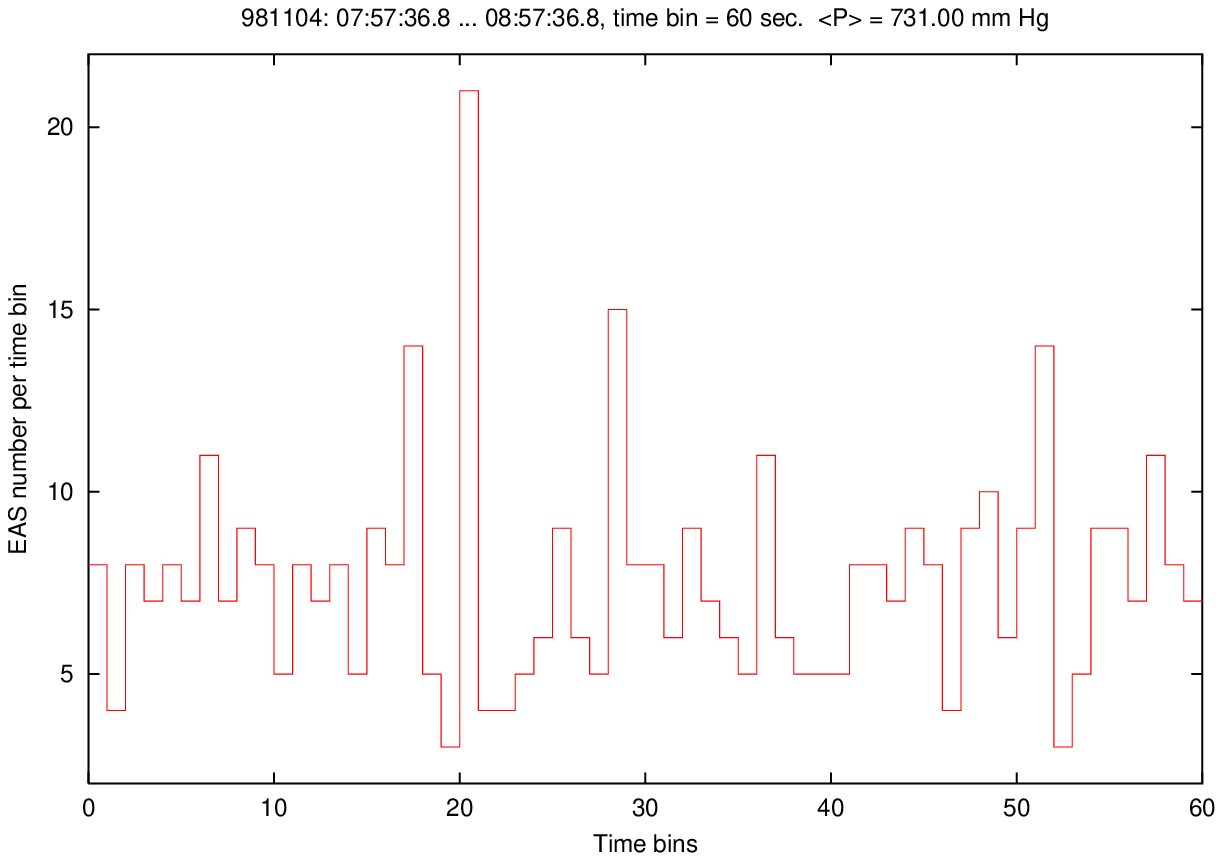} \\[1cm]
\includegraphics[width=12cm]{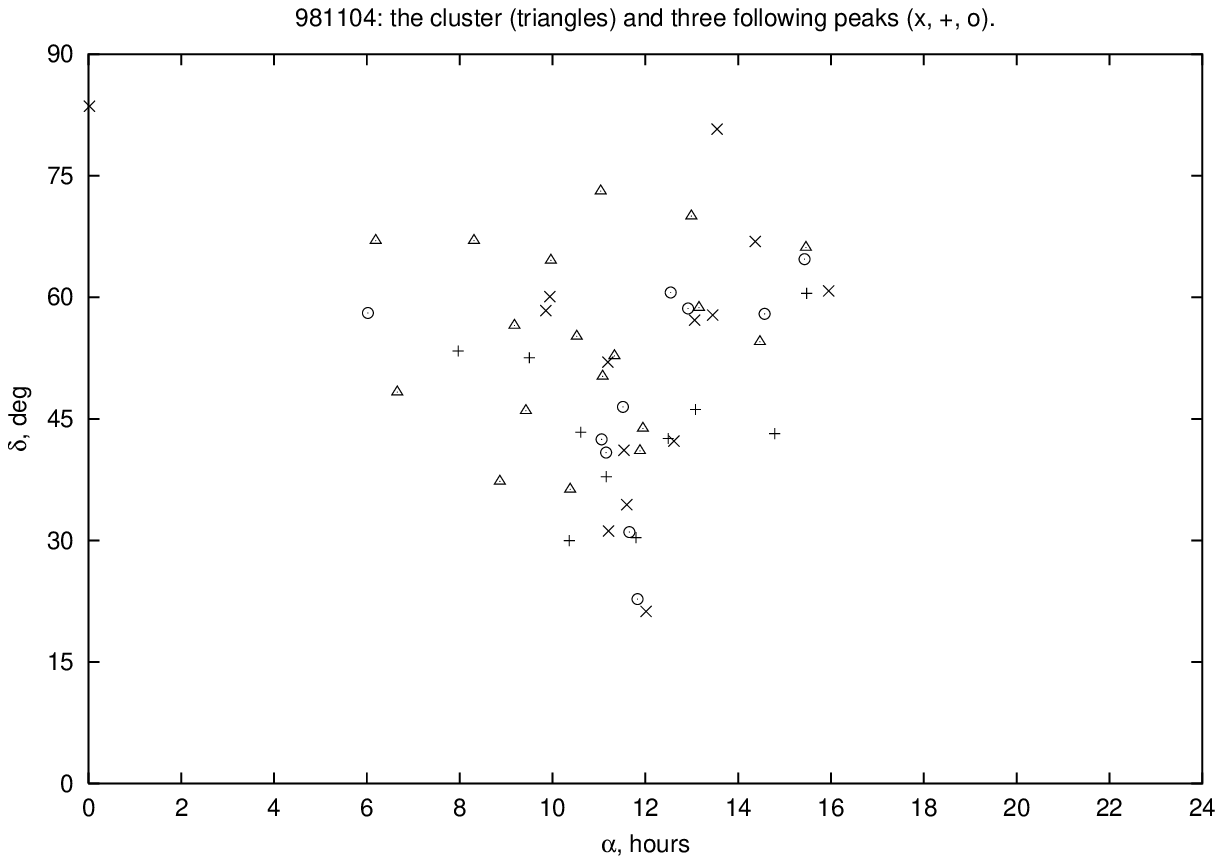}
\end{center}
\caption{%
An event registered on November~4, 1998, cf. Fig.~\ref{Fig:981104}.
The top plot: the count rate during an hour time interval.
The bottom plot: arrival directions of EAS that constitute the
cluster~($\vartriangle$)
and three peaks located at 29th, 37th, and 52th bins
($\times$, $+$, and $\circ$ respectively).}
\label{Fig:981104after}
\end{figure}

     It should be marked that as a rule the clusters represent
     single ``bursts" of the count rate in a sense that there are no
     other bursts around.
     In view of this fact, an event with some kind of an ``afterglow"
     seems to be especially interesting.
     It has been registered on November~4, 1998 (one of the events
     with a ``simple embedding," see above).
     As it is clearly seen from Fig.~\ref{Fig:981104after}, this cluster
     is accompanied by a sequence of less intensive ``bursts."
     This figure also shows arrival directions of EAS that form the
     cluster and three following peaks (located at 29th, 37th, and 52th
     bins).
     One can see that the cluster has showers that are close to the
     showers in the first and the third peaks; these peaks also have EAS
     with close arrival directions.
     It is also interesting that time intervals between the peak  that
     contains the cluster and three highest peaks that follow it are
     equal to~7, 7, and 14~minutes.

     One can also see a ``burst" that precedes the cluster.
     It also contains showers that have arrival directions close
     to some of the EAS in the following peaks.
     In our opinion, this may be a manifestation of some
     astrophysical event.

\begin{figure}
\begin{center}
\includegraphics[width=12cm]{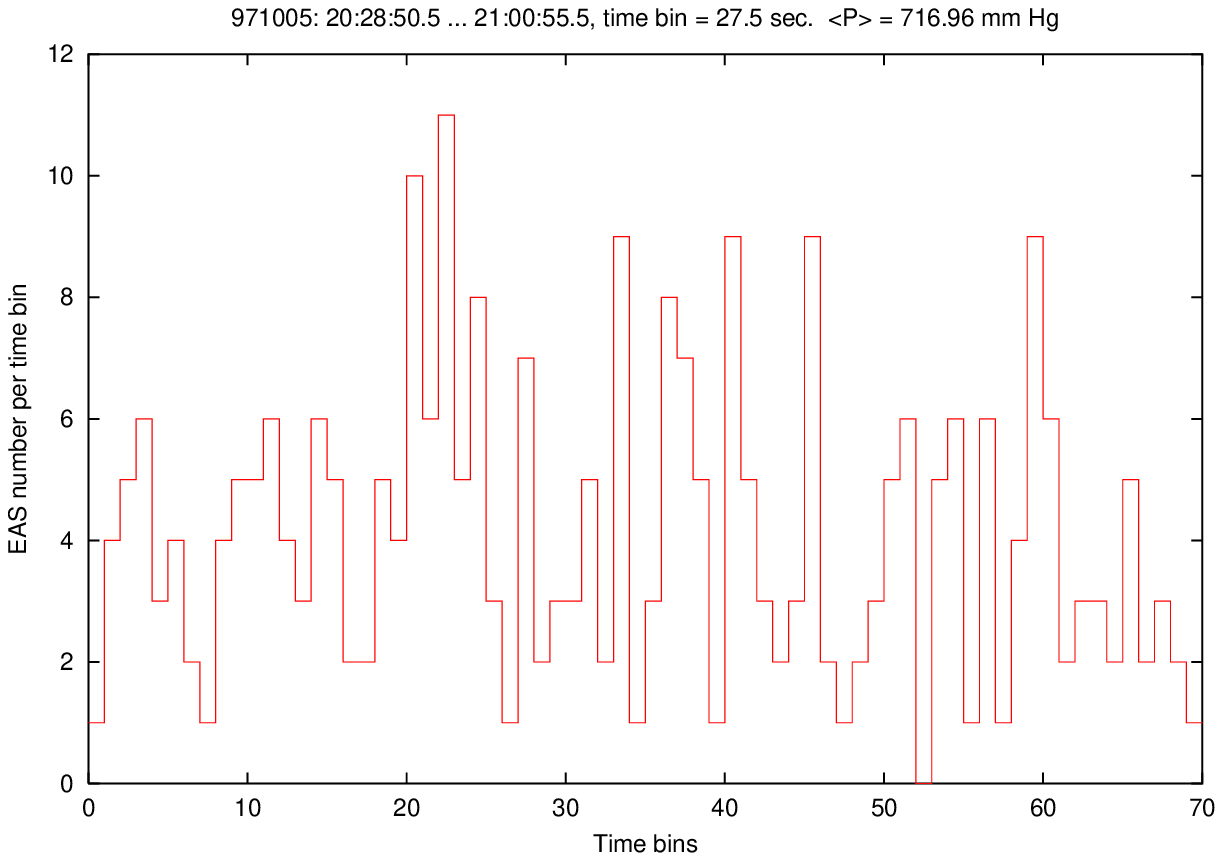} \\[1cm]
\includegraphics[width=12cm]{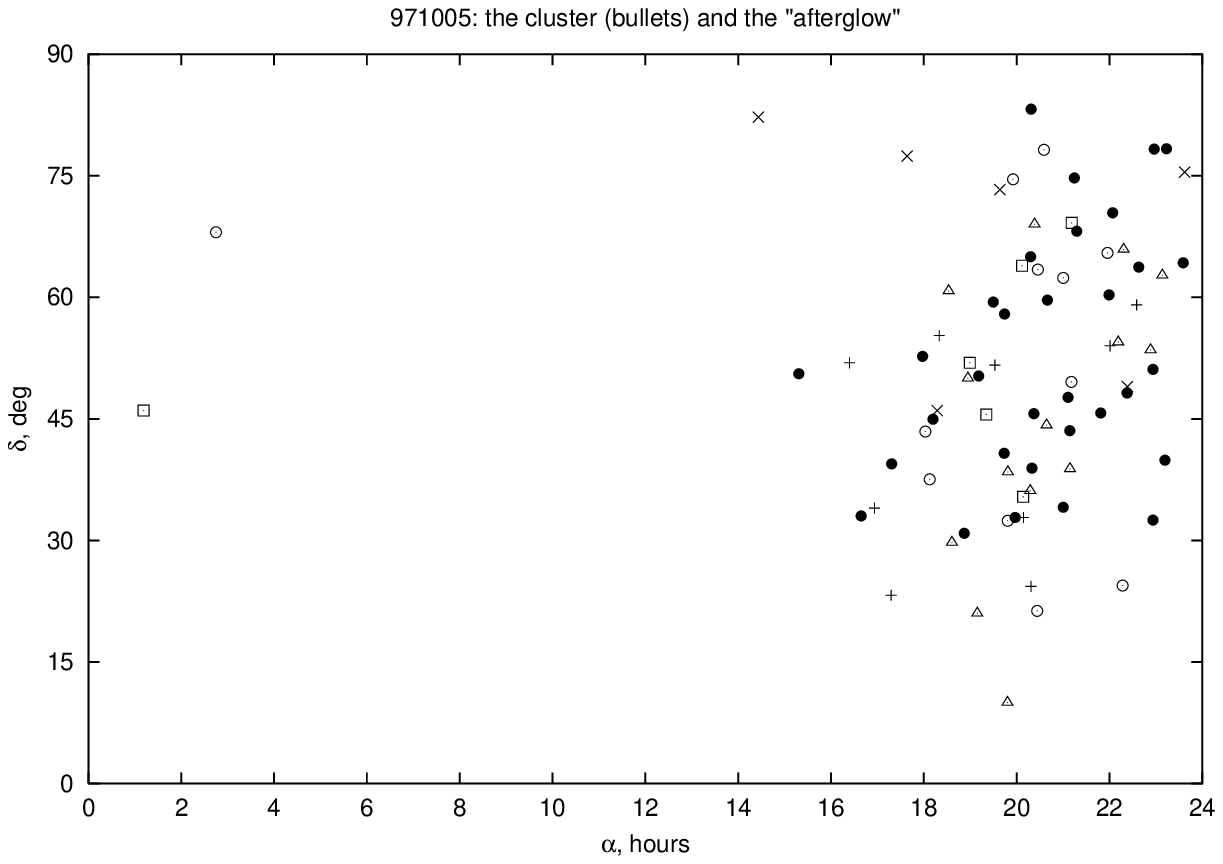}
\end{center}
\caption{%
An event registered on October~5, 1997, cf.\ Fig.~\ref{Fig:971005}.
The top plot: the count rate during a 32-minute time interval;
the cluster occupies bins No.~20--25.
The bottom plot: arrival directions of EAS that constitute the
cluster~($\bullet$) and those that belong to bins No.~28~($\times$),
34~($+$), 41~and 42~($\circ$), 46~($\Box$), 60~and 61~($\vartriangle$).
}
\label{Fig:971005after}
\end{figure}

	A similar phenomenon can be observed for the event registered
	on October~5, 1997, see Table~2a and Fig.~\ref{Fig:971005}.
	This time, the ``afterglow" is observed if one studies the
	``interior structure" of the cluster, see Fig.~\ref{Fig:971005after}.
	Namely, if we consider the count rate at a time interval that contains
	the cluster using comparatively small time bins (approximately in
	the range 25--55~sec) then we observe a sequence of peaks that
	follow the cluster.
	It is remarkable that some of EAS that form these peaks have
	arrival directions close to those of the EAS that form the cluster
	(arrival directions were determined for~32 of 40~showers in the
	cluster).
	Namely, as one can see from Fig.~\ref{Fig:971005after}, at least ten
	of the EAS in the cluster have close counterparts from the following
	peaks: there are six pairs, three triplets, and a quadruplet.
	It is interesting that if we add arrival directions of EAS that
	form peaks at bins No.~37--39 then we find only one more pair.
	It is also important that if we compare arrival directions of EAS
	that form the cluster with arrival directions of the following
	150~EAS (thus covering all 9-shower peaks shown in
	Fig.~\ref{Fig:971005after}) then we find that almost all
	showers with arrival directions close to that of the cluster
	belong to the selected peaks.
	In our opinion, the observed sequence of short ``bursts"
	can be caused by a common astrophysical process.

\begin{figure}
\begin{center}
\includegraphics[width=12cm]{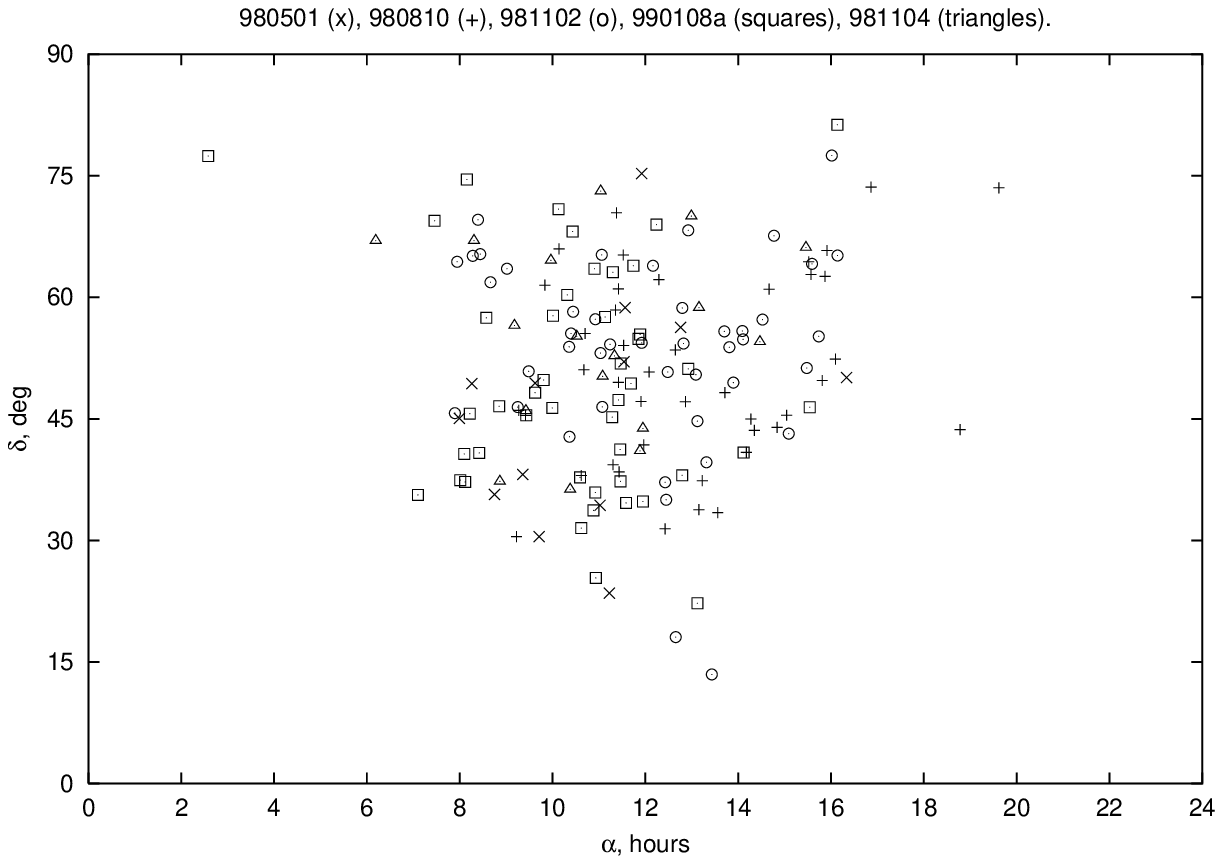} \\
\includegraphics[width=12cm]{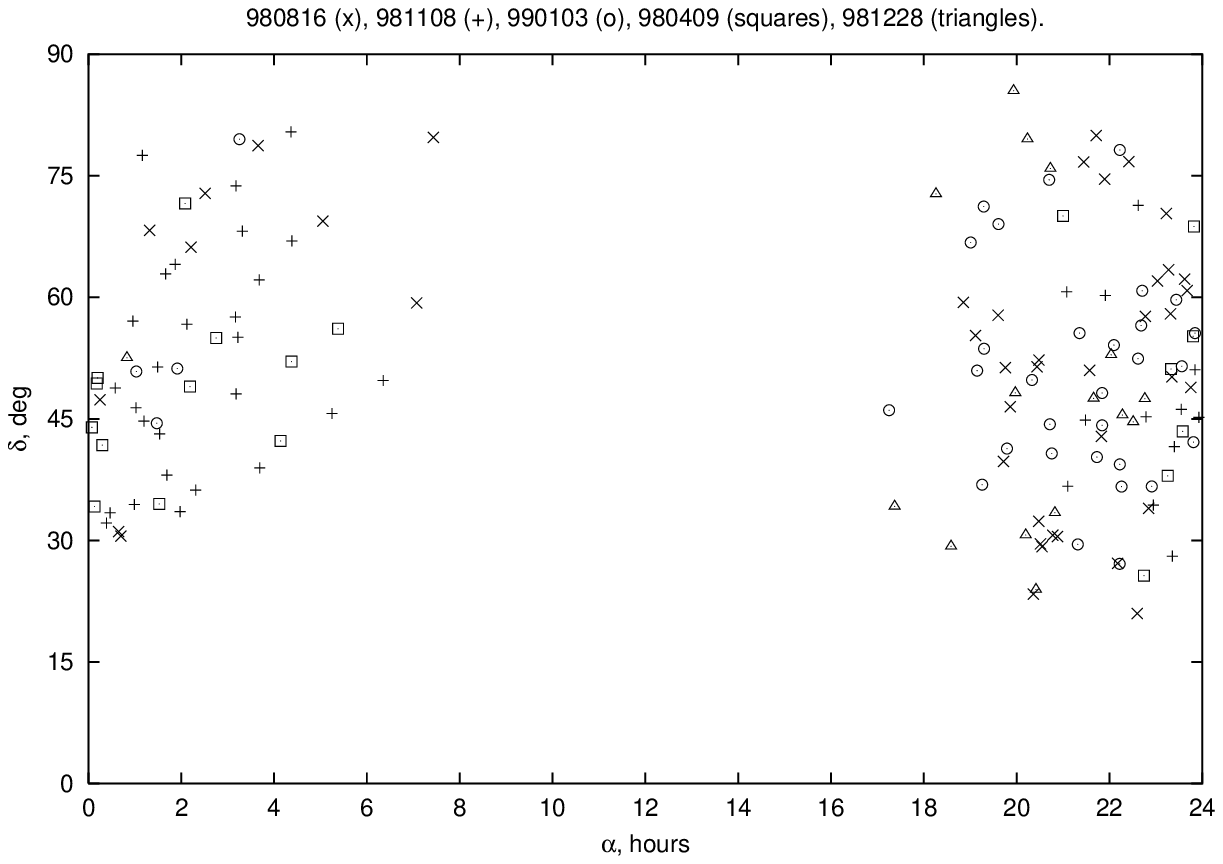}
\end{center}
\caption{Close arrival directions of EAS in different clusters.
}
\label{Fig:different}
\end{figure}

     Another interesting phenomenon revealed by our analysis is the fact
     that showers that form \emph{different} clusters can also have close
     arrival directions.
     For instance, we have found numerous showers with nearly coincident
     arrival directions in the cluster events registered on
     01.05.98, 10.08.98, 02.11.98, 04.11.98, 08.01.99(a) and
     09.04.98, 16.08.98, 08.11.98, 28.12.98, 03.01.99,
     see Fig.~\ref{Fig:different}.
     Notice in particular that there are five showers with very close
     arrival directions in the upper plot (around $\alpha = 9.43$~hour,
     $\delta=46^\circ$).
     In our opinion, this finding can give a clue to the search of
	point sources of cosmic rays with the energy $\sim10^{15}$~eV.

	Finally, a few remarks concerning the electron number~$N_e$ in
	EAS that form clusters are in order.
	As we have already mentioned above, the mean value~$\bar N_e$ for
	the whole data set under consideration approximately equals
	$1.2\times10^5$.
	For the clusters presented in Tables~2a--c,
	$\bar N_e$ has the same order of magnitude.
	The most energetic showers were found within clusters registered
     on November~11, 1998 and January~6, 1999(b) with
     $N_e \approx 10^6$ particles.
	Slightly less powerful showers with $N_e \approx 5\div7\times10^5$
	were detected within clusters registered on December~24 and~28, 1998,
	November~8, 1998, and January~3, 1999.


\section{Discussion}\label{sec:disc}

     It is interesting to compare the results of our investigation
     with results obtained at other experimental EAS arrays.
     Unfortunately, we have found very few results of similar
     investigations undertaken during the same period of data collection.
     The authors of~\cite{SLC_Japan} studied sequences
     of EAS arrival times that demonstrated chaotic behavior.
     Among the data they present, we find two events that were
     registered during the days which are also present in our data
     set (namely, those registered on 15.05.98 and 01.08.98).
     A similar investigation presented in~\cite{Chaos_Japan} also
     contains an event observed on a day present in our data (11.04.98).
     A comparison did not reveal any correlation between these events
     and our results of clusters search.
     This seems to be quite natural since the energy of the events under
     consideration is comparatively low.
     Finally, the results presented by the LAAS group~\cite{LAAS01}
     contain seven successive air shower (SAS) events observed during the
     period covered by our data set.
     SAS events are defined as sequences of EAS detected within a short
     time interval and thus may be called clusters in our terminology.
     It is interesting that two of the SAS events have arrival time
     comparatively close to that of the clusters we have found on
     August~16, 1998 and January~8, 1999.
     A question whether there may be some correlation between these
     events needs to be studied in details.

\begin{figure}[!t]
\begin{center}
\includegraphics[width=14cm]{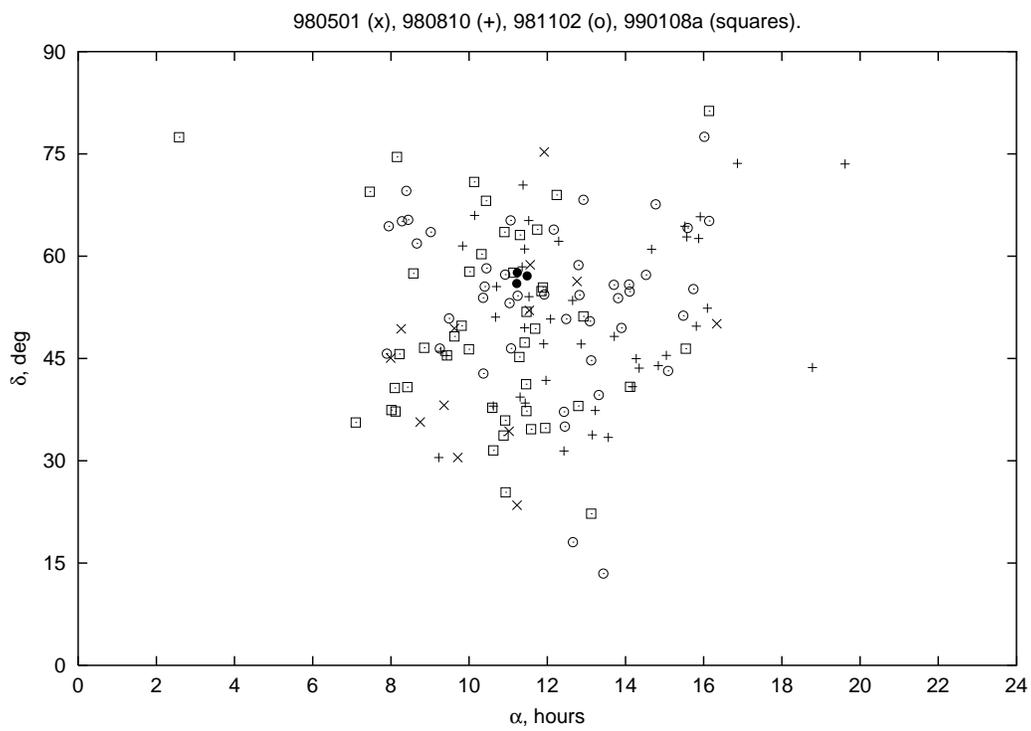} \\
\end{center}
\caption{Arrival directions of EAS in four clusters
and the C2 cluster of AGASA events~($\bullet$)~\cite{AGASA}.
}
\label{Fig:AGASA}
\end{figure}

     Besides this, we have studied ultrahigh energy events
     ($>4\times10^{19}$~eV) registered with the AGASA array
     either~\cite{AGASA}.
     Again, two of the AGASA events were registered during the days
     that belong to our data set (04.04.98 and 27.10.98).
     For these days, our data contain a number of EAS
     with arrival directions close to the arrival directions of
     the corresponding AGASA event, but there are no coincidences
     between these events and EAS clusters we selected, since these
     days do not contain EAS clusters.
     Still, we have found that there is some interesting correlation
     between arrival direction of the C2 triplet of the AGASA events
     and arrival directions of showers in several EAS clusters,
     see Fig.~\ref{Fig:AGASA}.
     It is worth recalling that this AGASA triplet is located near the
     supergalactic plane in the direction of the Ursa-Major~II cluster
     of galaxies and near such astrophysical objects as NGC~3642,
     Mrk~40, and Mrk~171~\cite{AGASA}.

     Finally, we have compared our EAS clusters with the gamma-ray
     bursts (GRBs) registered by BATSE~\cite{BATSE}.
     In total, we have found 78 GRBs in the BATSE Catalog that
     have a positive value of declination~$\delta$ and which
     were registered during the days present in our data set.
     (One of them has also been analyzed within Project
     GRAND~\cite{GRAND}, namely the one registered on April~20, 1998.)
     Nine of these GRBs were observed during the days when EAS clusters
     were registered, though four of them have arrived before the
     corresponding cluster.

     To obtain a more detailed understanding of possible correlation
     between GRBs and EAS, we have analyzed our data looking for EAS that
     have arrival times close to the corresponding GRB and arrival
     directions that deviate from the GRB coordinates at most
     by~$6^\circ$ in declination~$\delta$ and by $6/\cos(\delta)$ in
	right ascension~$\alpha$ (the division by the cosine was used
	to have an equivalent area of the window for all~$\delta$).
     We have found that for 62~GRBs there is at least one EAS
	with a close arrival direction and registered during the same day.
	(It is necessary to mention that though the arrival direction is
	usually determined for the majority of registered EAS, there are
	always a number of showers with an unknown arrival direction.
	Thus the number of EAS with arrival directions close to the
	corresponding GRB may be bigger.)
     For~31 GRBs there are more than 60~EAS with close arrival
     directions, and 16~GRBs have more than 100 close EAS (up to~137).
     Among them, there are 11 GRBs that have one or more EAS
     counterparts within a 200-second interval.
     For six of these events, the deviation in coordinates
     does not exceed 2--3$^\circ$.
     Recall that the geometry of the EAS-1000 prototype array allows
     one to determine arrival directions of EAS with an error of the
     order of~$5^\circ$.
     Hence the difference in arrival directions of these GRBs and the
     corresponding EAS is really small.

     It is interesting to note that for the above mentioned 62~GRBs
     the corresponding sequences of showers often contain a number of
     doublets and even triplets of EAS registered within a short
     period of time.
     This finding gives rise to another problem to be studied, namely
     to analyze the distribution of arrival times of these EAS.
     This may give useful information for the search of point sources
     of both EAS and GRBs.

     Concerning the EAS clusters discussed above, we have found that the
     cluster registered on January~2, 1999 contains a shower with the
     arrival direction close to that one of the GRB No.~7293:  the GRB
     has $\alpha=277^\circ$, $\delta=39^\circ$, while the corresponding
     EAS has $\alpha=271^\circ$, $\delta=45^\circ$.
     On the other hand, there are no EAS that have arrival directions close
     to that of the GRB and registered at the time of the GRB observation.

     The most interesting GRB/EAS-cluster pair is the one registered on
     December~28, 1998.
     The first shower in the EAS cluster has arrived approximately
     one minute before the trigger time of the GRB No.~7285.
     In this connection, recall that the BATSE Catalog, besides other
     information, provides ST90 (start time for the T90 interval relative
     to the trigger time) and T90 (the duration of time after ST90 which
     includes~90\% of the total observed counts in the GRB).
     For the GRB No.~7285, ST$90=-77.824$~sec and T$90=156.096$~sec with
     the uncertainty equal to~$1.590$~sec.
     Thus this GRB begins by 14.6~sec earlier than the EAS cluster, and
     the time interval of the GRB observation contains the time interval
     of the EAS cluster registration.

     In connection with this pair, let us also mention that the
     GRB No.~7285 has the equatorial coordinates $\alpha = 286^\circ$
     and $\delta = 31^\circ$.
     In its turn, the EAS cluster contains a shower with
     $\alpha = 279^\circ$ and $\delta = 29^\circ$.
     Taking into account the precision of the EAS--1000 prototype array,
     we assume that these arrival directions are close.

     We would also like to mention a situation that takes place with the
     EAS cluster registered on 11.11.98 (see Table~2b and the discussion
     above) and the GRB No.~7207.
     This GRB was registered more than eight hours later than the EAS
     cluster (the trigger time equals 09:43:40.90).
	The EAS data set does not contain any shower with an arrival direction
	close to this GRB within an hour time interval.
	As for the cluster, it consists of 136~EAS, and the arrival directions
	are known for~111 of them (see Fig.~\ref{Fig:dirsB}).
	To compare, 353~EAS were registered within the hour 01:00\dots02:00
	(Moscow local time), the arrival directions were determined for~299
	of them, 13~of them are close to that of the GRB, and 8~belong to the
	cluster.
	Thus we see some excess of EAS with arrival directions close to
	the GRB within a cluster.
	It is interesting, whether this can be a manifestation of an
	astrophysical process.

     Thus we think that there may be certain correlation between
     GRBs and EAS clusters and even single EAS generated by cosmic
     rays with the energy $\sim10^{15}$~eV.
     A more detailed analysis of the arrival directions and
     moments of registration of EAS and the corresponding GRBs
     is planned to be carried out to verify this hypothesis.


\section{Conclusion}

     The analysis of arrival times of extensive air showers
     registered with the EAS--1000 prototype array has revealed a
     number of events that may be identified as clusters.
     The discovery of these events gives rise to numerous questions in
     both statistics and phenomenology.
     First of all, it is a question about possible sources and
     physical reasons of cluster appearance, and their
     connection with other astrophysical phenomena.
     The presented results demonstrate that there may be a correlation
     between some GRBs, ultrahigh energy events, and clusters of EAS with
     mean electron number of the order of~$10^5$ particles.

     One of the most promising approaches in future study of EAS
     arrival times and in particular the clusters we have already
     found is an application of methods of the nonlinear
     time series analysis (see,
     e.g.,~\cite{EckmannRuelle,Theiler,KaS,Sch99,MP}).
     A preliminary investigation based on fractal analysis approach has
     revealed that some of EAS clusters demonstrate signs of chaotic
     dynamics.
     These results will be presented elsewhere~\cite{We:chaos}.

     A separate problem is connected with the distribution of arrival
     times of showers that form a cluster.
     As it was shown above, the position of the highest count rate
     may vary in different clusters.
     By the moment, we did not find a statistical distribution that
     is valid for all clusters.
     A study of this question is in the plan of our future research,
     especially in connection with the clusters that contain
     a sufficiently big number of EAS.

     Finally, our investigation has also revealed a number of
     superclusters---EAS clusters that have duration more than
     30~min.
     As a rule, they do not contain short embedded clusters.
     But two of the clusters given in Tables~2b and~2c belong to the
     corresponding superclusters.
     A detailed analysis of superclusters may be the subject of a
     separate investigation.


\section*{Acknowledgments}

     We gratefully acknowledge numerous useful discussions with
     A.~V.~Igoshin, A.~V.~Shirokov, and V.~P.~Sulakov who have helped
     us a lot with the data set we used.
     We also thank O.~V.~Vedeneev who took an active participation at
     the first part of this investigation.
     M.Z. thanks Jonathan Drews for his invaluable advice concerning
     plotting in Octave.

     This work was done with financial support of the Federal
     Scientific-Technical Program for 2001, ``Research and design in
     the most important directions of science and techniques for
     civil applications," subprogram ``High Energy Physics," and
     by Russian Foundation for Basic Research grant No.\
     99--02--16250.

     Only free, open source software was used for this investigation.


\nonfrenchspacing
\newpage

\begin{figure}
\centerline{
\includegraphics[width=8.4cm]{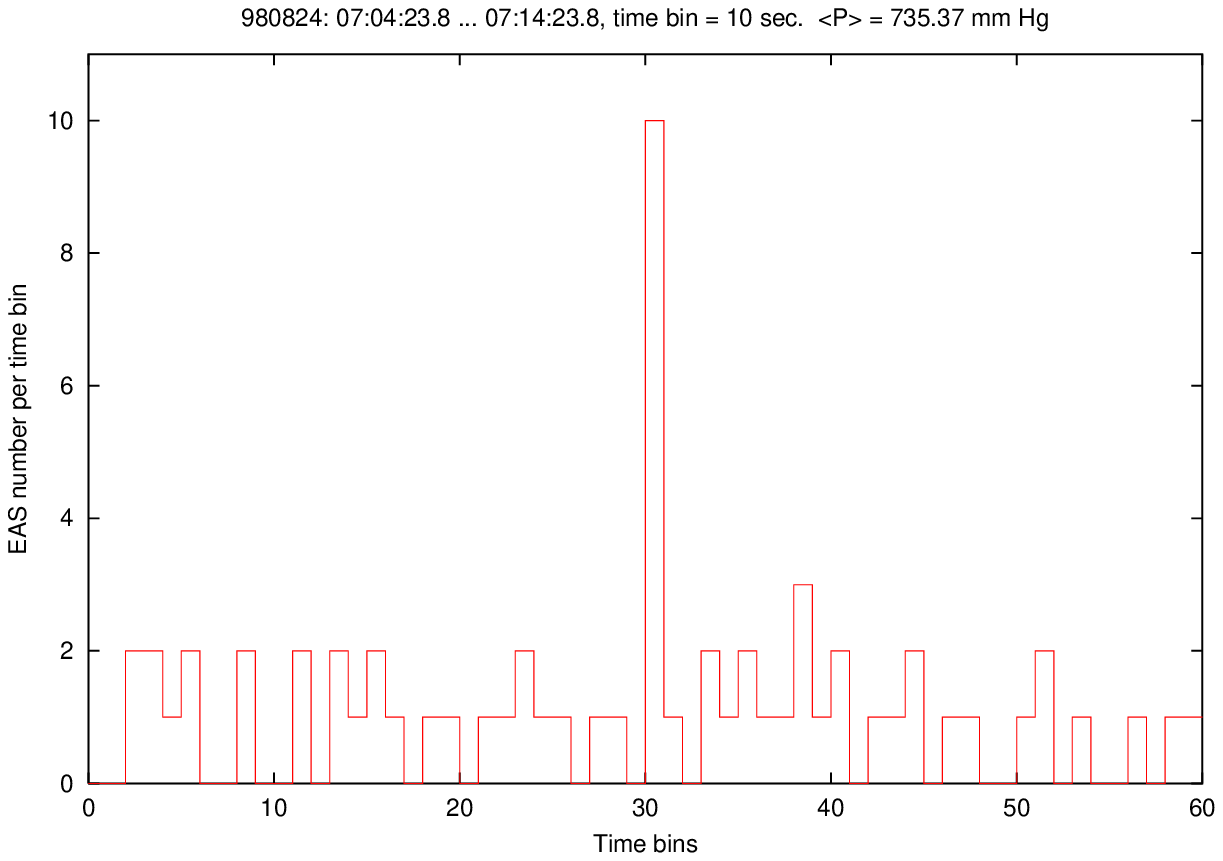}
\includegraphics[width=8.4cm]{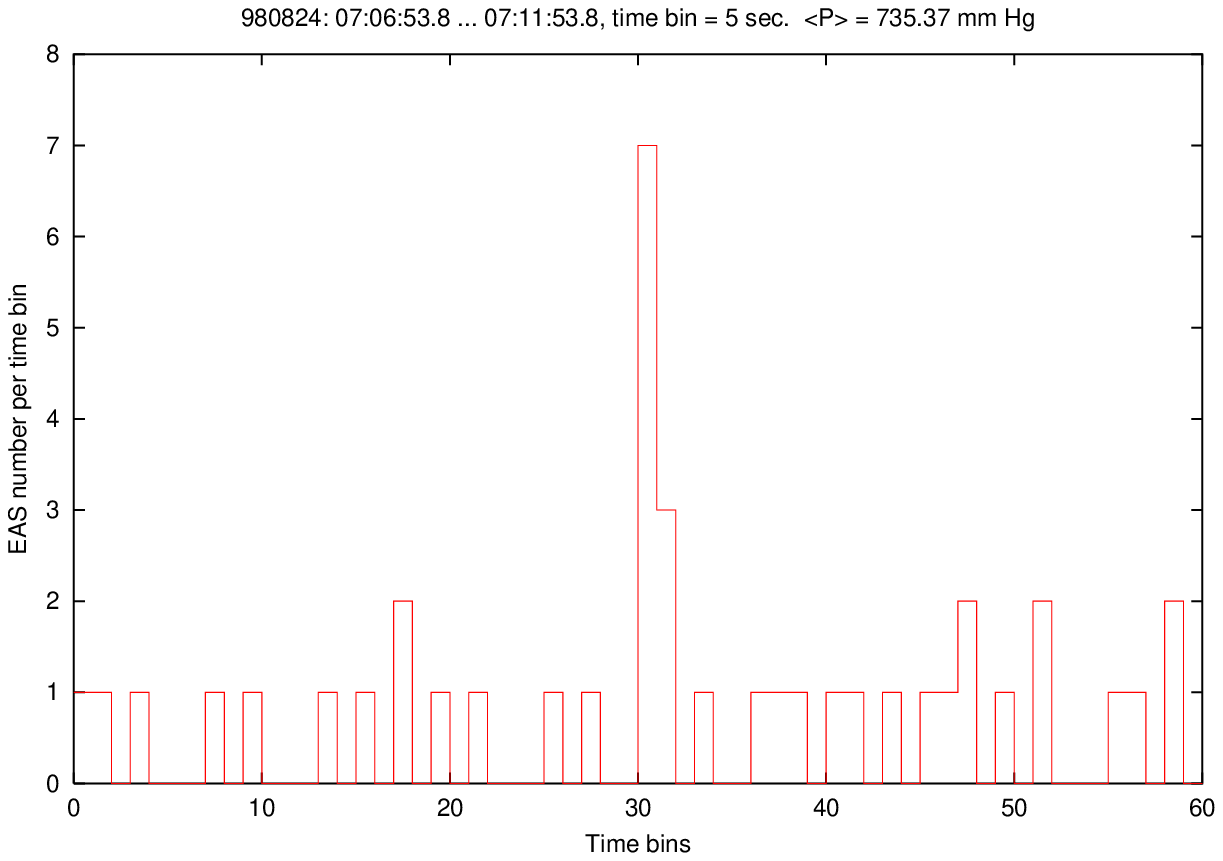}
}
\centerline{
\includegraphics[width=8.4cm]{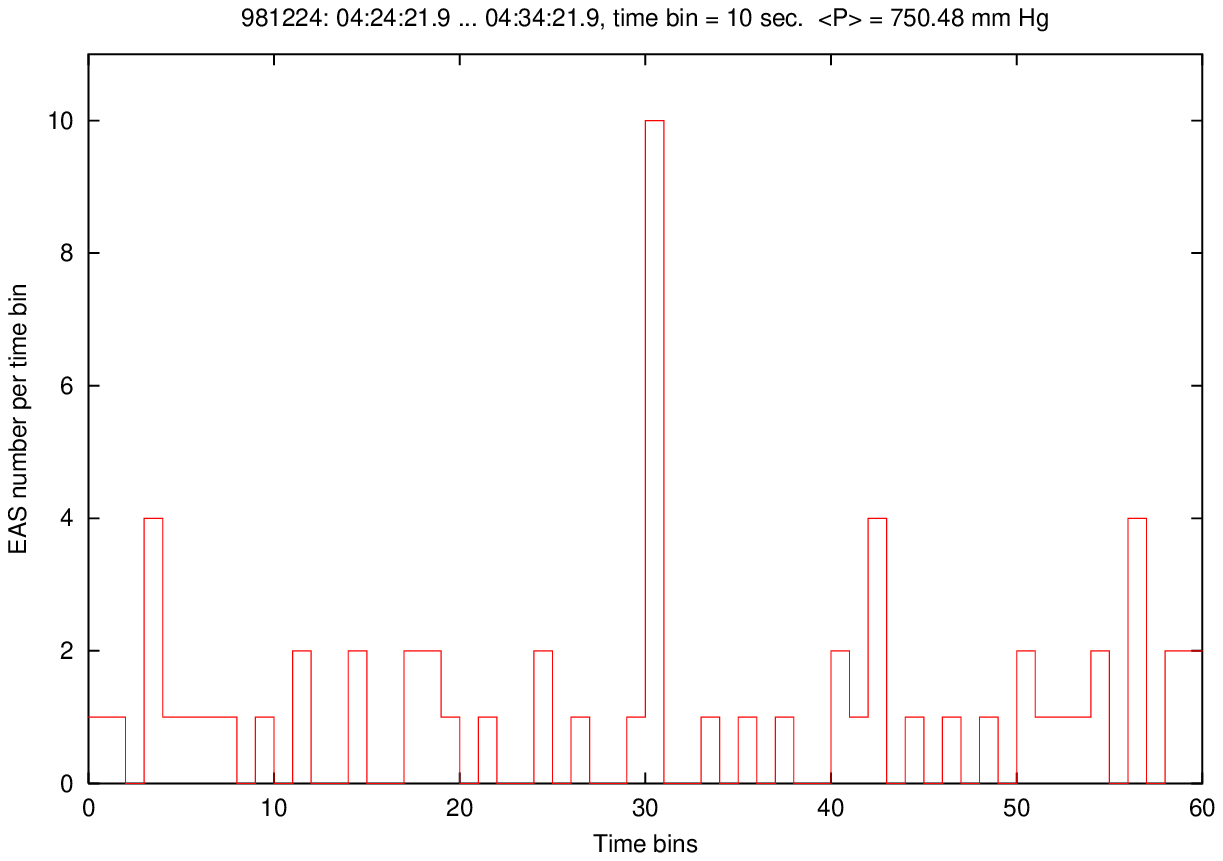}
\includegraphics[width=8.4cm]{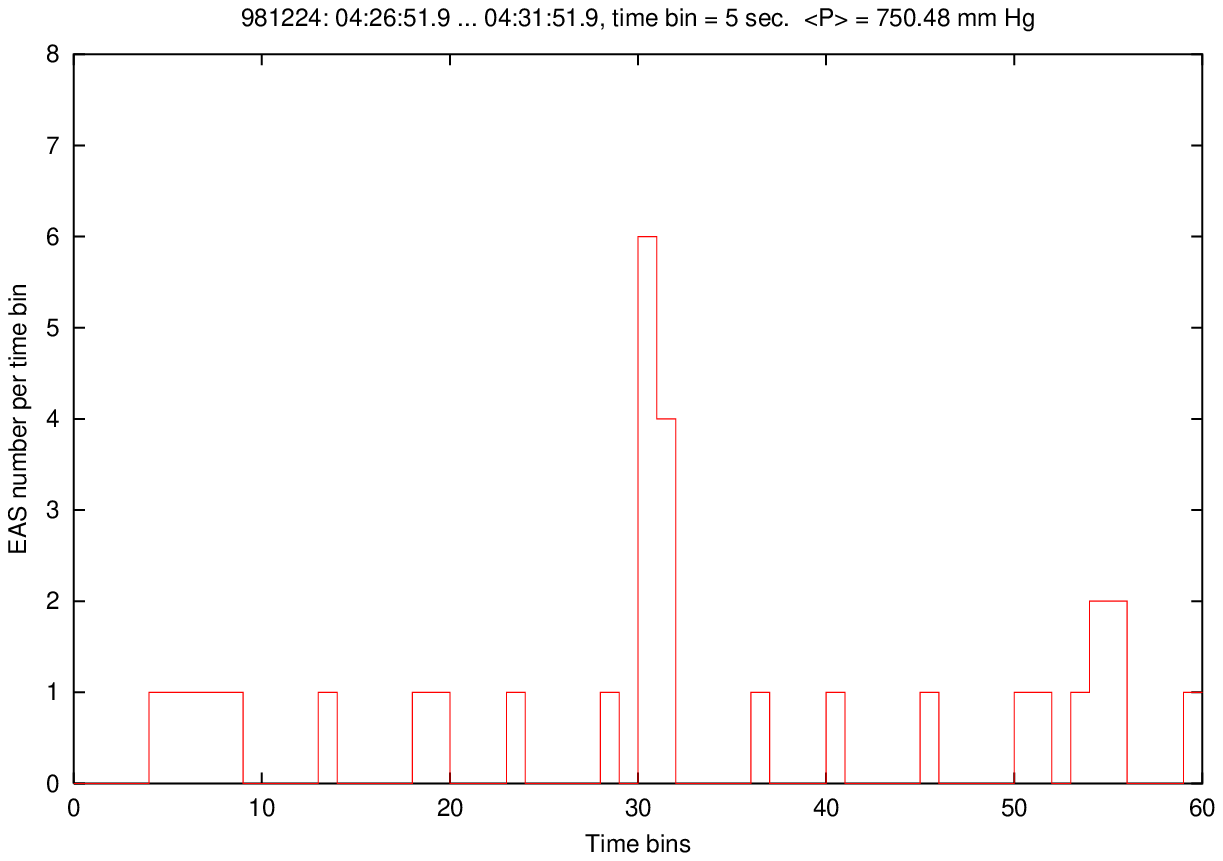}
}
\caption{The count rate during time intervals that contain
clusters registered on August~24, 1998 (top row) and
December~24, 1998 (bottom row), see Table~2a.
The right column shows the 5-second structure of the events.
}
\label{Fig:980824,981224}
\end{figure}
\clearpage
\null
\vfill
\clearpage

\clearpage
\clearpage
\null
\clearpage
\clearpage
\newpage\null
\clearpage
\newpage\null
\begin{figure}[!h]
\centerline{
\includegraphics[width=8.4cm]{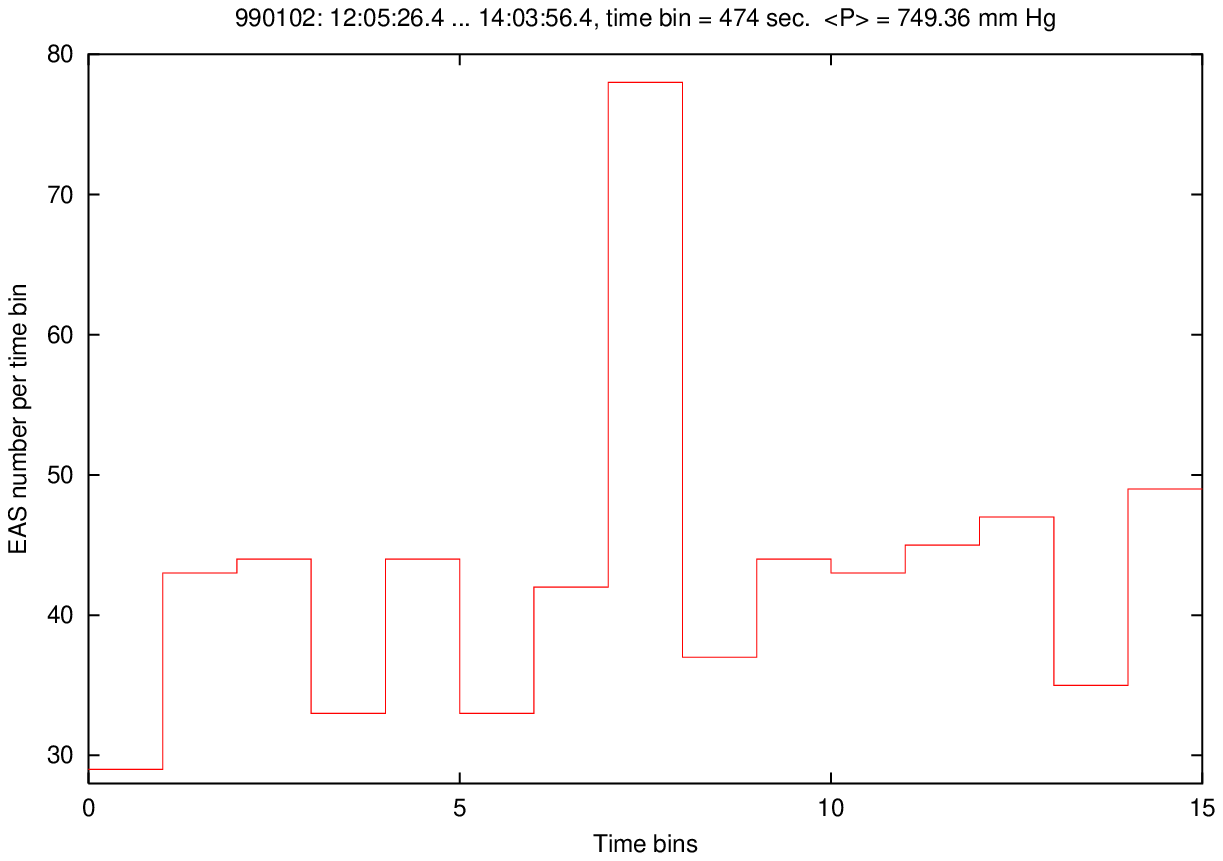}
\includegraphics[width=8.4cm]{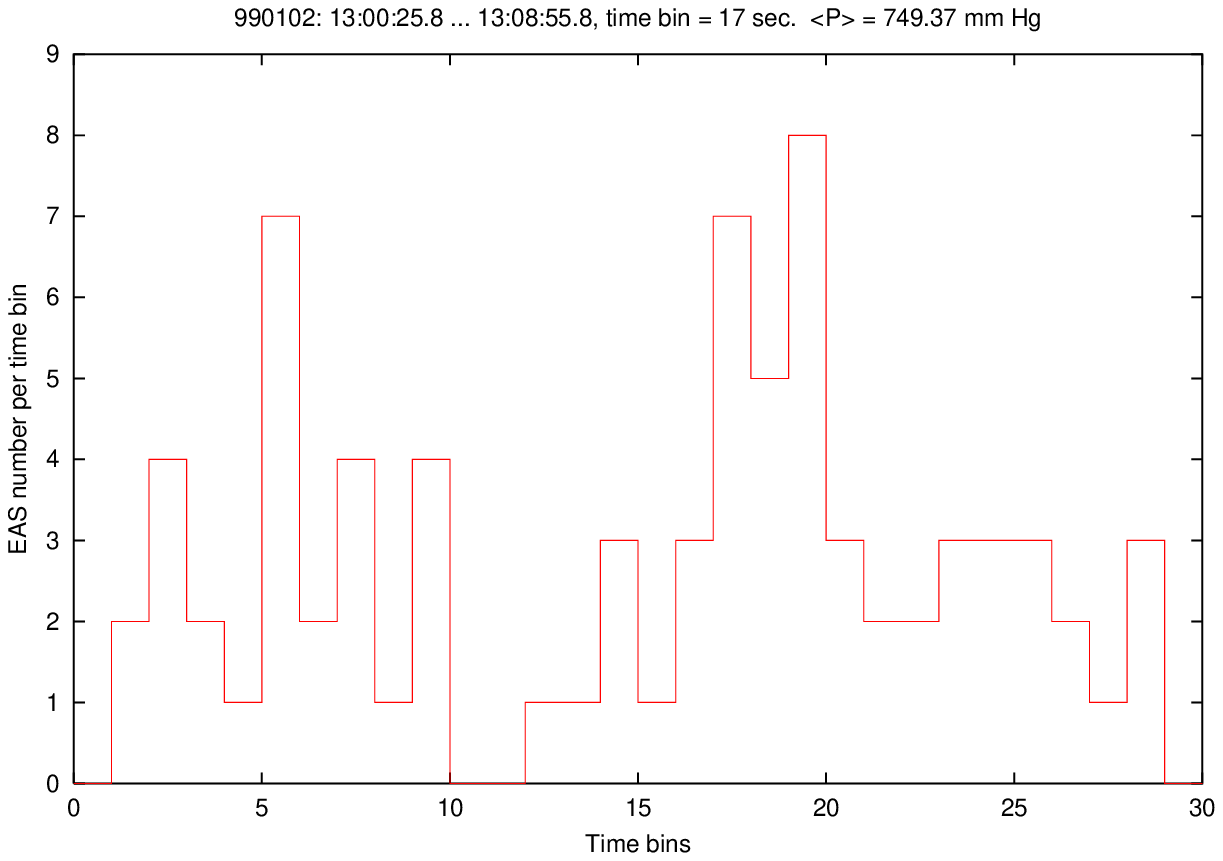}
}
\caption{The widest group of clusters selected for the
event registered on January~2, 1999 (left) and its
``interior structure" (right).
Notice that the first of the clusters selected for $T=4$~min
(see Fig.~\ref{Fig:990102c}) begins at the 6th bin of the right plot,
while the second one begins at the 13th bin.
Bins No.~18--20 represent the first of the clusters selected for
$T=45$, 50~sec (see the right plot in Fig.~\ref{Fig:990102a}).
}
\label{Fig:990102d}
\end{figure}
\newpage
\null
\clearpage
\clearpage

\newpage
\newpage
\newpage
\newpage
\newpage

\end{document}